\documentclass[a4paper,fleqn,usenatbib,useAMS]{mnras}

\usepackage[T1]{fontenc}
\usepackage{ae,aecompl}

\usepackage{hyperref}
\usepackage{graphicx}	
\usepackage{amsmath}	
\usepackage{amssymb}	
\usepackage{bm}            

\def\url#1{\expandafter\string\csname #1\endcasname}
\newlength{\colwidth}
\newcommand{\persec}                    {\,{\rm s}^{-1}}

\newcommand{\peryr}                    {\,{\rm yr}^{-1}}
\newcommand{\Myr}                      {\,{\rm Myr}}

\newcommand{\Gyr}                      {\,{\rm Gyr}}

\newcommand{\pc}                        {\,{\rm pc}}
\newcommand{\kpc}                      {\,{\rm kpc}}
\newcommand{\Mpc}                      {\,{\rm Mpc}}

\newcommand{\pkpc}                      {\,{\rm pkpc}}

\newcommand{\ckpc}                      {\,{\rm ckpc}}
\newcommand{\cMpc}                      {\,{\rm cMpc}}
\newcommand{\Msun}                    {\,{\rm M}_\odot}

\newcommand{\cmsquared}              {{\rm cm}^{-2}}
\newcommand{\cmcubed}              {{\rm cm}^{-3}}
\newcommand{\Zsun}                     {\,{\rm Z}_\odot}

\newcommand{\K}                          {\,{\rm K}}

\newcommand{\Msunyrkpcsq} {\,{\rm M}_\odot\,{\rm yr}^{-1}\,{\rm kpc}^{-2}}

\newcommand{\gadget}         {\textsc{gadget3}}

\newcommand{\HI}         {H\,\textsc{i}}
\newcommand{\HII}         {H\,\textsc{ii}}
\newcommand{\Hmol}       {${\rm H}_2$}

\def\lsim{\mathrel{\lower0.6ex\hbox{$\buildrel {\textstyle <}
 \over {\scriptstyle \sim}$}}}

\title[EAGLE: atomic hydrogen in galaxies]{The EAGLE simulations: atomic hydrogen associated with galaxies}

\author[R.A.Crain et al.]{Robert A. Crain$^{1}$\thanks{E-mail: r.a.crain@ljmu.ac.uk (RAC)},
  Yannick M. Bah\'e$^{2}$, Claudia del P. Lagos$^{3,4}$, Alireza Rahmati$^{5}$,
  \newauthor Joop Schaye$^{6}$, Ian G. McCarthy$^{1}$, Antonino Marasco$^{7}$, Richard G. Bower$^{8}$, 
  \newauthor Matthieu Schaller$^{8}$, Tom Theuns$^{8}$ and Thijs van der Hulst$^{7}$
\\
$^{1}$Astrophysics Research Institute, Liverpool John Moores University, 146 Brownlow Hill, Liverpool, L3 5RF\\
$^{2}$Max-Planck-Institut f\"ur Astrophysik, Karl-Schwarzschild Str. 1, 85748 Garching, Germany\\
$^{3}$International Centre for Radio Astronomy Research (ICRAR), M468, University of Western Australia, 35 Stirling Hwy, Crawley, WA 6009, Australia.\\
$^{4}$Australian Research Council Centre of Excellence for All-sky Astrophysics (CAASTRO), 44 Rosehill Street Redfern, NSW 2016, Australia.\\
$^{5}$Institute for Computational Science, University of Z\"urich, Winterthurerstrasse 190, CH-8057 Z\"urich, Switzerland\\
$^{6}$Leiden Observatory, Leiden University, PO Box 9513, 2300 RA Leiden, The Netherlands\\
$^{7}$Kapteyn Astronomical Institute, Postbus 800, 9700 AV, Groningen, The Netherlands\\
$^{8}$Institute for Computational Cosmology, Department of Physics, University of Durham, South Road, Durham, DH1 3LE
}
   
\date{Accepted XXX. Received YYY; in original form ZZZ}

\pubyear{2015}

\begin{document}
\label{firstpage}
\pagerange{\pageref{firstpage}--\pageref{lastpage}}
\maketitle

\begin{abstract}
We examine the properties of atomic hydrogen (\HI) associated with galaxies in the EAGLE simulations of galaxy formation. EAGLE's feedback parameters were calibrated to reproduce the stellar mass function and galaxy sizes at $z=0.1$, and we assess whether this calibration also yields realistic \HI\ properties. We estimate the self-shielding density with a fitting function calibrated using radiation transport simulations, and correct for molecular hydrogen with empirical or theoretical relations. The `standard-resolution' simulations systematically underestimate \HI\ column densities, leading to an \HI\ deficiency in low-mass ($M_\star < 10^{10}\Msun$) galaxies and poor reproduction of the observed \HI\ mass function. These shortcomings are largely absent from EAGLE simulations featuring a factor of 8 (2) better mass (spatial) resolution, within which the \HI\ mass of galaxies evolves more mildly from $z=1$ to $0$ than in the standard-resolution simulations. The largest-volume simulation reproduces the observed clustering of \HI\ systems, and its dependence on \HI-richness. At fixed $M_\star$, galaxies acquire more \HI\ in simulations with stronger feedback, as they become associated with more massive haloes and higher infall rates. They acquire less \HI\ in simulations with a greater star formation efficiency, since the star formation and feedback necessary to balance the infall rate is produced by smaller gas reservoirs. The simulations indicate that the \HI\ of present-day galaxies was acquired primarily by the smooth accretion of ionized, intergalactic gas at $z\simeq1$, which later self-shields, and that only a small fraction is contributed by the reincorporation of gas previously heated strongly by feedback. \HI\ reservoirs are highly dynamic: over $40$ percent of \HI\ associated with $z=0.1$ galaxies is converted to stars or ejected by $z=0$.
\end{abstract}

\begin{keywords}
cosmology: theory -- galaxies:formation -- galaxies: evolution -- galaxies: haloes
\end{keywords}



\section{Introduction}
\label{sec:introduction}

A comprehensive theory of galaxy evolution requires an understanding not only of the assembly and evolution of stellar discs and spheroids, but also of the co-evolution of these components with interstellar and circumgalactic gas. Besides being subject to gravity and hydrodynamical forces, gas participating in the non-linear process of galaxy formation can be strongly influenced by radiative processes, and thus spans many orders of magnitude in (column) density, temperature and ionization fraction. This diversity of conditions poses a challenge both to the development of numerical simulations capable of reproducing galaxies with realistic stellar and gaseous components, and to the synthesis of multiwavelength observations into an holistic view of the connection between galaxies and circumgalactic gas.

Galaxy formation simulations indicate that the majority of the gas that accretes on to haloes and subsequently fuels star formation (SF), arrives without shock heating to temperatures $\gg 10^4\K$ near the virial radius \citep[e.g.][]{Keres_et_al_05,Dekel_et_al_09,van_de_Voort_et_al_11a}. Cosmological accretion therefore potentially delivers significant quantities of neutral gas to galaxies \citep[e.g.][]{van_de_Voort_et_al_12,Faucher-Giguere_et_al_15}, making atomic hydrogen (\HI, which dominates the neutral hydrogen budget except at the highest column densities) a valuable probe of the gas within galaxies and their circumgalactic medium. Examination of this gas is typically achieved via analysis of the absorption lines of hydrogen and low-ionization metals in quasar spectra but, in the local Universe, the spatial distribution of high column density \HI\ can also be mapped with the hyperfine 21 cm emission line.

Wide-area 21 cm surveys such as the \HI\ Parkes All-Sky Survey \citep[HIPASS,][]{Meyer_et_al_04_short} and the Arecibo Fast Legacy Alfa Survey \citep[ALFALFA,][]{Giovanelli_et_al_05_short} have revealed the population of \HI\ sources in the local Universe, demonstrating that the \HI\ masses of galaxies are, like their stellar masses, well-described by a Schechter function \citep{Zwaan_et_al_05a,Haynes_et_al_11_short}. Targeted follow-up campaigns such as the GALEX-Arecibo-SDSS Survey \citep[GASS,][]{Catinella_et_al_10_short,Catinella_et_al_13_short} and the Bluedisks project \citep[][]{Wang_et_al_13}, adopting selection functions based on optical measurements to mitigate the intrinsic bias of `blind' (i.e. untargeted) surveys towards \HI-rich galaxies, have since revealed the typical \HI\ properties of galaxies and characterized the impact of environment on \HI\ properties \citep{Cortese_et_al_11,Fabello_et_al_12}. Deep targeted surveys, such as HALOGAS \citep{Heald_et_al_11_short}, are beginning to unveil the properties of low column density \HI\ in the outskirts of galaxies.

Deep, blind 21 cm surveys are beginning to push beyond the very local Universe. The Blind Ultra Deep \HI\ Survey \citep{Jaffe_et_al_13} exploits the upgraded Westerbork Synthesis Radio Telescope (WSRT) to examine the \HI\ content of galaxies in two massive galaxy clusters at $z\simeq 0.2$, thus probing more extreme environments than possible with GASS. The COSMOS \HI\ Large Extragalactic Survey \citep[CHILES,][]{Fernandez_et_al_13_short} exploits the upgraded Very Large Array to produce an \HI\ `deep field' spanning $z=0 - 0.45$, probing the \HI\ content of galaxies over an inverval during which the cosmic star formation rate (SFR) density is observed to decrease by a factor of $2.5-3$ \citep[e.g.][]{Hopkins_and_Beacom_06}. Moreover, in advance of the construction of the Square Kilometre Array (SKA), its pathfinder facilities will soon begin to conduct unprecedentedly deep and wide surveys of the 21 cm emission line: the DINGO \citep{Meyer_09} and WALLABY surveys \citep{Koribalski_12} using the Australian SKA Pathfinder and the LADUMA survey \citep{Holwerda_Blyth_and_Baker_12} using MeerKAT. The installation of the Apertif phased array feeds \citep{Oosterloo_et_al_09} on the WSRT also promises to markedly enhance the 21 cm survey capacity in the Northern hemisphere \citep{Verheijen_et_al_09}.

Owing to the dynamic and potentially complex nature of gas flows into and out of galaxies \citep[e.g.][]{Keres_et_al_05,Oppenheimer_et_al_10,van_de_Voort_and_Schaye_12,Muratov_et_al_15}, detailed examination of the evolution of the \HI\ content of galaxies and their immediate environments requires cosmological hydrodynamic simulations, particularly so if seeking to understand the build-up of these reservoirs. Several authors have examined the \HI\ column density distribution function (CDDF) in hydrodynamical simulations \citep[e.g.][]{Popping_et_al_09,Altay_et_al_11,Altay_et_al_13,Bird_et_al_13,R13a,R13b} and the two-point correlation function of \HI\ absorbers \citep{Popping_et_al_09}. \citet{Duffy_et_al_12} and \citet{Dave_et_al_13} presented detailed examinations of the impact of numerical resolution and subgrid physics on the evolving \HI\ properties of galaxies in cosmological volumes; similar exercises werre performed by \citet{Walker_et_al_14} and \citet{Stinson_et_al_15} using zoomed simulations of individual galaxies. \citet{Cunnama_et_al_14} and \citet{Rafieferantsoa_et_al_15} have examined the impact of physical processes related to the cosmological environment on the \HI\ content of galaxies.

Until recently, cosmological hydrodynamical simulations were compromised and/or limited in scope by their inability to produce a realistic population of galaxies, often instead yielding galaxies that were too massive, too compact, too old, and characterized by a galaxy stellar mass function (GSMF) that differed markedly from those observed by galaxy redshift surveys \citep[e.g.][]{Crain_et_al_09_short,Lackner_Ostriker_and_Joung_12}. This study examines the properties of cosmic \HI\ within the `Evolution and Assembly of GaLaxies and their Environments' project \citep{S15,C15}, a suite of cosmological hydrodynamical simulations of the $\Lambda$-cold dark matter ($\Lambda$CDM) cosmogony within which the ill-constrained parameters governing the efficiency of feedback processes have been explicitly calibrated to ensure reproduction of the stellar masses and sizes of galaxies observed at low-redshift. Simulations adopting this calibration also reproduce the observed present-day colours of galaxies \citep{Trayford_et_al_15_short} and the evolution of i) the stellar mass function \citep{Furlong_et_al_15_short}, ii) galaxy sizes \citep{Furlong_et_al_16} and iii) galaxy colours \citep{Trayford_et_al_16}. The EAGLE simulations follow larger simulation volumes than, at comparable or superior resolution to, previous studies focusing on \HI\ in galaxies. They also adopt an advanced hydrodynamics algorithm and do not disable physical processes, such as radiative cooling or hydrodynamic forces, when modelling feedback associated with the formation of stars or the growth of black holes.

The primary aim of this study is to examine the degree to which reproducing key aspects of the present-day stellar component of galaxies also results in them exhibiting realistic \HI\ reservoirs. Such reservoirs represent an instructive test of simulations, because gas exhibiting a high neutral fraction represents the interface between fluid elements whose evolution is governed solely by direct integration (i.e. by the gravity, hydrodynamics and radiative cooling algorithms), and those additionally subject to source and sink terms specified by phenomenological `subgrid routines'. We also explore the sensitivity of \HI\ reservoirs to the efficiency of energetic feedback, the photoionisation rate of the metagalactic ultraviolet/X-ray background (UVB), and numerical resolution. An additional aim is to quantify the contribution of smooth accretion and mergers to the build-up of present-day \HI\ reservoirs, and to chart the history and fate of gas comprising \HI\ reservoirs associated with galaxies at different epochs. These questions cannot be authoritatively addressed with extant observations, and are well-suited to address with numerical simulations.

Neutral gas is the focus of a number of companion studies using the EAGLE simulations: \citet{Rahmati_et_al_15} show that the simulations reproduce the evolution of the cosmic \HI\ mass density, and both the observed \HI\ CDDF and the covering fraction of galaxies at $z \ge 1$. \citet{Bahe_et_al_16_short} found that EAGLE reproduces the mass, radial distribution and morphology of galactic \HI\ in a sample of galaxies similar to those selected by the GASS survey. \citet{Lagos_et_al_15_short,Lagos_et_al_16_short} demonstrate that EAGLE reproduces the \Hmol\ properties of galaxies across a range of redshifts, and elucidated the role of \Hmol\ in establishing the Fundamental Plane of SF. In a companion study, we show that the simulations reproduce key observed relations between the \HI\ content of galaxies and popular diagnostics for describing the cosmic environment of galaxies \citep{Marasco_et_al_16}.

This paper is structured as follows. The details of the simulations, the methods by which galaxies and haloes are identified, and the schemes for partitioning hydrogen into its ionized, atomic and molecular components, are described in Section \ref{sec:methods}. The \HI\ properties of the simulations across many orders of magnitude in column density are explored via the \HI\ CDDF in Section \ref{sec:CDDF}. Key results concerning the mass and physical properties of \HI\ associated with galaxies, and the \HI\ mass function (HIMF), are presented in Section \ref{sec:HI_in_gals}. The clustering of \HI\ sources, and its dependence on the \HI-richness of the sources, is presented in Section \ref{sec:clustering}. Section \ref{sec:reservoir_accretion_and_evolution} explores the nature of the accretion processes by which present-day \HI\ reservoirs are established, and also examines the evolving temperature-density probability distribution functions of the gas comprising \HI\ reservoirs at several epochs. The results are summarized and discussed in Section \ref{sec:summary}.

\section{Methods}
\label{sec:methods}

This section provides an overview of the EAGLE simulations used here, including a brief description of the subgrid physics routines, and the numerical methods used in their analysis. The latter includes a description of the algorithms for identifying galaxies and their host haloes, and of the schemes used to identify neutral hydrogen and then partition it into its atomic and molecular components.

\subsection{Simulations and subgrid physics}
\label{sec:simulations}

\begin{table*} 
\begin{center}
  \caption{Key parameters of the simulations. Those highlighted in bold differentiate a simulation from the corresponding reference (Ref) case. The top section describes the largest EAGLE simulation, Ref-L100N1504, the second section corresponds to box size variations of Ref, the third section corresponds to L025N0376 simulations incorporating parameter variations with respect to Ref, and the bottom section corresponds to the high-resolution simulations that enable strong and weak convergence tests. Columns are: the side length of the simulation volume ($L$) and the particle number per species per dimension ($N$), the normalization of the SF law ($A$) from equation (\ref{eq:KSlaw}), the asymptotic maximum and minimum values of $f_{\rm th}$, the density term denominator ($n_{\rm H,0}$) and exponent ($n_{\rm n}$) from equation (\ref{eq:fth(Z,n)}), the subgrid accretion disc viscosity parameter \citep[$C_{\rm visc}$, from equation 9 of][]{S15}, and the temperature increment of stochastic AGN heating \citep[$\Delta T_{\rm AGN}$, equation 12 of][]{S15}.}
\label{tbl:sims}
\begin{tabular}{l r r l l l l l l l l}
\hline
\hline
Identifier         & $L$        & $N$          &  $A$ (Eq. \ref{eq:KSlaw}) &   $f_{\rm th,max}$  & $f_{\rm th,min}$ & $n_{\rm H,0}$ & $n_{\rm n}$      & $C_{\rm visc}/2{\rm \pi}$ & $\Delta T_{\rm AGN}$  \\
                   & [\cMpc]    &              & [$\Msunyrkpcsq$]          &                    &                & [$\cmcubed$]  &              &                 & $\log_{10}$ [K]       \\
\hline                                                            
Ref-L100N1504      & $100$      & 1504         & $1.515\times 10^{-4}$      & $3.0$             & $0.3$          & $0.67$       & $2/\ln{10}$     & $10^0$      & $8.5$ \\
\hline                                                           
\textit{Box size variations} \\                                  
Ref-L025N0376      & $\bm{25}$  &  376         & $1.515\times 10^{-4}$      & $1.0$             & $1.0$          & $0.67$      & $2/\ln{10}$      & $10^0$      & $8.5$ \\
Ref-L050N0752      & $\bm{50}$  &  752         & $1.515\times 10^{-4}$      & $3.0$             & $0.3$          & $0.67$      & $2/\ln{10}$      & $10^0$      & $8.5$ \\
\hline                                            
\textit{Subgrid variations of Ref-L025N0376} \\                               
KSNormLo-L025N0376 & $25$       &  376         & \bm{$4.791\times 10^{-5}$} & $3.0$             & $0.3$          & $0.67$      & $2/\ln{10}$      & $10^0$      & $8.5$   \\
KSNormHi-L025N0376 & $25$       &  376         & \bm{$4.791\times 10^{-4}$} & $3.0$             & $0.3$          & $0.67$      & $2/\ln{10}$      & $10^0$      & $8.5$            \\
WeakFB-L025N0376   & $25$       &  376         & $1.515\times 10^{-4}$      & $\bm{1.5}$        & $\bm{0.15}$    & $0.67$      & $2/\ln{10}$      & $10^0$      & $8.5$   \\
StrongFB-L25N0376  & $25$       &  376         & $1.515\times 10^{-4}$      & $\bm{6.0}$        & $\bm{0.6}$     & $0.67$      & $2/\ln{10}$      & $10^0$      & $8.5$   \\
\hline                                                                                                           
\textit{High-resolution simulations}  \\                                                                         
Ref-L025N0752      & $25$       &  $\bm{752}$  & $1.515\times 10^{-4}$      & $3.0$             & $0.3$          & $0.67$      & $2/\ln{10}$      & $10^0$      & $8.5$ \\
Recal-L025N0752    & $25$       &  $\bm{752}$  & $1.515\times 10^{-4}$      & $3.0$             & $0.3$          & $\bm{0.25}$ & $\bm{1/\ln{10}}$ & $\bm{10^3}$ & $\bm{9.0}$ \\
\hline
\end{tabular}
\end{center}
\end{table*}

The EAGLE simulations \citep{S15,C15} are a suite of hydrodynamical simulations of the formation, assembly, and evolution of galaxies in the $\Lambda$CDM cosmogony. The simulations were evolved by a version of the $N$-body TreePM smoothed particle hydrodynamics (SPH) code \gadget, last described by \citet{Springel_05}, featuring modifications to the hydrodynamics algorithm and the time-stepping criteria, and the incorporation of subgrid routines governing the phenomenological implementation of processes acting on scales below the resolution limit of the simulations. The updates to the hydrodynamics algorithm, collectively referred to as `Anarchy' \citep[Dalla Vecchia in prep.; see also appendix A of][]{S15}, comprise an implementation of the pressure-entropy formulation of SPH of \citet{Hopkins_13}, the artificial viscosity switch proposed by \citet{Cullen_and_Dehnen_10}, an artificial conduction switch similar to that proposed by \citet{Price_08}, the $C^2$ smoothing kernel of \citet{Wendland_95}, and the time-step limiter of \citet{Durier_and_Dalla_Vecchia_12}. The impact of each of these individual developments on the resulting galaxy population realized in the EAGLE simulations is explored in the study of \citet{Schaller_et_al_15b_short}.

Element-by-element radiative cooling and photoionization heating for 11 species (H, He and 9 metal species) is treated using the scheme described by \citet{Wiersma_Schaye_and_Smith_09}, in the presence of a spatially uniform, temporally evolving radiation field due to the cosmic microwave background and the metagalactic UVB from galaxies and quasars, as modelled by \citet{Haardt_and_Madau_01}. Gas with density greater than the metallicity-dependent threshold advocated by \citet{Schaye_04}, and which is within 0.5 decades of a Jeans-limiting temperature floor (see below), is eligible for conversion to a collisionless stellar particle. The probability of conversion is proportional to the particle's SFR, which is a function of its pressure, such that the simulation reproduces by construction the \citet{Kennicutt_review_98} SF law \citep{Schaye_and_Dalla_Vecchia_08}. Each stellar particle is assumed to represent a simple stellar population with the initial mass function (IMF) of \citet{Chabrier_03}, and the return of mass and metals from stellar populations to the interstellar medium (ISM) is achieved using the scheme described by \citet{Wiersma_et_al_09}, which tracks the abundances of the same 11 elements considered when computing the radiative cooling and photoionization heating rates. The simulations also incorporate routines to model the formation and growth of BHs via gas accretion and mergers with other BHs \citep{Springel_Di_Matteo_and_Hernquist_05,Rosas_Guevara_et_al_15_short,S15}, and feedback associated with SF \citep{Dalla_Vecchia_and_Schaye_12} and the growth of BHs \citep{Booth_and_Schaye_09,S15}, via stochastic gas heating. Implemented as a single heating mode, the active galactic nucleus (AGN) feedback nevertheless mimics quiescent `radio'-like and vigorous `quasar'-like AGN modes when the BH accretion rate is a small ($\ll 1$) or large ($\sim 1$) fraction of the Eddington rate, respectively \citep[][]{McCarthy_et_al_11}. 

The simulations lack both the resolution and physics required to model the cold, dense phase of the ISM. Gas is therefore subject to a polytropic temperature floor, $T_{\rm eos}(\rho_{\rm g})$, that corresponds to the equation of state $P_{\rm eos} \propto \rho_{\rm g}^{4/3}$, normalised to $T_{\rm eos} = 8000\K$ at $n_{\rm H} \equiv X_{\rm H}\rho/m_{\rm H} = 10^{-1}\,\cmcubed$, where $X_{\rm H}$ is the hydrogen mass fraction. The exponent of $4/3$ ensures that the Jeans mass, and the ratio of the Jeans length to the SPH kernel support radius, are independent of the density \citep{Schaye_and_Dalla_Vecchia_08}. This is a necessary condition to limit artificial fragmentation. Gas heated by feedback to $\log_{10} T > \log_{10} T_{\rm eos}(\rho_{\rm g}) + 0.5$ is ineligible for SF, irrespective of its density. The use of a Jeans-limiting temperature floor is of particular relevance to the application of post-processing schemes for partitioning hydrogen into its ionized, atomic and molecular phases, discussed in \S\ref{sec:partitioning}.

\citet{S15} argue that cosmological simulations lack at present the resolution and physics required to calculate, ab-initio, the efficiency of the feedback processes that regulate galaxy growth. The subgrid efficiencies of such processes adopted by EAGLE are therefore calibrated to reproduce well-characterized observables. That governing AGN feedback is assumed to be constant, and is calibrated to ensure that the simulations reproduce the observed present-day relation between the mass of central BHs and the stellar mass of their host galaxy \citep[see also][]{Booth_and_Schaye_09}. The subgrid efficiency of feedback associated with SF is a smoothly varying function of the metallicity and density of gas local to newly formed stellar particles. The efficiency function of the EAGLE Reference model (`Ref'), introduced by \citet{S15}, is calibrated to ensure reproduction of the present-day GSMF, and the size-mass relation of disc galaxies, at the resolution of the largest simulation. \citet{C15} demonstrate that both calibration criteria are necessary conditions to ensure the emergence of a realistic galaxy population. 

The simulations adopt the cosmological parameters inferred by the \citet{Planck_2014_paperI_short}, namely $\Omega_{\rm m} =  0.307$, $\Omega_\Lambda = 0.693$, $\Omega_{\rm b} = 0.04825$, $h = 0.6777$ and  $\sigma_8 =  0.8288$. The `standard-resolution' EAGLE simulations have particle masses corresponding to a volume of side $L=100~{\rm comoving~Mpc}$ (hereafter~$\cMpc$) realized with $2 \times 1504^3$ particles (an equal number of baryonic and dark matter particles), such that the initial gas particle mass is $m_{\rm g}=1.81 \times 10^6\Msun$, and the mass of dark matter particles is $m_{\rm dm}=9.70 \times 10^6\Msun$. The Plummer-equivalent gravitational softening length is fixed in comoving units to $1/25$ of the mean interparticle separation ($2.66~{\rm comoving~kpc}$, hereafter~$\ckpc$) until $z=2.8$, and in proper units ($0.70~{\rm proper~kpc}$, hereafter~$\pkpc$) thereafter. The standard-resolution simulations therefore (marginally) resolve the Jeans scales at the SF threshold in the warm ($T\simeq 10^4\K$) ISM. High-resolution simulations adopt particle masses and softening lengths that are smaller by factors of 8 and 2, respectively. The SPH kernel size, specifically its support radius, is limited to a minimum of one-tenth of the  gravitational softening scale.

The fiducial simulation studied here is the Ref model realized within a volume of side $L=100\cMpc$ (Ref-L100N1504). Standard-resolution simulations realizing Ref in volumes of side $L=25$ and $50\cMpc$ (Ref-L025N0376 and Ref-L050N0752, respectively) are also included to enable an assessment of the convergence behaviour of various properties as the box size is varied, and to facilitate comparison with L025N0376 simulations that incorporate variations of the Ref subgrid parameters. The adjusted parameters relate to the efficiencies of SF, and the feedback associated with it. Specifically, the former is the normalization, $A$, of Kennicutt-Schmidt law, 
\begin{equation}
  \Sigma_{\rm SFR} = A \left( \frac{\Sigma_{\rm g}}{1~\Msun \pc^{-2}}\right)^n,
\label{eq:KSlaw}
\end{equation}
where $\Sigma_{\rm SFR}$ and $\Sigma_{\rm g}$ are the surface densities of the SFR and the star-forming gas mass, respectively, and an exponent of $n=1.4$ is adopted \citep{Kennicutt_review_98}. The Ref model adopts $A=1.515\times 10^{-4}\Msunyrkpcsq$ \citep[the value quoted by][adjusted from a Salpeter to Chabrier IMF]{Kennicutt_review_98}, whilst the models KSNormLo and KSNormHi adopt values that are lower and higher by $0.5\,{\rm dex}$, respectively (see also Table \ref{tbl:sims}). Note that, at fixed pressure, the SFRs of gas particles in the simulation are linearly proportional to $A$ \citep[see equation 20 of][]{Schaye_and_Dalla_Vecchia_08}. The efficiency of feedback associated with SF is varied via the asymptotic values ($f_{\rm th,max}$ and $f_{\rm th,min}$) of the function describing the energy budget associated with SF feedback:
\begin{equation}
f_{\rm th}(n_{\rm H},Z) = f_{\rm th,min} + \frac{f_{\rm th,max} - f_{\rm th,min}}
{1 + \left (\frac{Z}{0.1\Zsun}\right )^{n_Z} \left (\frac{n_{\rm H}}{n_{{\rm H},0}}\right )^{-n_n}},
\label{eq:fth(Z,n)}
\end{equation}
where $n_{\rm H}$ and $Z$ are the density and metallicity a stellar particle inherits from its parent gas particle, and the parameters $n_{\rm Z}$ and $n_{\rm n}$ describe how rapidly the efficiency function varies in response to changes of the metallicity and density, respectively. The effect of such a dependence is discussed in detail by \citet{C15}. The models WeakFB and StrongFB vary the Ref values of both $f_{\rm th,max}$ and $f_{\rm th,min}$ by factors of $0.5$ and $2$, respectively.

Finally, to facilitate convergence testing, two high-resolution $L=25\cMpc$ simulations are also examined. Strong convergence\footnote{The concept of strong and weak convergence was introduced by \citet{S15}.} is enabled by comparison of the Ref model realized at standard resolution with the Ref-L025N0752 simulation, whilst weak convergence is enabled by comparison with the Recal-L025N0752 simulation. The latter incorporates feedback efficiency parameters for both SF ($n_{\rm H,0}$ and $n_{\rm n}$ of equation \ref{eq:fth(Z,n)}) and AGN feedback \citep[$C_{\rm visc}$ and $\Delta T_{\rm AGN}$ - see equations 9 and 12 of][]{S15} that are recalibrated to ensure reproduction of the calibration diagnostics at high resolution. Key parameters of the subgrid prescriptions used by all models examined here are specified in Table \ref{tbl:sims}.

\subsection{The identification of galaxies and assignment of their properties}
\label{sec:id_galaxies}

Galaxies and their host haloes are identified by a multi-stage process, beginning with the application of the friends-of-friends (FoF) algorithm to the dark matter particle distribution, with a linking length of $b=0.2$ times the mean interparticle separation. Gas, star and BH particles are associated with the FoF group, if any, of their nearest neighbour dark matter particle. The SUBFIND algorithm \citep{Springel_et_al_01,Dolag_et_al_09} is then used to identify self-bound substructures, or subhaloes, within the full particle distribution (gas, stars, BHs and dark matter) of FoF haloes. Subhaloes can in principle be comprised of any combination of baryonic and/or dark matter particles. The subhalo comprising the particle with the minimum gravitational potential, which is almost exclusively the most massive subhalo, is defined as the central subhalo (hosting the central galaxy), the remainder being satellite subhaloes (hosting satellite galaxies). The coordinate of the particle with the minimum potential also defines the position of the halo, about which is computed the spherical overdensity mass, $M_{200}$, for the adopted density contrast of $200$ times the critical density. Satellite subhaloes separated from their central galaxy by less than the minimum of $3\pkpc$ and the stellar half-mass radius of the central galaxy are merged into the central; this step eliminates a small number of low-mass subhaloes dominated by single, high-density gas particles or BHs.

When aggregating the properties of galaxies that are typically measured from optical diagnostics, for example stellar masses ($M_\star$) and SFRs ($\dot{M}_\star$), only those bound particles within a spherical aperture of radius $30\pkpc$ centred on the potential minimum of the subhalo are considered. \citet[][their fig. 6]{S15} demonstrate that this practice yields stellar masses comparable to those recovered within a projected circular aperture of the Petrosian radius. In order to place the atomic hydrogen masses ($M_{\rm HI}$) of galaxies on an equivalent footing to 21 cm measurements, a larger aperture of $70\pkpc$ is used. \citet{Bahe_et_al_16_short} demonstrate that this yields similar masses to those recovered via direct mimicry of the beam size and bandwidth of the Arecibo L-band Feed Array \citep[ALFA;][]{Giovanelli_et_al_05_short} at the median redshift, $z=0.037$, of the GASS survey. Since our motivation is to examine the (evolving) \HI\ properties of galaxies and the means by which this gas is acquired, we adopt this simple method of associating \HI\ to galaxies rather than the observationally-motivated technique of \citet{Bahe_et_al_16_short}.

\subsection{Partitioning hydrogen into ionized, atomic and molecular components}
\label{sec:partitioning}

The EAGLE simulations self-consistently model each fluid element's mass fraction in the form of hydrogen, helium, and nine metal species. The self-consistent partitioning of the hydrogen component into its ionized, atomic and molecular forms, however, requires the use of detailed radiation transport (RT) and photo-chemical modelling and, ideally, explicit consideration of the cold, dense, ISM. RT in large cosmological volumes is computationally expensive, and infeasible for the simulations studied here. Moreover, EAGLE lacks the resolution and physics required to model the cold ISM. Therefore, a two-stage approximation scheme is used to partition the mass of each particle in the form of hydrogen between \HII, \HI, and \Hmol. This scheme, or elements of it, also features in the analyses of EAGLE simulations presented by \citet{Rahmati_et_al_15}, \citet{Lagos_et_al_15_short,Lagos_et_al_16_short} and \citet{Bahe_et_al_16_short}.

\subsubsection{Partitioning hydrogen into neutral and ionized components}
\label{sec:HII_to_HI}

Hydrogen is first partitioned into its neutral (i.e. \HI+\Hmol) and ionized (\HII) components. Consideration of the key ionizing mechanisms is therefore necessary, namely collisional ionization at high temperature, and photoionization by the metagalactic UVB radiation. In general the latter, which is imposed by the simulations, dominates the ionization of cosmic hydrogen, particularly at $z \gtrsim 1$ \citep{R13a}. Radiation from sources within and/or local to galaxies is not considered explicitly here, since the detailed characterization of their impact would require analysis of a suite of very high resolution RT simulations incorporating a model of the multiphase ISM. The effects of such sources are likely significant \citep[e.g.][]{Miralda_Escude_05,Schaye_06} and difficult to estimate, since they can act both to reduce the \HI\ mass of galaxies by ionizing \HI\ to \HII, and to increase it by dissociating \Hmol\ into \HI\ (the latter effect is accounted for implicitly, however, via the use of the empirical and theoretical schemes used to partition neutral hydrogen into \HI\ and \Hmol, see \S\ref{sec:HI_to_H2}). \citet{R13b} analysed RT simulations without a multiphase ISM and concluded that local sources likely reduce the abundance of high column density systems significantly \citep[but see also][]{Pontzen_et_al_08_short}. We therefore caution that neglect of local radiation represents a potentially significant systematic uncertainty on \HI\ masses that we are unable to authoritatively characterize here.

The collisional ionization rate is a function only of temperature, and is parametrised using the relations collated by \citet{Cen_92}. The effective photoionization rate is specified as a function of the mass-weighted density and the redshift-dependent photoionization rate of the metagalactic background, $\Gamma_{\rm UVB}(z)$, 
using the redshift-dependent fitting function of \citet[][see their table A1]{R13a}. This function was calibrated using TRAPHIC \citep{Pawlik_and_Schaye_08} RT simulations, which account for the self-shielding of gas against the UVB and recombination radiation. To maintain consistency with the thermodynamics of the EAGLE simulations, we compute the effective photoionization rate assuming the same \citet{Haardt_and_Madau_01} model of $\Gamma_{\rm UVB}(z)$ assumed by the \citet{Wiersma_Schaye_and_Smith_09} radiative heating and cooling tables. However, both observational constraints and theoretical models of the amplitude and spectral shape of the UVB are uncertain by a factor of a few, with recent studies indicating that the amplitude is likely lower than the value of $\Gamma_{\rm UVB}(z=0) = 8.34\times 10^{-14}\persec$ specified by the \citet{Haardt_and_Madau_01} model \citep[e.g.][]{Faucher-Giguere_et_al_09,Haardt_and_Madau_12,Becker_and_Bolton_13}. Since a weaker UVB enables hydrogen to self-shield at a lower (column) density, this uncertainty may influence the \HI\ masses of low (stellar) mass galaxies. We therefore examine in following sections the effect of reducing $\Gamma_{\rm UVB}$ by a factor of 3 on the \HI\ CDDF (\S\ref{sec:CDDF}), the $M_{\rm HI}-M_\star$ relation (\S\ref{sec:M_HI-Mstar}) and the HIMF (\S\ref{sec:HIMF}) at $z=0$.

The temperatures of high-density particles near to the Jeans-limiting temperature floor (see \S\ref{sec:simulations}) are not physical, but rather reflect the mass-weighted pressure of the multiphase ISM they represent. Therefore, for the purposes of calculating their ionization states, the temperature of star-forming particles is assumed to be $T=10^4\K$, characteristic of the warm, diffuse phase of the ISM.

\begin{figure*}
\includegraphics[width=\columnwidth]{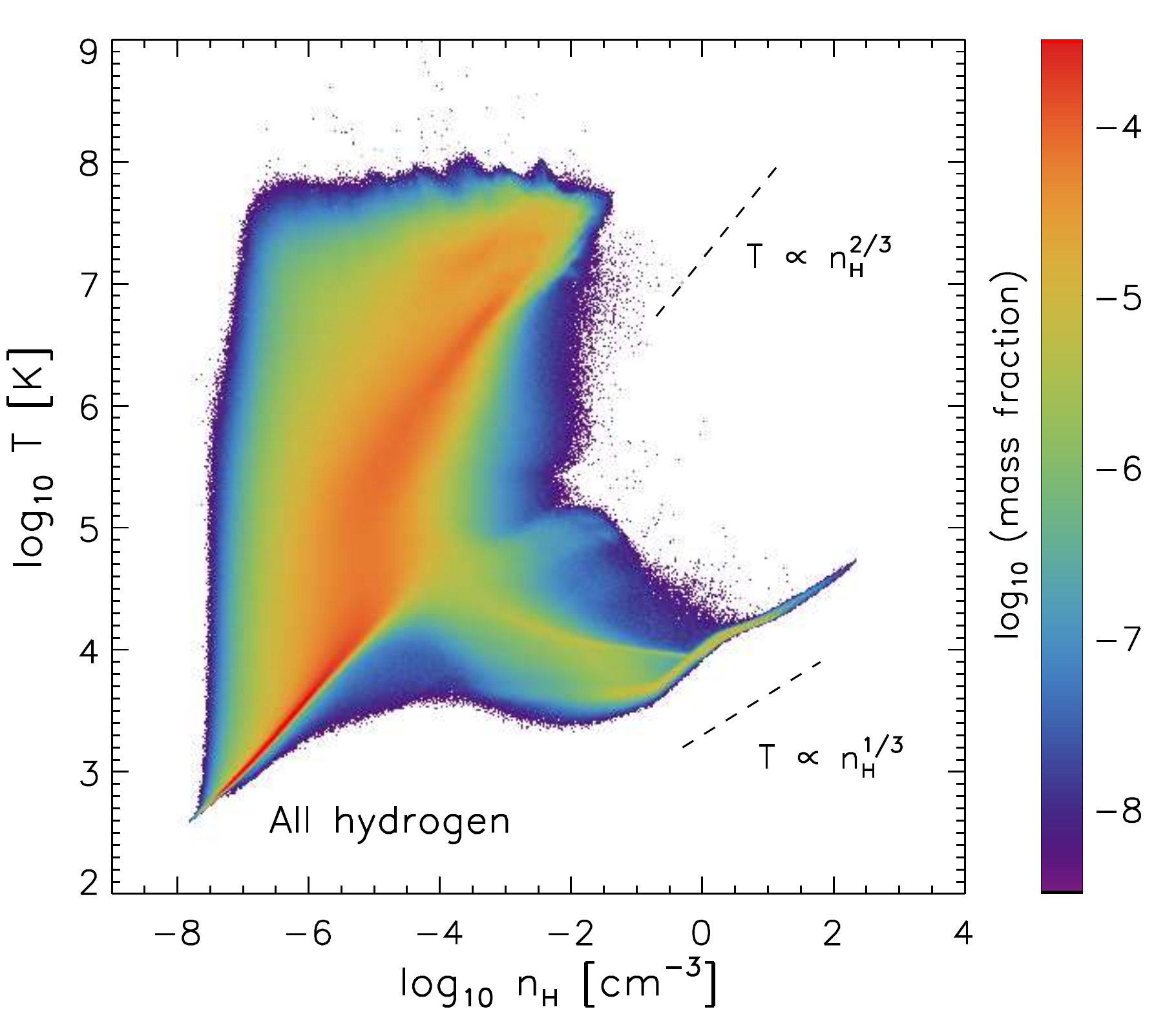}
\includegraphics[width=\columnwidth]{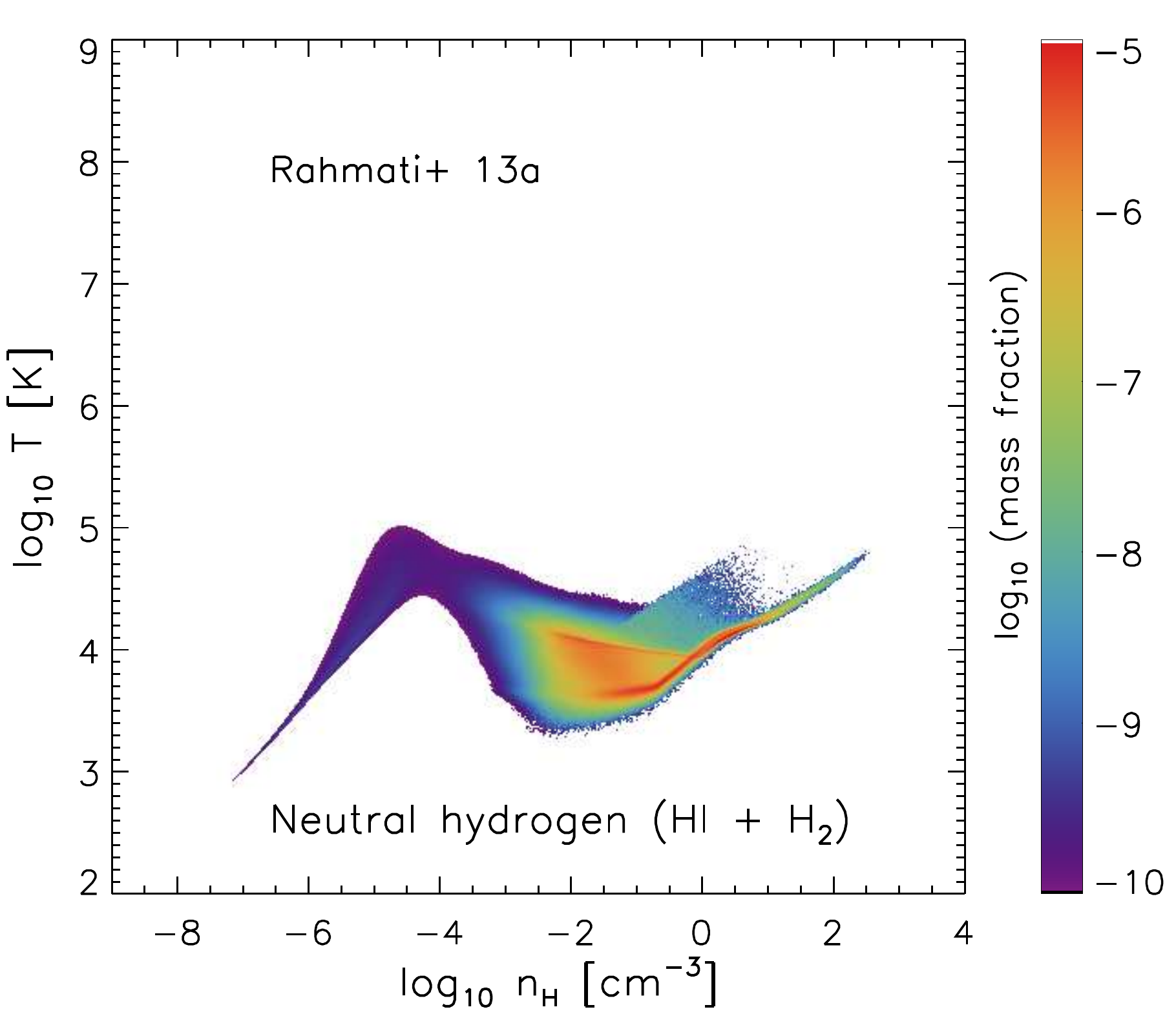}
\includegraphics[width=\columnwidth]{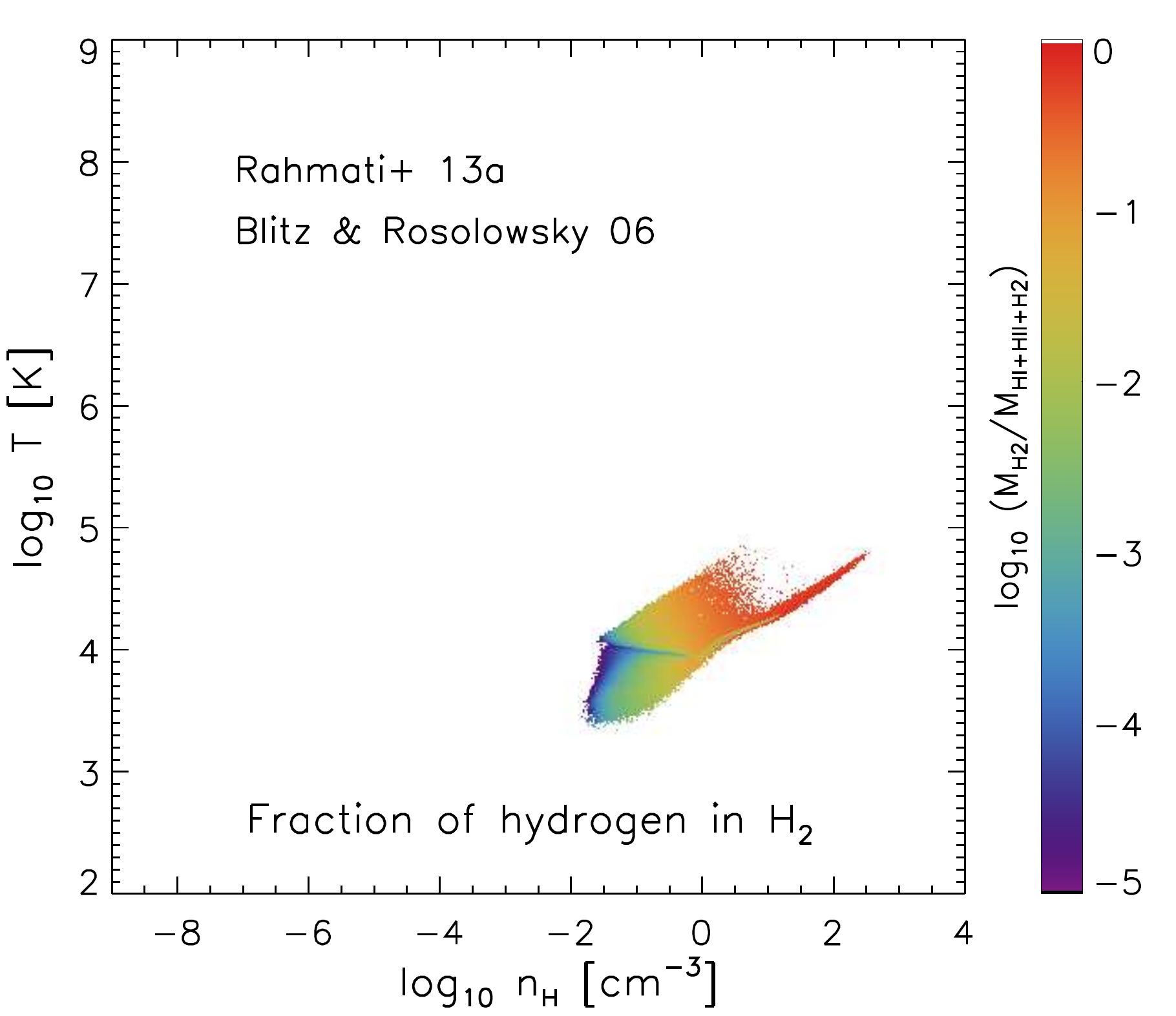}
\includegraphics[width=\columnwidth]{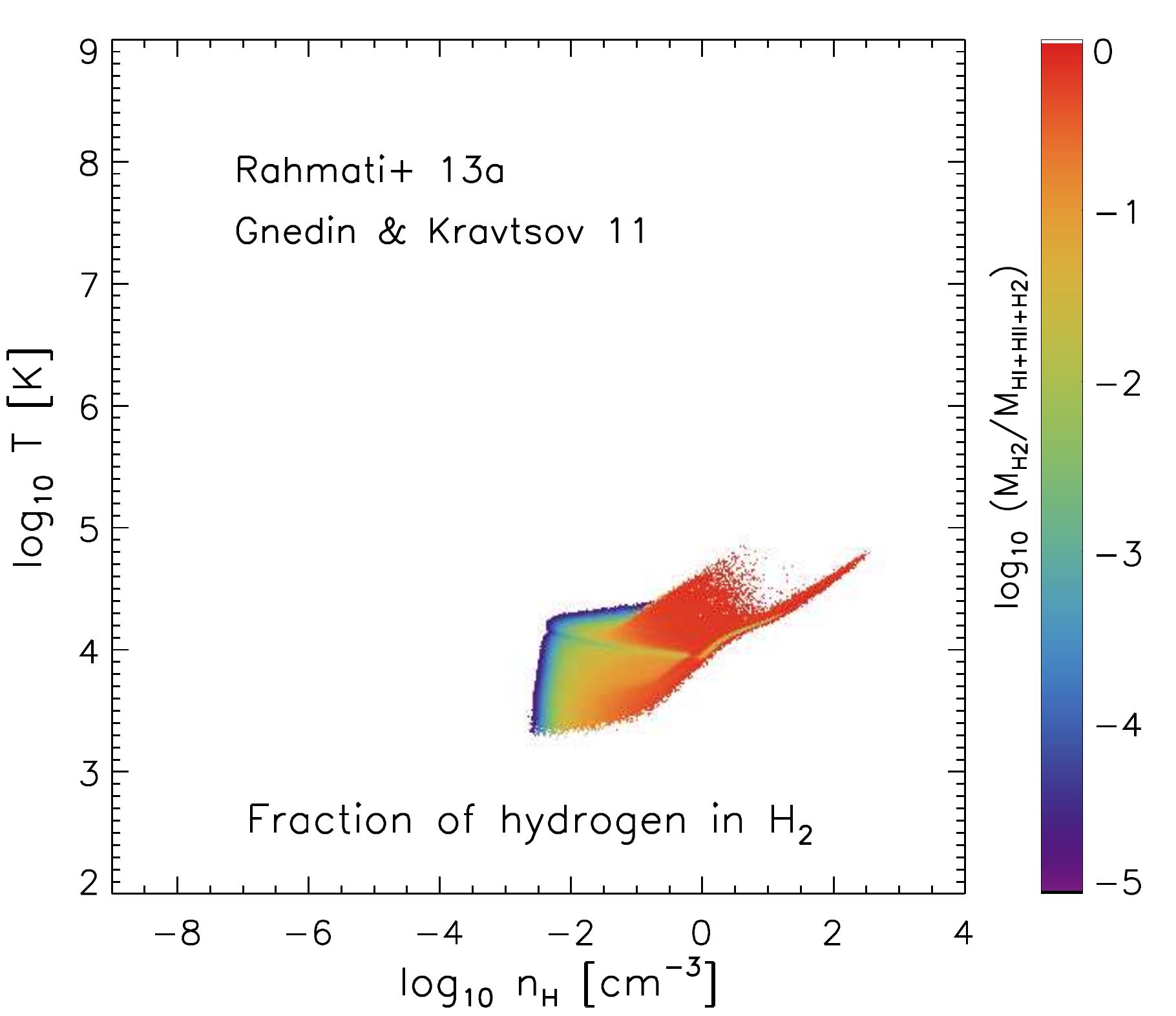}
\caption{Temperature-density phase space distribution of hydrogen in the Ref-L100N1504 simulation at $z=0$. \textit{Upper-left}: The mass-weighted 2D PDF  of all hydrogen, irrespective of its ionization state. This panel exhibits the common features of a temperature-density diagram: the photoionized intergalactic medium at low temperatures ($T \lesssim 10^4\K$) and densities ($n_{\rm H} \lesssim 10^{-5}\,\cmcubed$ at $z=0$), the plume of shocked, high-entropy intergalactic and intracluster gas at high temperature ($T\gg10^5\K$) and intermediate density ($n_{\rm H}\sim 10^{-5}-10^{-1}\cmcubed$), cold streams penetrating haloes at low temperature ($T\lesssim 10^5\K$) and intermediate density, and the Jeans-limited ISM at low temperature and high density ($n_{\rm H}\gtrsim 10^{-1}\cmcubed$). Dashed lines denote the $T\propto n_{\rm H}^{2/3}$ relation described by adiabatic gas, and the $T\propto n_{\rm H}^{1/3}$ scaling of the Jeans-limiting temperature floor. \textit{Upper right}: The 2d-PDF weighted by the neutral (\HI\ + \Hmol) hydrogen mass, highlighting that most is found at densities and temperatures characteristic of the CGM and the ISM at $z=0$. The lower panels show the molecular hydrogen (\Hmol) mass fraction as a function of thermodynamic phase, when partitioning neutral gas into atomic and molecular components with the empirical \citet{BR06} pressure law (\textit{left}) and the theoretical \citet{GK11} prescription (\textit{right}).}
\label{fig:phase_diagrams}
\end{figure*}

\subsubsection{Partitioning neutral hydrogen into atomic and molecular components}
\label{sec:HI_to_H2}

In the absence of explicit RT modelling of the multiphase ISM, the partitioning of neutral hydrogen into \HI\ and \Hmol\ requires the use of either an empirical relation or a theoretical prescription. As with the partitioning of ionized and neutral components, this procedure is subject to systematic uncertainty. In the context of EAGLE, this partitioning has been shown to be significant by the studies of \citet{Bahe_et_al_16_short}, focusing on atomic hydrogen, and \citet{Lagos_et_al_15_short,Lagos_et_al_16_short}, focusing on the molecular component. The latter adopted the theoretically-motivated prescriptions of \citet[][hereafter GK11]{GK11} and \citet{Krumholz_13} to estimate the mass fractions of the two components, finding that their application to EAGLE galaxies results in the reproduction of many observed molecular hydrogen scaling relations. However, their use also underestimates the observed \HI\ mass of massive galaxies ($M_\star > 10^{10}\Msun$) in the standard-resolution Ref simulations \citep{Bahe_et_al_16_short}. The consistency with observational measurements is, however, considerably better in high-resolution EAGLE simulations.

\citet{Bahe_et_al_16_short} demonstrated that the total neutral hydrogen mass, $M_{\rm HI}+M_{\rm H_2}$, of massive ($M_\star > 10^{10}\Msun$) EAGLE galaxies is compatible with observational constraints. Motivated by the observed scaling of the molecular to atomic hydrogen surface density ratio ($R_{\rm mol} \equiv \Sigma_{\rm H_2}/\Sigma_{\rm HI}$) with the inferred mid-plane pressure of galaxy discs \citep[e.g.][]{Wong_and_Blitz_02,BR06}, they adopted an empirical scaling relation between the molecular hydrogen fraction and the pressure of gas particles; similar approaches have also been applied to hydrodynamical simulations in similar studies \citep[e.g.][]{Popping_et_al_09,Altay_et_al_11,Duffy_et_al_12,Dave_et_al_13,R13a,R13b,Bird_et_al_15}. Based on observational measurements of 11 nearby galaxies spanning a factor of 5 in metallicity and three decades in ISM mid-plane pressure, \citet[][hereafter BR06]{BR06} recovered the empirical relation $R_{\rm mol} = \left(P/P_0\right)^\alpha$, where $P_0/k_{\rm B} = 4.3\times 10^4\,\cmcubed\K$ and $\alpha=0.92$. This relation can be used to compute the mass fraction of the neutral hydrogen in atomic and molecular forms for each gas particle, as a function of their pressure. Although more recent studies suggest that the power law exponent $\alpha$ is a mildly-varying function of galaxy mass \citep{Leroy_et_al_08}, we retain a single power-law for simplicity. \citet{Bahe_et_al_16_short} demonstrated that the use of the \citet{Leroy_et_al_08} best-fitting parameters for massive disc galaxies has a negligible impact on the \HI\ masses and profiles of $M_\star > 10^{10}\Msun$ EAGLE galaxies. For consistency with the metallicity-dependent SF density threshold adopted by the simulations, which is motivated by an analysis of the conditions within the ISM conducive to the formation of a cold ($T \ll 10^4\K$) phase \citep[e.g.][]{Schaye_04}, only gas particles with a non-zero SFR are assigned a non-zero \Hmol\ fraction. \citet{Bahe_et_al_16_short} demonstrate that the application of this prescription to massive EAGLE galaxies yields atomic hydrogen-to-stellar mass ratios that are consistent with those reported by the GASS survey, and \HI\ surface density profiles consistent with those reported by the Bluedisk survey \citep{Wang_et_al_13}, for standard- and high-resolution EAGLE simulations alike. This prescription is therefore used here as the fiducial method for partitioning neutral hydrogen into \HI\ and \Hmol.

\subsubsection{The distribution of hydrogen in temperature-density space.}
\label{sec:temp_dens}

To highlight the conditions of the hydrogen partitioned into \HI\ and \Hmol, we show in Fig. \ref{fig:phase_diagrams} the distribution in temperature-density space of all cosmic hydrogen (i.e. gas within galaxies and the intergalactic medium), and partitioned subsets thereof, within the Ref-L100N1504 simulation at $z=0$. The upper panels are mass-weighted, two-dimensional probability distribution functions (2D PDFs) such that the sum over all pixels is unity, the lower panels show per-pixel mass fractions. 

The upper-left panel shows the 2D PDF of cosmic hydrogen (\HII, \HI\ and \Hmol) and highlights the common features of a temperature-density phase diagram \citep[see also e.g.][]{Theuns_et_al_98}. The majority of cosmic hydrogen resides in the narrow strip at low density ($n_{\rm H} \lesssim 10^{-5}\,\cmcubed$) and low temperature ($T \lesssim 10^5\K$) corresponding to the photoionized intergalactic medium \citep[IGM; e.g.][]{Hui_and_Gnedin_97,Theuns_et_al_98}. At low temperatures but higher densities ($n_{\rm H} \gtrsim 10^{-5}\,\cmcubed$), efficient radiative cooling inverts the $T-\rho$ relation, enabling net cooling. There are well-defined tracks within this phase; the upper track, with a mildly-negative $T-\rho$ gradient, is associated with metal-poor inflows from the IGM \citep{van_de_Voort_et_al_12} whilst the lower track is typically metal rich circumgalactic medium (CGM) previously ejected from the ISM by feedback \citep{Rahmati_et_al_16}.

As discussed in \S\ref{sec:simulations}, gas is subject to a Jeans-limiting temperature floor corresponding to the equation of state, $P_{\rm eos} \propto \rho_{\rm g}^{4/3}$, or $T \propto n_{\rm H}^{1/3}$. This feature is visible as the narrow strip extending from $n_{\rm H} \gtrsim 10^{-1}\,\cmcubed$, and the lower dashed line (arbitrarily normalised) illustrates the imposed slope. The density of star-forming particles close to the pressure floor represents a mass-weighted mean of unresolved ISM phases, and their temperature reflects their effective pressure rather than their physical temperature. Deviation from a pure power law form of the relation (i.e. broadening, and the `kink' at $n_{\rm H}\sim 1\,\cmcubed$) is a consequence of variations in the mean molecular weight of particles, and the transfer of internal energy by the artificial conduction scheme. The upper dashed line shows the $T\propto n_{\rm H}^{2/3}$ relation characteristic of adiabatic gas, highlighting that the hot ($T\gg 10^5\K$), high-entropy gas associated with the shock-heated intergalactic and intracluster media do not cool efficiently via radiative processes.

The upper-right panel shows the 2D PDF of neutral hydrogen (\HI\ + \Hmol) modelled by the \citet{R13a} redshift-dependent fitting function. Although the majority of cosmic hydrogen resides in the diffuse IGM, the recombination time of this phase is long and its ionization rate is relatively high, rendering its neutral fraction so low that the (largely-photoionized) IGM contributes little to the overall cosmic budget of neutral gas (or equivalently \HI). Most neutral hydrogen therefore resides at temperatures and densities characteristic of the circumgalactic medium and the star-forming ISM \citep[see also][]{Duffy_et_al_12,Bird_et_al_15,Rahmati_et_al_15}. 

The lower panels show the mass fraction of the hydrogen in \Hmol, partitioning neutral hydrogen into \HI\ and \Hmol\ using the fiducial scheme based on the BR06 pressure law (\textit{lower-left}) and the theoretically-motivated prescription of GK11 \citep[\textit{lower-right}; a detailed description of the implementation of this scheme in EAGLE is given by][]{Lagos_et_al_15_short}. In both cases, dense gas close to the Jeans-limiting temperature floor exhibits a molecular fraction close to unity; this behaviour extends to lower densities for the GK11 prescription than the BR06 pressure law. In general, the molecular fraction indicated by the GK11 prescription is greater at all densities and temperatures for which the neutral fraction is significant \citep{Bahe_et_al_16_short}. Note that, since the molecular fraction, $M_{\rm H_2}/M_{\rm H}$, is typically less than 0.1 except at densities $n_{\rm H} \gtrsim 1\cmcubed$, the upper-right panel of Fig. \ref{fig:phase_diagrams} would look similar if the mass fraction of atomic (as opposed to neutral) hydrogen were shown instead.

\section{The HI column density distribution function}
\label{sec:CDDF}

Prior to an examination of the properties of atomic hydrogen associated with galaxies, it is appropriate to establish the degree to which the simulations reproduce the observed present-day \HI\ CDDF, i.e. $f(N_{\rm HI},z)$ the number of systems per unit column density (${\rm d}N_{\rm HI}$) per unit absorption length, ${\rm d}X = {\rm d}z[H_0/H(z)](1+z)^2$. Note that ${\rm d}X = {\rm d}z$ at $z=0$. This quantity is often traced by Lyman-$\alpha$ absorption against bright background sources, enabling systems with column densities many orders of magnitude lower than those revealed by 21 cm emission to be probed. At such low column densities, the gas dynamics are dominated by the gravitational collapse of the cosmic large scale structure, enabling simulations to be tested in this relatively simple regime \citep[e.g.][]{Altay_et_al_11,Altay_et_al_13,Bird_et_al_13,R13a}. The \HI\ CDDF of Ref-L100N1504 from $z=5$ to $z=1$ was presented by \citet[][see their Fig. 2]{Rahmati_et_al_15}, who showed that the simulation accurately reproduces the \HI\ CDDF from $\log_{10} N_{\rm HI}\,[\cmsquared] \sim 16\,$ to $22$.

\begin{figure}
  \includegraphics[width=\columnwidth]{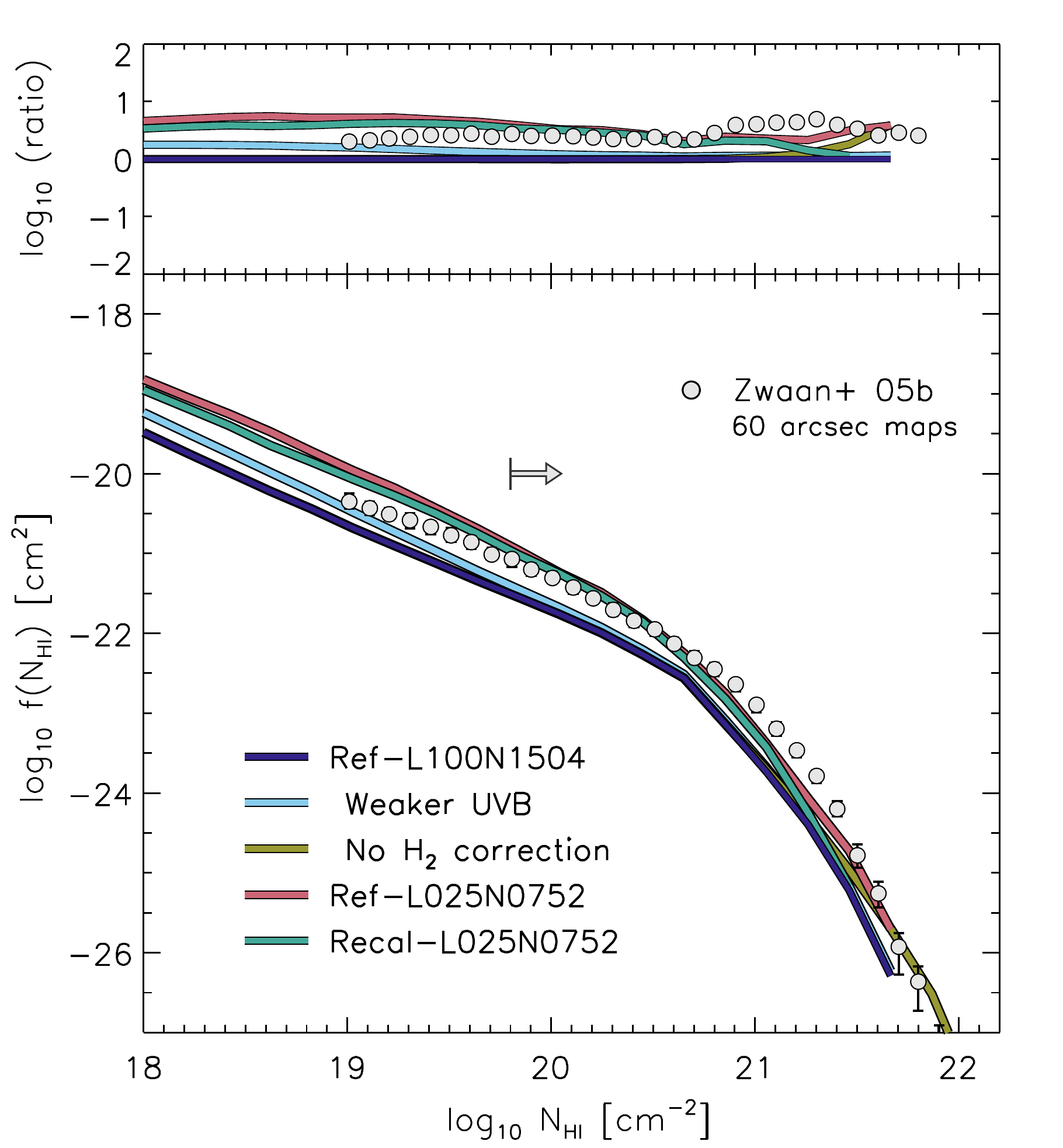}
  \caption{The $z=0$ \HI\ CDDF of the Ref-L100N1504 (dark blue curve), Ref-L025N0752 (red) and Recal-L025N0752 (green) simulations. The CDDF of Ref-L100N1504 is also shown assuming a UVB photoionization rate a factor of 3 lower than the fiducial value (cyan), and without correcting the column density for \Hmol\ (gold). Open circles denote the observational measurements of \citet{Zwaan_et_al_05b}, and the right-facing arrow denotes their lower sensitivity limit. The upper panel shows the ratio of the CDDFs inferred from the simulations and observations with respect to the fiducial Ref-L100N1504 case. Differences due to numerical resolution are significantly greater than those relating to the recalibration of subgrid parameters at high resolution, and those associated with partitioning hydrogen in the simulations into \HII, \HI\ and \Hmol.}
  \label{fig:CDDF}
\end{figure}

The \HI\ CDDF of the nearby Universe can be probed in the high-column density regime ($\log_{10} N_{\rm HI}\,[\cmsquared] \gtrsim 19$) via 21 cm emission. Gas at high column densities is sensitive to the physics of galaxy formation, making this comparison a demanding test of the simulations. We therefore extend the \citet{Rahmati_et_al_15} analysis of EAGLE to the present-epoch, constructing the CDDF by interpolating the particle distribution, using the SPH kernel, on to a 2D grid comprised of pixels with size $\Delta x = 5\kpc$; for Ref-L100N1504 this requires a grid of $20000^2$ pixels. We have conducted convergence tests varying the cell size and find that this spatial scale is sufficient to yield a CDDF that, conservatively, remains well converged at column densities as high as $\log_{10} N_{\rm HI}\,[\cmsquared] = 21$. By smoothing on to multiple planes in the depth axis, we are able to probe column densities as low as $\log_{10} N_{\rm HI}\,[\cmsquared] \sim 15$ without `contamination' from fore- and background structures with significant velocity offsets \citep[see also][]{Rahmati_et_al_15}. 

The main panel of Fig. \ref{fig:CDDF} shows the \HI\ CDDF of Ref-L100N1504 (dark blue curve), and those of the Ref-L025N0752 (red) and Recal-L025N0752 simulations to enable an assessment of the strong and weak convergence behaviour, respectively, of the CDDF. Overplotted symbols denote observational measurements of the CDDF by \citet{Zwaan_et_al_05b}, derived from WHISP survey data \citep{van_der_Hulst_van_Albada_and_Sancisi_01}; we use the 60 arcsec resolution measurements from that study since at the median source distance of WHISP, this beam size corresponds to $\simeq 5\kpc$, the same spatial scale on which we compute the CDDFs of the simulations. As noted by \citet{Zwaan_et_al_05b}, the observations themselves are not `converged' at this spatial scale \citep[see also][]{Braun_12}, but any comparison between the simulated and observed CDDFs measured on the same spatial scale is a fair test. The right-facing arrow on the figure denotes the approximate sensitivity limit of the WHISP maps, of $\log_{10} N_{\rm HI}\,[\cmsquared] = 19.8$.

The CDDF of Ref-L100N1504 is systematically offset to lower column density than that reported by \citet{Zwaan_et_al_05b}: at a fixed abundance of $\log_{10} f(N_{\rm HI})[\cmsquared]= -22\,(-24)$ the offset in column density is $-0.28\,(-0.22)\,{\rm dex}$. The high-resolution simulations exhibit higher column densities than Ref-L100N1504, particularly so for low column density systems with abundance $\log_{10} f(N_{\rm HI})[\cmsquared] >  -22$, and are hence in better agreement with the observations. For example the characteristic column density of Ref-L025N0752 and Recal-L025N0752 at $\log_{10} f(N_{\rm HI})[\cmsquared]= -22$ is offset from that of the fiducial Ref-L100N1504 case by $0.28\,{\rm dex}$ and $0.26,{\rm dex}$, respectively. Similar behaviour at $z \ge 1$ was reported by \citet{Rahmati_et_al_15}, who also demonstrated that the covering fraction of systems at fixed $N_{\rm HI}$ increases at the higher resolution of the L025N0752 simulations. The offset is not a consequence of the difference of the box size of L100N1504 and L025N0752 simulations, since the CDDF of Ref-L025N0376 (which for brevity is not shown) is indistinguishable\footnote{Strictly, strong and weak convergence tests should be performed by comparing the L025N0752 simulations with Ref-L025N0376 to exclude box size effects, but the similarity of the Ref-L025N0376 and Ref-L100N1504 CDDFs indicates that comparison with Ref-L100N1504 is adequate. This also applies to convergence tests of the $M_{\rm HI}-M_\star$ relation (\S\ref{sec:M_HI-Mstar}).} from that of Ref-L100N1504 for $\log_{10} N_{\rm HI}\,[\cmsquared] < 21.3$. The relatively poor resolution convergence of the CDDF is similar in both the weak and strong regimes, with the CDDFs of the Ref-L025N0752 and Recal-L025N0752 simulations being near-identical except at $\log_{10} N_{\rm HI}\,[\cmsquared] \gtrsim 21$. This indicates that although numerical resolution impacts upon the CDDF over a wide range of column densities, the recalibration of the feedback parameters necessary to ensure reproduction of the $z=0.1$ GSMF only influences the high column densities associated with galaxies and their immediate environments. 

\begin{figure*}
  \includegraphics[width=\columnwidth]{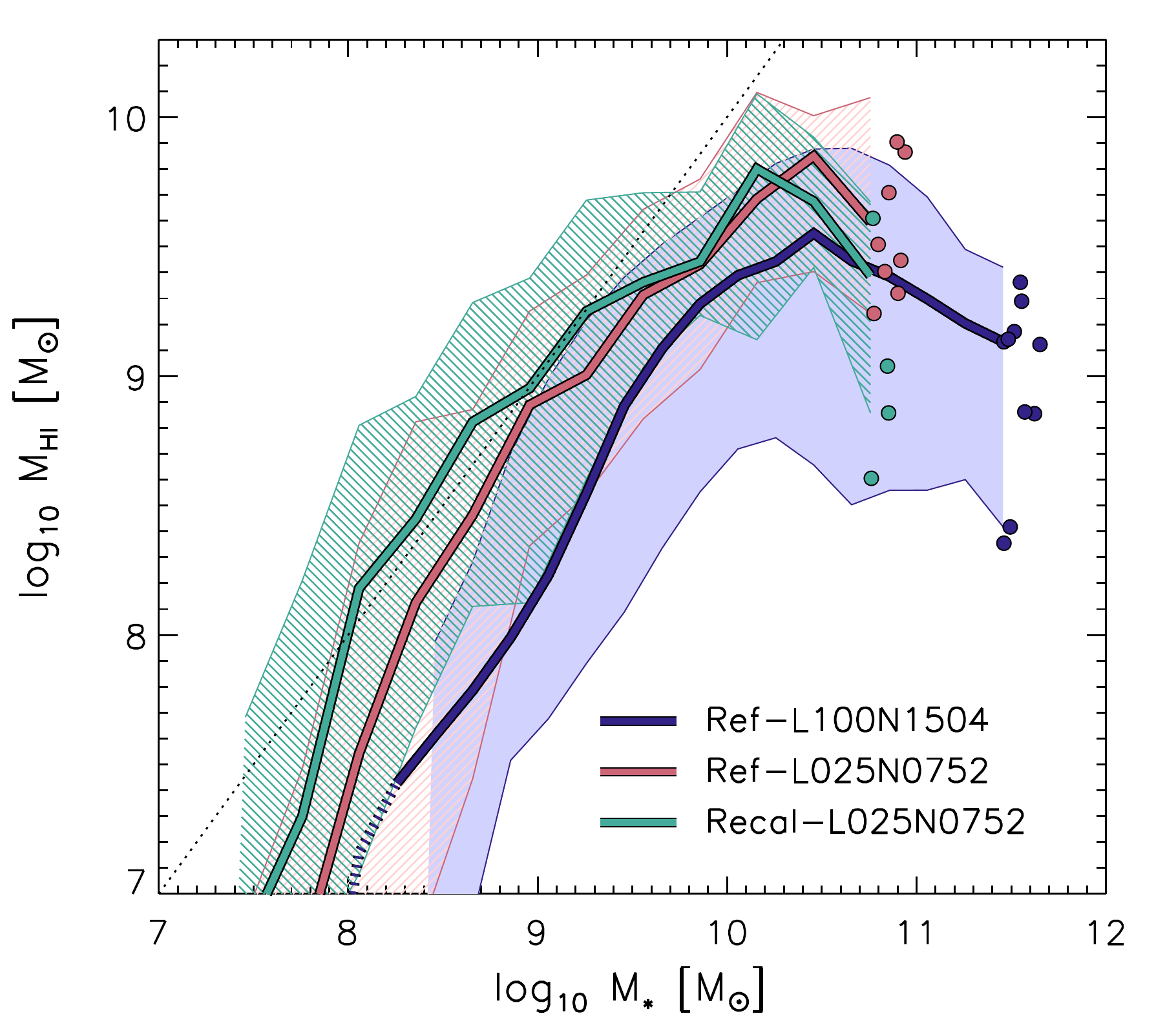}
  \includegraphics[width=\columnwidth]{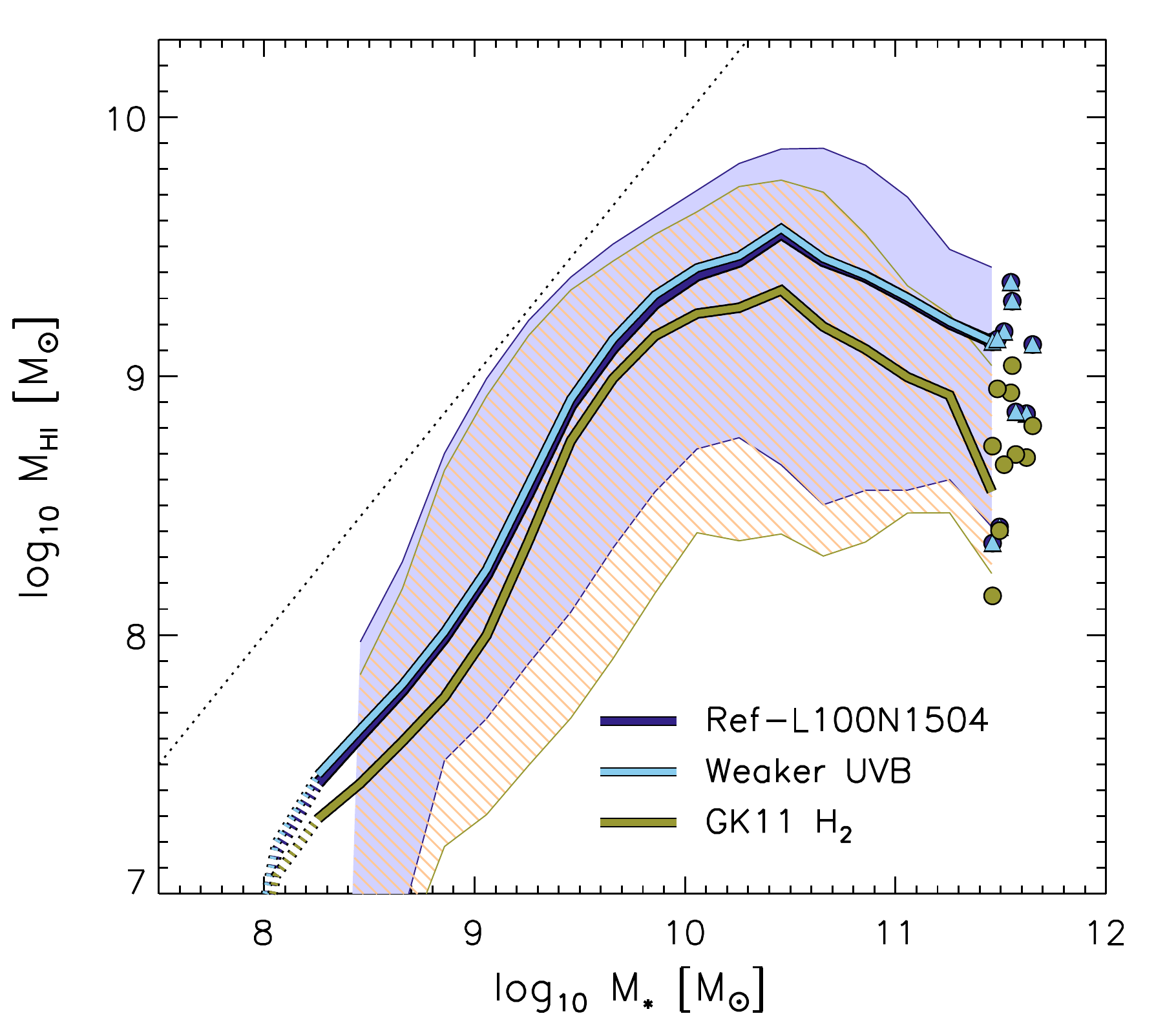}
  \caption{The mass of atomic hydrogen, $M_{\rm HI}$, associated with central galaxies as a function of their stellar mass, $M_\star$, at $z=0$. Curves show the median $M_{\rm HI}$ in bins of ${\rm d}\log_{10} M_\star = 0.2$ ($0.3$ for $L=25\cMpc$ simulations), and are drawn with a dotted linestyle where $M_\star < 100 m_{\rm g}$. Individual galaxies are plotted where bins contain fewer than 10 galaxies. Shaded regions denote the $1\sigma$ scatter about the median, and the overplotted dotted line traces the locus $M_{\rm HI} = M_\star$. \textit{Left}: The relation recovered when applying the fiducial BR06 scheme for partitioning hydrogen into atomic and molecular components (BR06) to Ref-L100N1504 (dark blue), Ref-L025N0752 (red), and Recal-L025N0752 (green). This enables an assessment of the strong and weak forms of convergence. \textit{Right}: Comparison of the fiducial relation from Ref-L100N1504 with those recovered under the assumption of a UVB photoionization rate, $\Gamma_{\rm UVB}$, a factor of 3 lower than the fiducial case (cyan), and when partitioning neutral hydrogen into atomic and molecular components using the GK11 scheme (gold). Owing to their similarity to the fiducial case, 
individual galaxies in the weaker UVB case are plotted as triangles for clarity.}
\label{fig:M_HI_convergence}
\end{figure*}

Fig. \ref{fig:CDDF} also enables an assessment of the systematic uncertainty of the CDDF stemming from the partitioning of hydrogen into \HII, \HI, and \Hmol. The cyan curve denotes the CDDF of Ref-L100N1504 recovered assuming a weaker UVB (a photoionization rate, $\Gamma_{\rm UVB}$, one-third of the of fiducial value), and the gold curve shows the effect of neglecting the BR06 correction for \Hmol, effectively treating all neutral gas as \HI. The weaker UVB only impacts the \HI\ CDDF noticeably at  $\log_{10} N_{\rm HI}\,[\cmsquared] \lesssim 20$, shifting systems of fixed abundance to slightly higher column density, e.g. by $0.20\,{\rm dex}$ at $\log_{10} f(N_{\rm HI})[\cmsquared]= -20$. Similar behaviour is seen at higher redshift \citep[see Fig. A1 of][]{Rahmati_et_al_15}. In contrast, the omission of a correction for \Hmol\ only affects high column densities, $\log_{10} N_{\rm HI}\,[\cmsquared] \gtrsim 21.3$. Although the effect on the CDDF is relatively small, the highest column densities dominate the \HI\ mass of galaxies and we show in \S\ref{sec:M_HI-Mstar} that the partitioning of neutral hydrogen can affect the $M_{\rm HI}-M_\star$ relation significantly.

We reiterate that radiation sources within and/or local to galaxies, which are not considered here, may also influence the abundance of the highest column density systems $\log_{10} N_{\rm HI}\,[\cmsquared] \gtrsim 21$ \citep{R13b}. For lower column densities, it is clear that numerical resolution is a more significant systematic uncertainty than the (re)calibration of the subgrid model parameters at high resolution, and the methods for partitioning hydrogen into \HII, \HI\ and \Hmol. We show in \S\ref{sec:HIMF} that the shortfall of high-$N_{\rm HI}$ systems seen at standard resolution is also manifest in the mass function of \HI\ sources, and explore the origin of the shortfall in \S\ref{sec:HIMF_Mstar}.

\section{The properties of atomic hydrogen associated with galaxies}
\label{sec:HI_in_gals}

This section begins with an examination of the atomic hydrogen mass, $M_{\rm HI}$, of galaxies as a function of their stellar mass, $M_\star$. A key difference with respect to the study of \citet{Bahe_et_al_16_short} is that here we examine the relation for all central galaxies, not only those similar to the GASS sample, and hence test the relation at lower stellar masses. The strong and weak resolution convergence behaviour of the $M_{\rm HI}-M_\star$ relation is explored, as is the impact of systematic uncertainties associated with partitioning hydrogen into \HII, \HI\ and \Hmol. Since the gas properties of galaxies were not considered during the calibration of the model parameters, the \HI\ masses of galaxies in the calibrated simulations can be considered predictions, and are compared with observational measurements. We also examine the redshift evolution of the $M_{\rm HI}-M_\star$ relation, and the impact of varying the efficiency of SF and its associated feedback with respect to the calibrated simulations, enabling assessment of the sensitivity of \HI\ masses to these elements of the EAGLE model. 

A comparison of the HIMF with those recovered by 21 cm emission line surveys is presented in \S\ref{sec:HIMF}. Note that, although the Ref and Recal models were calibrated at $z=0.1$, which is the approximate median redshift of the optical galaxy redshift surveys used for the calibration, the results presented here are generally derived from the $z=0$ output of each simulation, to reflect the lower median redshift of the current generation of 21 cm surveys. The origin of the relatively poor convergence behaviour of the \HI\ masses of low-stellar mass galaxies is explored in \S\ref{sec:HIMF_Mstar}.

\subsection{The HI mass of galaxies}
\label{sec:M_HI-Mstar}

The left-hand panel of Fig. \ref{fig:M_HI_convergence} shows the atomic hydrogen mass, $M_{\rm HI}$, of central galaxies, as a function of their stellar mass, $M_\star$, at $z=0$. Satellite galaxies are excluded from this analysis since they can be subject to environmental processes that can have a significant impact upon their atomic hydrogen reservoirs \citep[e.g.][]{Cortese_et_al_11,Fabello_et_al_12,Jaffe_et_al_15}. Results are shown for three simulations: Ref-L100N1504 (dark blue), Ref-L025N0752 (red) and Recal-L025N0752 (green), to facilitate strong and weak convergence tests. The right-hand panel illustrates the effect of adopting a weaker UVB (photoionization rate one-third of the fiducial value, cyan) and the use of the theoretically motivated GK11 scheme (rather than the empirical BR06 pressure law) for partitioning neutral hydrogen into atomic and molecular components (gold).

The \HI\ mass of galaxies in Ref-L100N1504 scales approximately linearly with stellar mass in the interval $10^{8} \lesssim M_\star \lesssim 10^{10.5}\Msun$. At lower masses, the relation cannot be reliably sampled by Ref-L100N1504, but the high-resolution simulations indicate that the relation steepens. At greater masses, the $M_{\rm HI}-M_\star$ relation turns over. Whilst a small part of this effect may be due to the reservoirs of the most \HI-massive galaxies extending beyond the $70\pkpc$ aperture that we adopt \citep[owing to the classical Malmquist bias, the most 21 cm-luminous sources may warrant the use of a larger aperture, but see appendix A2 of][]{Bahe_et_al_16_short}, the primary factor governing the massive end of the $M_{\rm HI}-M_\star$ relation of the simulations is the change of the massive galaxy population from gas-rich discs to dispersion-supported, gas-poor ellipticals \citep[as shown for EAGLE by][]{Furlong_et_al_15_short,Trayford_et_al_15_short}, which host massive central BHs that heat and eject cold gas via AGN feedback. The gas can therefore be quickly depleted by both SF and feedback in these systems. The higher characteristic pressure of gas confined by the deep potential of a massive galaxy also exhibits a higher \Hmol\ fraction at the expense of \HI. In addition, the cold gas reservoir has a longer time-scale for replenishment in massive galaxies, because the cooling of gas on to the galaxy is dominated by cooling out of a quasi-hydrostatic hot halo \citep[e.g.][]{van_de_Voort_et_al_11a}. Examination of the $M_{\rm HI}-M_{200}$ and $M_{\rm H2}-M_{200}$ relations for central galaxies (for brevity, not shown) of Ref-L050N0752 and a matched simulation in which the AGN feedback is disabled indicates that the decline is primarily driven by the regulatory action of AGN.

The median  $M_{\rm HI}-M_\star$ relation is offset to higher $M_{\rm HI}$ in the high-resolution simulations with respect to Ref-L100N1504. The offset can be almost one decade in $M_{\rm HI}$ at $M_\star \simeq 10^{8.5}\Msun$, whilst for $M_\star \gtrsim 10^{9.5}\Msun$ the offset between the Ref-L100N1504 and the Ref-L025N0752 simulations declines to $<0.3\,{\rm dex}$. This convergence behaviour is consistent with that of the \HI\ CDDF, with systems of fixed $f(N_{\rm HI})$, particularly the rarer high column density systems that are clearly associated with galaxies \citep[e.g.][]{Rahmati_et_al_15}, being shifted to higher $N_{\rm HI}$ in the high-resolution simulations. As in that case, the offset is not a consequence of the simulation box size, since the $M_{\rm HI}-M_\star$ relations of Ref-L100N1504 and Ref-L025N0376 (for brevity, not shown) are very similar.

The convergence behaviour of the $M_{\rm HI}-M_\star$ relation is qualitatively similar to that of the relationship between the gas-phase metallicity and stellar mass of galaxies ($Z_{\rm gas}-M_\star$), with greater $M_{\rm HI}$ corresponding to lower $Z_{\rm g}$ \citep[see also Fig. 9 of][]{Lagos_et_al_16_short}. The high-resolution simulations therefore yield galaxies of lower metallicity and higher \HI\ mass at fixed $M_\star$. \citet[][their Fig. 13]{S15} noted that $Z_{\rm gas}-M_\star$ was the only major scaling relation for which the discrepancy between Ref-L100N1504 and observational measurements is substantially greater than the associated observational uncertainty. We show later in \S\ref{sec:obs_measurements} that this shortcoming is a consequence of galaxies with $M_\star \lesssim 10^{10}\Msun$ in Ref-L100N1504 exhibiting unrealistically low cold gas fractions. 

Comparison of the two high-resolution simulations highlights that the stronger outflows of the Recal model (with respect to Ref), which are a consequence of the recalibration of the subgrid feedback efficiencies necessary to reproduce the GSMF at high resolution, further increase the \HI\ mass (and reduce the metallicity - see S15) of galaxies at fixed $M_\star$. The difference between Ref-L100N1504 and Recal-L025N0752 is possible because the simulations were calibrated to reproduce the low-redshift GSMF, not the $M_{\rm HI}-M_\star$ relation, and it is clear that calibrating to reproduce the former does not guarantee reproduction of the latter. The higher resolution simulations exhibit systematially greater \HI\ masses at fixed $M_\star$, and the cause of this offset is explored in detail in \S\ref{sec:HIMF_Mstar}. It is also noteworth that at $M_\star \simeq 10^{10}\Msun$, the $1\sigma$ scatter in $M_{\rm HI}$ at fixed $M_\star$ is significantly lower in the high-resolution simulations than in Ref-L100N1504\footnote{The scatter about the median $M_{\rm HI}-M_\star$ relation, at fixed resolution, is well-converged with box size from $L=25\cMpc$ to $L=100\cMpc$ over the stellar mass range adequately sampled by the $L=25\cMpc$ volumes.}.

\citet{Bahe_et_al_16_short} demonstrated that a factor of 3 reduction in the photoionization rate of the UVB does not impact significantly upon the \HI\ content of massive EAGLE galaxies ($M_\star > 10^{10}\Msun$). It is clear from the right-hand panel of Fig. \ref{fig:M_HI_convergence} that this is the case for all galaxies resolved by Ref-L100N1504; the median $M_{\rm HI}-M_\star$ relation of the weak UVB case is near-identical to the fiducial case; for this reason, the scatter about the former is not shown. The systematic shift of low column density systems to slightly greater $N_{\rm HI}$ when adopting a weaker UVB (Fig. \ref{fig:CDDF}) does not therefore translate into a significant increase of the \HI\ masses of galaxies.

The choice of scheme for partitioning neutral hydrogen into \HI\ and \Hmol\ components is significant, however. The GK11 scheme yields systematically-lower $M_{\rm HI}$ for galaxies of all $M_\star$; the offset with respect to the fiducial BR06 scheme is small in low-mass galaxies,  $\lesssim 0.2\,{\rm dex}$ for $M_\star \lesssim 10^{10}\Msun$, but grows as large as $0.4\,{\rm dex}$ for galaxies of $M_\star \simeq 10^{11.5}\Msun$, whose cold gas reservoirs are typically at high pressure. As is clear from Fig. \ref{fig:phase_diagrams}, this is a consequence of the GK11 scheme specifying a systematically higher molecular fraction than BR06 at fixed density \citep[see also][]{Bahe_et_al_16_short}. This shift highlights that uncertainty on the column density of high column density systems translates into a significant uncertainty on the \HI\ masses of galaxies \citep[see also e.g.][]{Rahmati_et_al_15}, and serves as a reminder that our neglect of radiation sources within and/or local to galaxies, which likely influence high column density systems, represents a potentially significant source of unquantified systematic uncertainty.

\begin{figure}
\includegraphics[width=\columnwidth]{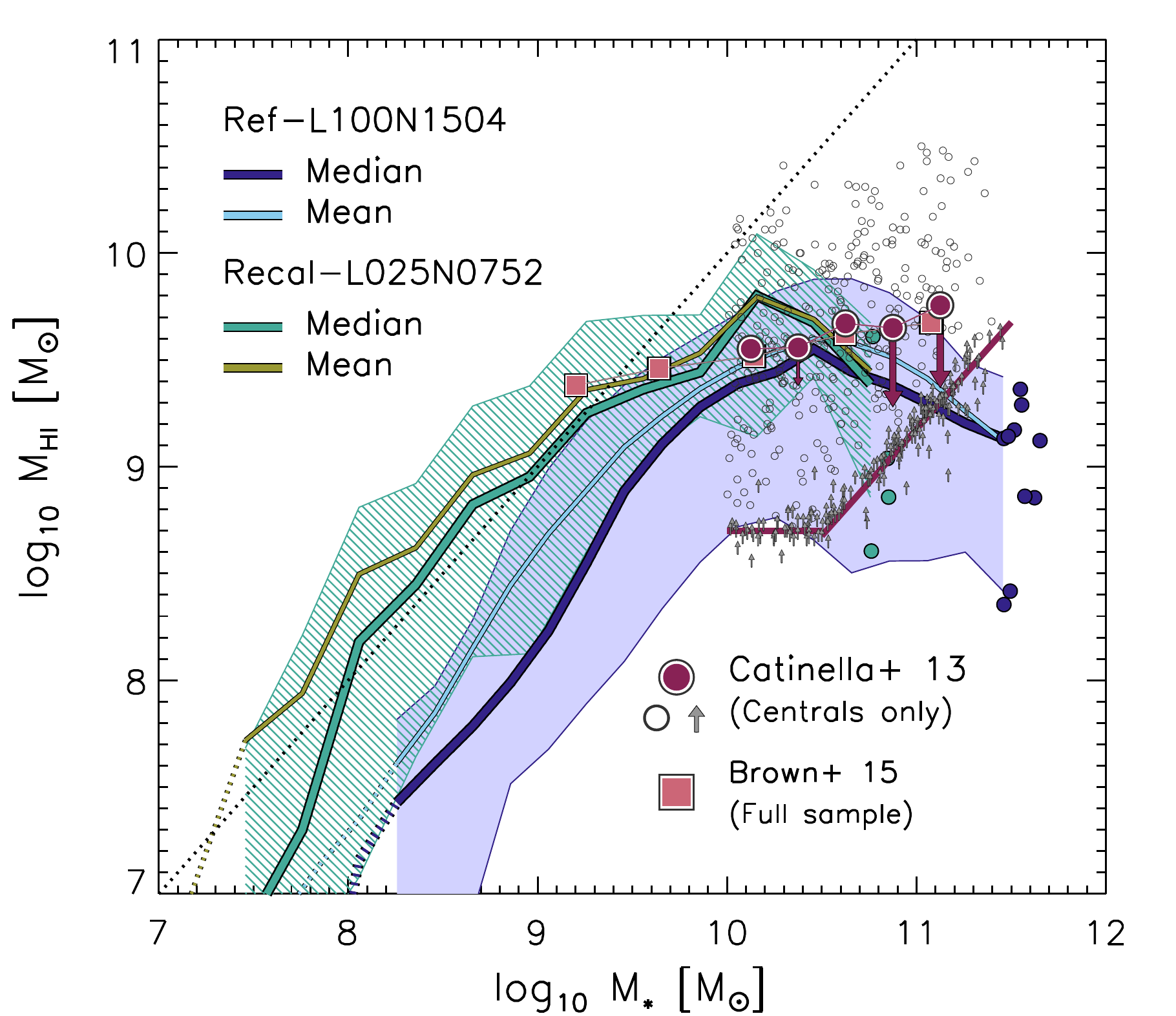}
\caption{Comparison of the $M_{\rm HI}-M_\star$ relations of Ref-L100N1504 and Recal-L025N0752 at $z=0$, with measurements from 21 cm surveys. The median of $M_{\rm HI}$ (thick curves) is shown for comparison with Fig. \ref{fig:M_HI_convergence}, and the linear mean is shown (thin curves) for comparison with observational data. GASS survey detections of central galaxies with $M_\star > 10^{10} \Msun$ are plotted as open circles, upper limits are drawn as grey arrows. The binned mean of these data are denoted by maroon circles, and are connected to their binned median by maroon arrows. In both cases, non-detections are assigned the GASS survey detection threshold \HI\ mass, which is denoted by the maroon lines. Red squares represent the binned mean of measurements based on 21 cm stacking about optical sources, enabling \HI\ masses to be probed to $M_\star \sim 10^9\Msun$. At these stellar masses, galaxies in Ref-L100N1504 are deficient in \HI\ with respect to observational measurements.}
\label{fig:M_HI_obs}
\end{figure}

\subsubsection{Confrontation with observational measurements}
\label{sec:obs_measurements}

At present, untargeted 21 cm surveys such as ALFALFA \citep{Giovanelli_et_al_05_short} and HIPASS \citep{Barnes_et_al_01_short,Meyer_et_al_04_short} lack the sensitivity to detect galaxies deficient in \HI\ beyond the very local Universe. This biases their detections significantly towards systems that, at fixed stellar mass, are uncharacteristically \HI-rich. To combat this, the GASS 
survey \citep{Catinella_et_al_10_short,Catinella_et_al_13_short} targeted $\simeq 800$ relatively massive ($M_\star > 10^{10}\Msun$) galaxies with the Arecibo radio telescope, to assemble a stellar mass-selected sample of galaxies with both optical and 21 cm measurements. Since the majority of the GASS sources also feature in the galaxy group catalogue of \citet{Yang_et_al_12}, which is based on analysis of the seventh data release of the Sloan Digital Sky Survey \citep[SDSS-DR7;][]{Abazajian_et_al_09}, \citet{Catinella_et_al_13_short} were able to label the galaxies as centrals and satellites, enabling the direct confrontation of central galaxies in EAGLE with observed counterparts. The GASS catalogue comprises 750 galaxies with \HI\ and stellar mass measurements in the redshift interval $0.025 < z < 0.05$; of these galaxies, 482 are centrals that \citet{Catinella_et_al_13_short} report as unaffected by source confusion stemming from the $\simeq 3.5$ arcmin beam of the ALFA instrument.

The GASS galaxies are plotted as open circles in $M_{\rm HI}-M_\star$ space in Fig. \ref{fig:M_HI_obs}; where a 21 cm detection was not obtained, the upper limit on $M_{\rm HI}$ is plotted as an upward arrow. Maroon circles denote the linear mean $M_{\rm HI}$ of GASS galaxies in bins of $M_\star$, and these data are connected to the median by maroon arrows. Both diagnostics were computed by assigning the threshold \HI\ mass of the GASS survey to non-detections, and we have verified that assigning $M_{\rm HI} = 0$ instead to non-detections introduces a systematic shift that is significantly smaller than the difference between the binned mean and median values shown in Fig. \ref{fig:M_HI_obs}. Since the simulations are able to probe the \HI\ masses of galaxies much less massive than the GASS lower limit of $M_\star = 10^{10}\Msun$, we also show with red squares the linear mean $M_{\rm HI}-M_\star$ relation recovered by \citet{Brown_et_al_15}, who recently spectrally-stacked 21 cm data about the coordinates of 25,000 optical sources in the overlap of the ALFALFA and SDSS-DR7 survey footprints. Although this precludes the explicit separation of central and satellite galaxies, potentially biasing $M_{\rm HI}$ low at fixed $M_\star$, it enables examination of galaxies as low as $M_\star \simeq 10^9\Msun$, an order of magnitude less massive than GASS. 

The figure shows both the median $M_{\rm HI}$, (thick curves) as plotted in Fig. \ref{fig:M_HI_convergence}, and the linear mean $M_{\rm HI}$ (thin curves), for comparison with the observational measurements, of Ref-L100N1504 (dark blue, light blue) and Recal-L025N0752 (green, gold). Shaded regions denote the $1\sigma$ scatter about the median curves, as per Fig. \ref{fig:M_HI_convergence}. The correspondence between the mean relations of GASS and Ref-L100N1504 simulation is good; over the mass interval $10^{10} < M_\star < 10^{11} \Msun$, the mean $M_{\rm HI}$ of simulated galaxies at fixed $M_\star$ differs from that inferred from GASS by $<0.14\,{\rm dex}$; \citet{Bahe_et_al_16_short} showed previously that the median relations also differ by $<0.2\,{\rm dex}$. The median $M_{\rm HI}$ of more massive galaxies in Ref-L100N1504 declines below the GASS detection limit. 

At the lower masses accessible only via 21 cm spectral stacking, galaxies in Ref-L100N1504 appear to be deficient in \HI. At $M_\star \simeq 10^9$, the mean $M_{\rm HI}$ of simulated galaxies differs from that inferred by \citet{Brown_et_al_15} by $-0.55\,{\rm dex}$. The $M_{\rm HI}-M_\star$ relation of the high-resolution Recal-L025N0752 simulation compares more favourably; over the interval $M_\star = 10^{9} - 10^{10} \Msun$ (the small volume of this simulation precludes meaningful comparison with more massive galaxies), the offset of the mean $M_{\rm HI}$ at fixed $M_\star$ with respect to that recovered by \citet{Brown_et_al_15} is $-0.08\,{\rm dex}$. The deficiency of \HI\ masses with respect to observational measurements is commensurate with the offset of the column density of \HI\ absorption systems at fixed density (Fig. \ref{fig:CDDF}).

\citet{Bahe_et_al_16_short} noted, as is clear from Fig. \ref{fig:M_HI_obs}, that the simulations do not exhibit scatter in $M_{\rm HI}$ at fixed $M_\star$ as great as measured by GASS. Most notably, the most HI-rich observed galaxies are absent from the simulations; the $50^{\rm th}$ and $84^{\rm th}$ percentiles of the measurements are separated by up to $0.7\,{\rm dex}$, in contrast to $< 0.44\,{\rm dex}$ for Ref-L100N1504. They suggested that the deficiency of HI-rich galaxies in EAGLE might be a consequence of the finite resolution of the simulations; the internal energy injected into the ISM by individual feedback heating events is linearly proportional to the baryon particle mass, $m_{\rm g}$, and in the case of the standard-resolution simulations they are a factor of $\sim 10^4$ more energetic than individual SNe. Locally \HI-rich patches of EAGLE's galaxy discs are therefore efficiently heated and/or ejected by feedback events that are much more stochastic than in real galaxies. We return to this issue in \S\ref{sec:subgrid_sensitivity}.

\subsubsection{The redshift evolution of galaxy HI masses}

\begin{figure}
\includegraphics[width=\columnwidth]{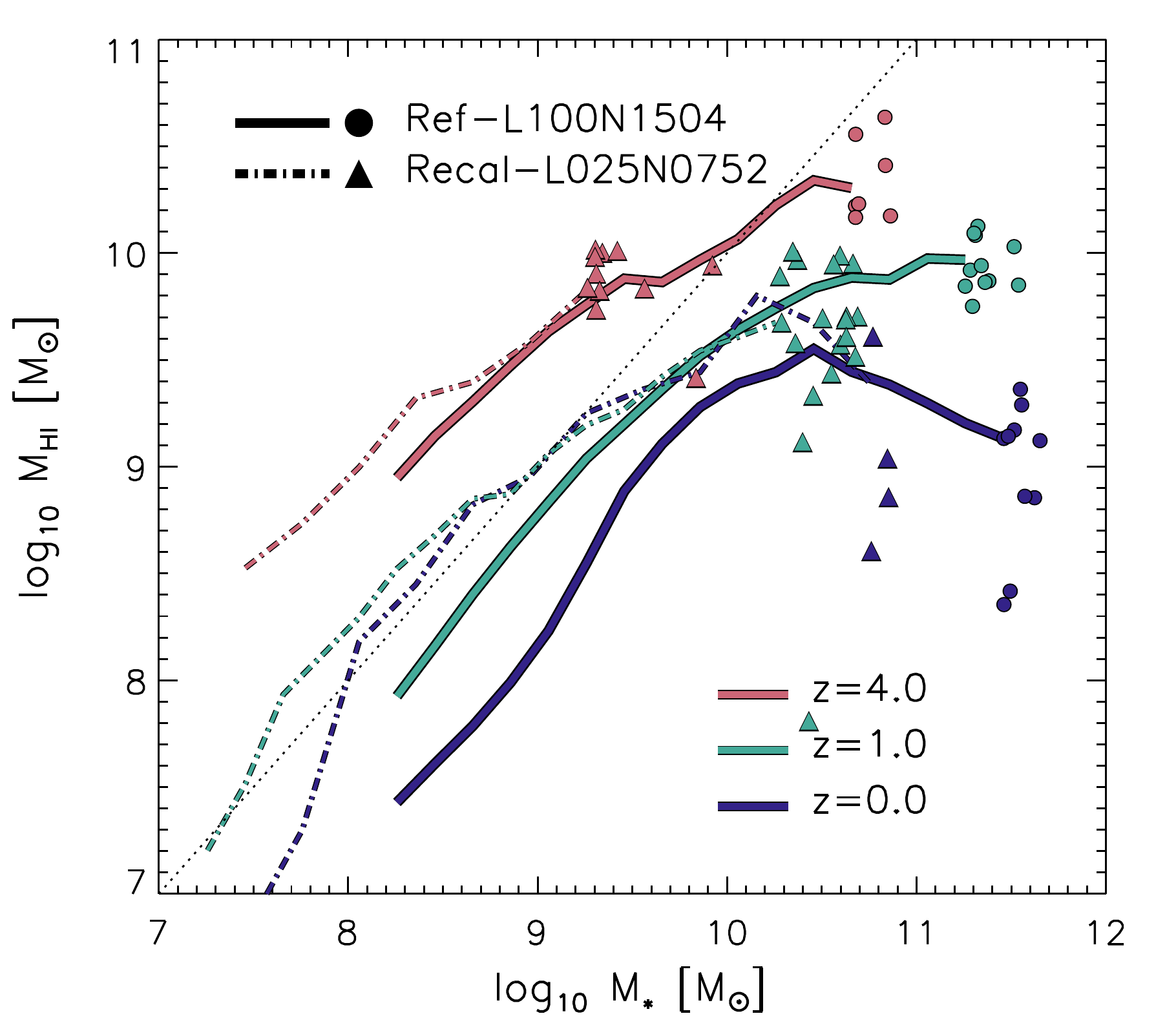}
\caption{The $M_{\rm HI}-M_\star$ relations of the Ref-L100N1504 (solid lines and circles) and Recal-L025N0752 (dot-dashed lines and triangles) simulations, at $z=4$ (red), $z=1$ (green) and $z=0$ (dark blue). For clarity, curves are not drawn for mass bins $M_\star < 100m_{\rm g}$. The evolution of the relation between $z=0$ and $z=1$ is markedly resolution-dependent: it evolves strongly in Ref-L100N1504 but barely at all in Recal-L025N0752 (and similarly for Ref-L025N0752). For $z > 1$, in both simulations, the \HI\ mass at fixed $M_\star$ evolves strongly, in concert with the evolution of the neutral gas mass.} 
\label{fig:M_HI_evol}
\end{figure}

Fig. \ref{fig:M_HI_evol} shows the $M_{\rm HI}-M_\star$ relations of central galaxies in the Ref-L100N1504 (solid lines and circles) and Recal-L025N0752 (dot-dashed lines and triangles) simulations, at $z=4$ (red), $z=1$ (green) and $z=0$ (blue). It highlights that the weak convergence of the relation is relatively good for $z\gtrsim 1$, with the poor convergence behaviour shown in Fig. \ref{fig:M_HI_convergence} developing at late epochs: galaxies at fixed $M_\star$ exhibit \HI\ masses that decline significantly between $z=1$ and $z=0$ in Ref-L100N1504, whilst counterparts in Recal-L025N0752 do not. Inspection of the overall gas fraction of galaxies (for brevity, not shown) indicates that the poor convergence is a consequence of the development, particularly at $z < 1$, of significant differences in the overall neutral hydrogen mass of galaxies (at fixed $M_\star$) between the two simulations, rather than differences in, for example, the molecular fractions.

Therefore, Recal-L025N0752 indicates that the typical \HI\ mass of galaxies does not evolve significantly over the approximate redshift interval spanned by, for example, the CHILES survey `deep field' ($z=0$ to $z=0.5$). \HI\ masses do however grow significantly at $z > 1$, at $M_\star = 10^9\Msun$ ($10^{10}\Msun$), the median \HI\ mass of galaxies at $z=4$ in Recal-L025N0752 is a factor of 4.2 (2.0) greater than at $z=0$. This evolutionary trend is consistent with the finding of \citet{Rahmati_et_al_15} that the covering fraction of strong \HI\ absorbers ($\log_{10} N_{\rm HI}\,[\cmsquared] > 17$), as a function of both their halo mass and specific SFR, also increases with redshift for $z \gtrsim 1$. These trends are primarily a reflection of the increased overall cold gas fraction of galaxies but, owing to the greater molecular fraction at high redshift \citep[see also][]{Duffy_et_al_12}, the neutral mass of galaxies grows more rapidly than the atomic mass: at $M_\star = 10^9\Msun$ ($10^{10}\Msun$) the median neutral hydrogen mass of galaxies in Recal-L025N0752 is a factor of 5.0 (3.1) greater at $z=4$ than at $z=0$.

Whilst the \citet{R13a} fitting function was calibrated against TRAPHIC RT simulations from $z=5$ to $z=0$, justifying its use at $z > 0$, we caution that prediction of the $M_{\rm HI}-M_\star$ relation at high redshift is subject to greater systematic uncertainty than at $z=0$. The impact of radiation sources within and/or local to galaxies on the column density of strong absorbers (and hence the \HI\ mass of galaxies), which is not modelled here, is greater at high redshift \citep{R13b}. Moreover, as also noted by \citet{Duffy_et_al_12}, the use of the BR06 pressure law to partition neutral hydrogen into \HI\ and \Hmol\ is less readily justified; the relation is a fit to observations of local galaxies, yet the atomic to molecular transition is likely sensitive to the characteristic (column) density and metallicity of the gas in and around galaxies \citep[e.g.][]{Schaye_01,Krumholz_and_McKee_08,Krumholz_McKee_and_Tumlinson_09}, both of which evolve with redshift. 

\subsubsection{Sensitivity to details of subgrid feedback models}
\label{sec:subgrid_sensitivity}

\begin{figure*}
\includegraphics[width=0.49\textwidth]{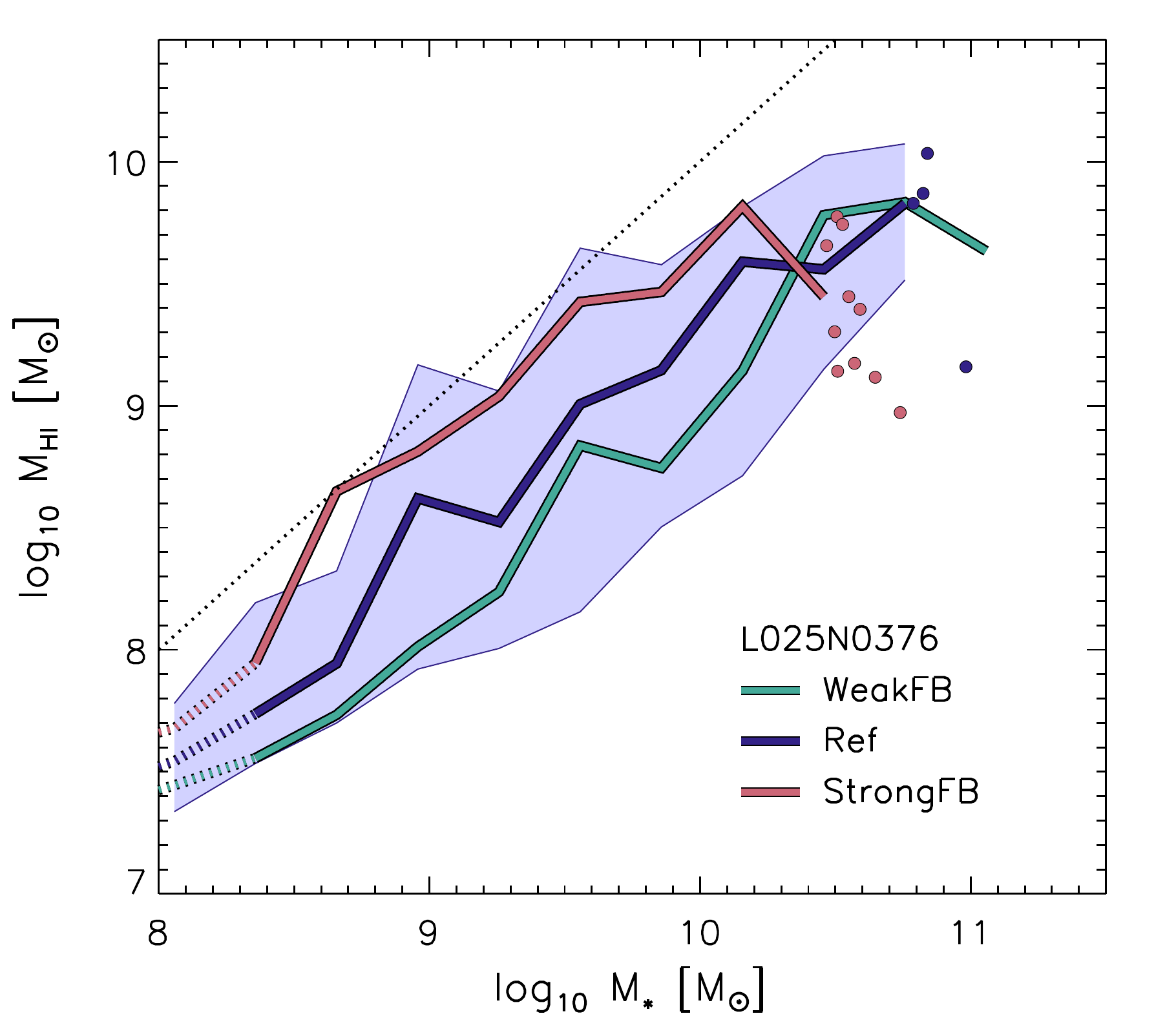}
\includegraphics[width=0.49\textwidth]{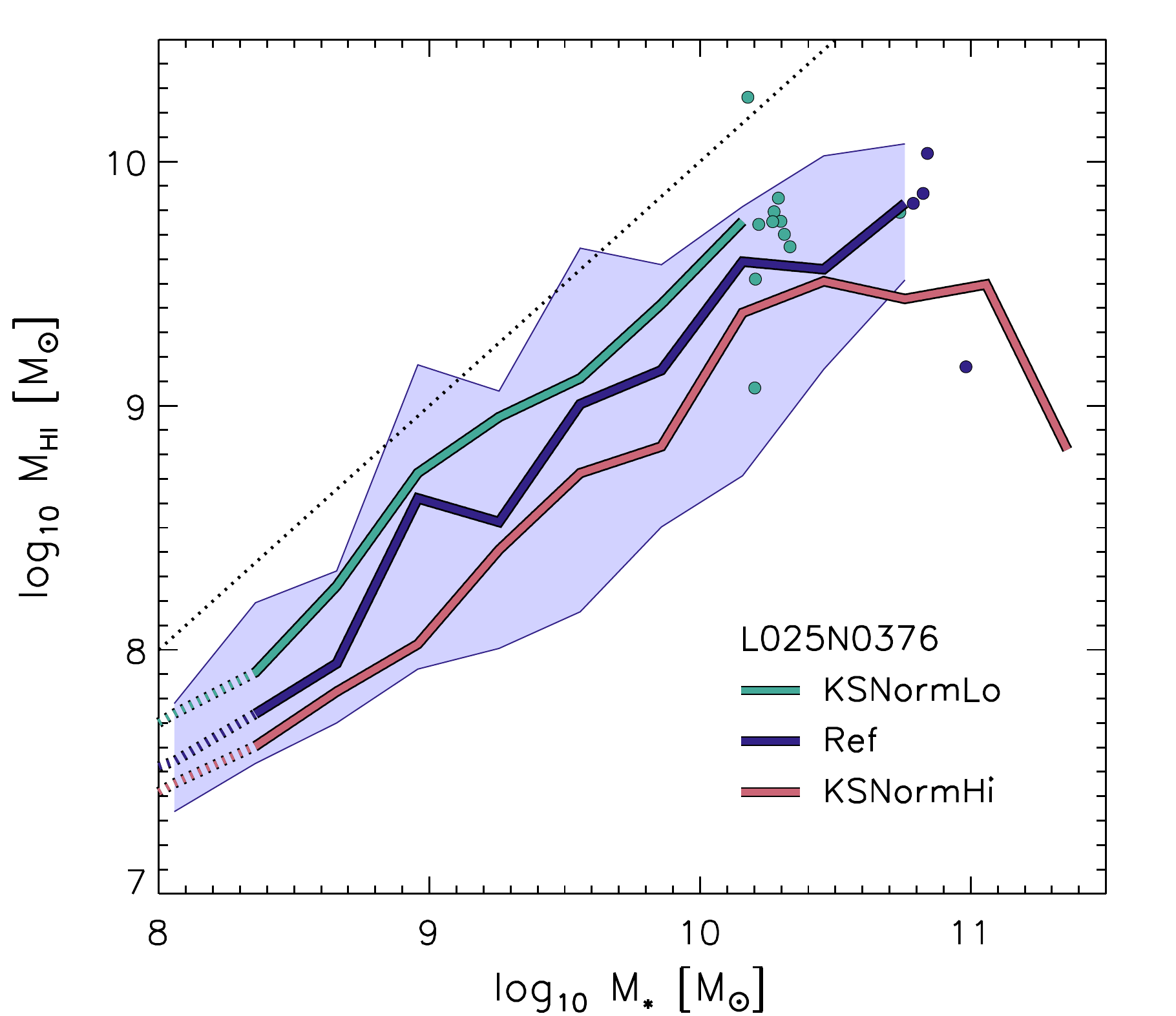}
\caption{The mass of \HI\ associated with \textit{star-forming} galaxies as a function of stellar mass at $z=0$, in L025N0376 simulations incorporating subgrid parameter variations of the EAGLE reference model. For clarity, the $1\sigma$ scatter is shown only for Ref in each panel. \textit{Left:} Comparison of the Ref model (realized using L025N0376 initial conditions) with models in which the asymptotic values of the SF feedback efficiency function differ by factors of 0.5 (WeakFB) and 2 (StrongFB). \textit{Right:} Comparison of Ref-L025N0376 with models in which the normalization of the Kennicutt-Schmidt law have been varied by $\pm0.5\,{\rm dex}$ (KSNormLo, KSNormHi). At fixed resolution, the mass of \HI\ associated with star-forming galaxies of fixed $M_\star$ increases with greater feedback efficiency and decreases with greater SF efficiency.}
\label{fig:M_HI_subgrid}
\end{figure*}

The framework of equilibrium models \citep[e.g][]{White_and_Frenk_91,Finlator_and_Dave_08,Schaye_et_al_10_short,Dave_Finlator_and_Oppenheimer_12,Lilly_et_al_13}, posits that star-forming galaxies exhibit a self-regulated equilibrium such that the cosmological gas inflow rate is balanced by the combined sinks of SF and ejective feedback (i.e. $\dot{M}_{\rm infall} = {\rm SFR} + \dot{M}_{\rm eject}$). This is a helpful means of understanding the sensitivity of the $M_{\rm HI}-M_\star$ relation to, for example, the assumed efficiency of SF feedback and the assumed SF law. The panels of Fig. \ref{fig:M_HI_subgrid} compare the $M_{\rm HI}-M_\star$ relation of Ref-L025N0376 with the relation from simulations in which the parameters governing the subgrid implementation of these aspects of the model are systematically varied from those adopted by Ref. In both panels, the median relation is plotted (and, in the case of the Ref model, the $1\sigma$ scatter about it) for \textit{star-forming} galaxies (i.e. specific star formation rate, ${\rm sSFR} \equiv {\rm SFR} / M_\star > 10^{-11}\peryr$), rather than central galaxies, as per Figs. \ref{fig:M_HI_convergence} and \ref{fig:M_HI_obs}, since the equilibrium framework is not applicable to quenched galaxies. However, since most central galaxies of $M_\star \lesssim 10^{11}\Msun$ are star-forming, plots featuring central galaxies exhibit similar trends. 

The left-hand panel compares Ref-L025N0376 with two simulations for which the efficiency of SF feedback at fixed density and metallicity is half (WeakFB-L025N0376) and twice (StrongFB-L025N0376) as efficient as in Ref (see also Table \ref{tbl:sims}). At fixed $M_\star$, $M_{\rm HI}$ is systematically offset by up to $\simeq 0.3\,{\rm dex}$ from that of Ref in the WeakFB and StrongFB simulations, with more efficient feedback resulting in greater \HI\ masses at fixed stellar mass. This result may appear counter-intuitive, since it is reasonable to suppose that more efficient feedback would suppress the development of \HI\ reservoirs. A qualitatively similar result was reported by \citet{Dave_et_al_13}; in their simulations with kinetic SF feedback, they found that imposing a higher mass loading in winds resulted in galaxies of fixed $M_\star$ being richer in atomic hydrogen. Similar responses to changes in the feedback efficiency were noted in the context of the \HI\ CDDF by \citet{Altay_et_al_13} and \citet{Rahmati_et_al_15}, the latter also noting a correlation between the \HI\ covering fraction of galaxies (at fixed $M_\star$) and the efficiency of feedback associated with SF.

The framework of equilibrium models elucidates the cause of this effect. \citet{C15} identified that, since the WeakFB-L025N0376 and StrongFB-L025N0376 simulations do not reproduce the low-redshift GSMF, the typical virial mass ($M_{200}$) of haloes associated with galaxies of a fixed $M_\star$ also differs significantly between these simulations (see Fig. 10b of that study). Since haloes of greater virial mass experience a greater specific infall rate at fixed redshift \citep[e.g.][]{Neistein_and_Dekel_08,Fakhouri_Ma_and_Boylan_Kolchin_10,Correa_et_al_15}, if a star-forming galaxy of fixed $M_\star$ resides in a more massive halo, it must assemble a more massive reservoir of star-forming gas to achieve the rates of SF and ejective feedback necessary to self-regulate. To first order, the $M_{\rm HI}-M_\star$ curves shift horizontally, not vertically, as the feedback efficiency is varied.

The simple, steady scaling of $M_{\rm HI}$ with feedback efficiency indicates that the gas reservoirs of EAGLE galaxies are finely regulated by feedback. It is tempting to speculate that the small scatter in $M_{\rm HI}$ at fixed $M_\star$ exhibited by EAGLE galaxies when compared with that measured by the GASS survey (Fig. \ref{fig:M_HI_obs}) is a consequence of the former self-regulating too finely; thus they are inhibited from (temporarily) accruing gas reservoirs considerably more massive than is typical for their stellar mass. This is also arguably corroborated the reduced scatter in the high-resolution L0025N0752 simulations (Fig. \ref{fig:M_HI_convergence}), for which the numerical sampling of the stochastic feedback is improved. However a detailed examination of such effects is beyond the scope of this study.

We stress that the offset in $M_{\rm HI}$ at fixed $M_\star$ is not a consequence of a change of the \Hmol\ mass fraction of the reservoirs. If a reservoir of fixed neutral mass is hosted by a more massive halo in response to an increased feedback efficiency, the characteristic pressure of the gas increases and so elevates the \Hmol\ fraction at the expense of the \HI\ fraction, contrary to the trend seen here. Therefore any such effect is clearly more than compensated by the change of infall rate. Nonetheless, inspection of the $M_{\rm HI}-M_{200}$ relations for star-forming galaxies (for brevity, not shown here), highlights that the \HI\ mass at fixed \textit{halo} mass is weakly dependent upon the SF feedback efficiency: haloes with $M_{200} \gtrsim 10^{11.5}\Msun$ have slightly less massive \HI\ reservoirs in the WeakFB simulation than their counterparts in the Ref and StrongFB simulations. This is likely a consequence of the haloes in the WeakFB simulation being unable to adequately regulate the cooling on to the galaxy of the gas delivered to the halo by cosmological infall. 

Since the SFR also enters the self-regulation continuity equation, and because the mass of gas heated by SF feedback is proportional to the SFR, the equilibrium gas mass is also governed in part by the adopted SF efficiency. The fiducial value of $A=1.515\times 10^{-4}\Msunyrkpcsq$ adopted by Ref is varied by $0.5\,{\rm dex}$ in the KSNormLo and KSNormHi simulations (Table \ref{tbl:sims}). The right-hand panel of Fig. \ref{fig:M_HI_subgrid} compares the $M_{\rm HI}-M_\star$ relation of star-forming galaxies drawn from these simulations with Ref-L025N0376. A lower normalization of the SF law results in a greater atomic hydrogen mass at fixed stellar mass, which is more intuitively understood: to achieve the SFR (and the rate of ejective feedback associated with this SF) necessary to balance a fixed gas infall rate, a more (less) massive cold gas reservoir must be assembled if the normalization of the Kennicutt-Schmidt SF law is decreased (increased); similar findings were reported by \citet{Haas_et_al_13b} from analysis of the OWLS simulations \citep{Schaye_et_al_10_short}. In contrast to variation of the feedback efficiency, to first order the $M_{\rm HI}-M_\star$ relation shifts vertically as the SF efficiency is varied. Whilst the SF law is well characterized by observations on the scale of galaxies, making it a less severe uncertainty than the efficiency of feedback, it should be borne in mind that the systematic uncertainty on the normalization of the SF law is dominated by the systematic uncertainty on the measurement of the IMF, such that variations of up to $0.3-0.5\,{\rm dex}$ are feasible.

No other EAGLE simulations incorporating variations of the Ref parameters exhibit variations of $M_{\rm HI}$ at fixed $M_\star$ as large as the (WeakFB,StrongFB) and (KSNormLo,KSNormHI) pairs. Modulo the impact of local radiation sources, the effect of which we are unable to authoritatively characterize here, the simulations indicate that the \HI-richness of a star-forming galaxy is governed primarily by the processes varied in these pairs. Firstly, the characteristic feedback efficiency associated with its SF history, which ultimately governs the cosmological baryonic infall rate on to the galaxy and the fraction of the infalling material that is retained. Secondly, the SF law which (on kpc scales) governs the rate at which \HI\ is converted into \Hmol\ and subsequently into stars.

\subsection{The HI mass function}
\label{sec:HIMF}

\begin{figure}
\includegraphics[width=0.49\textwidth]{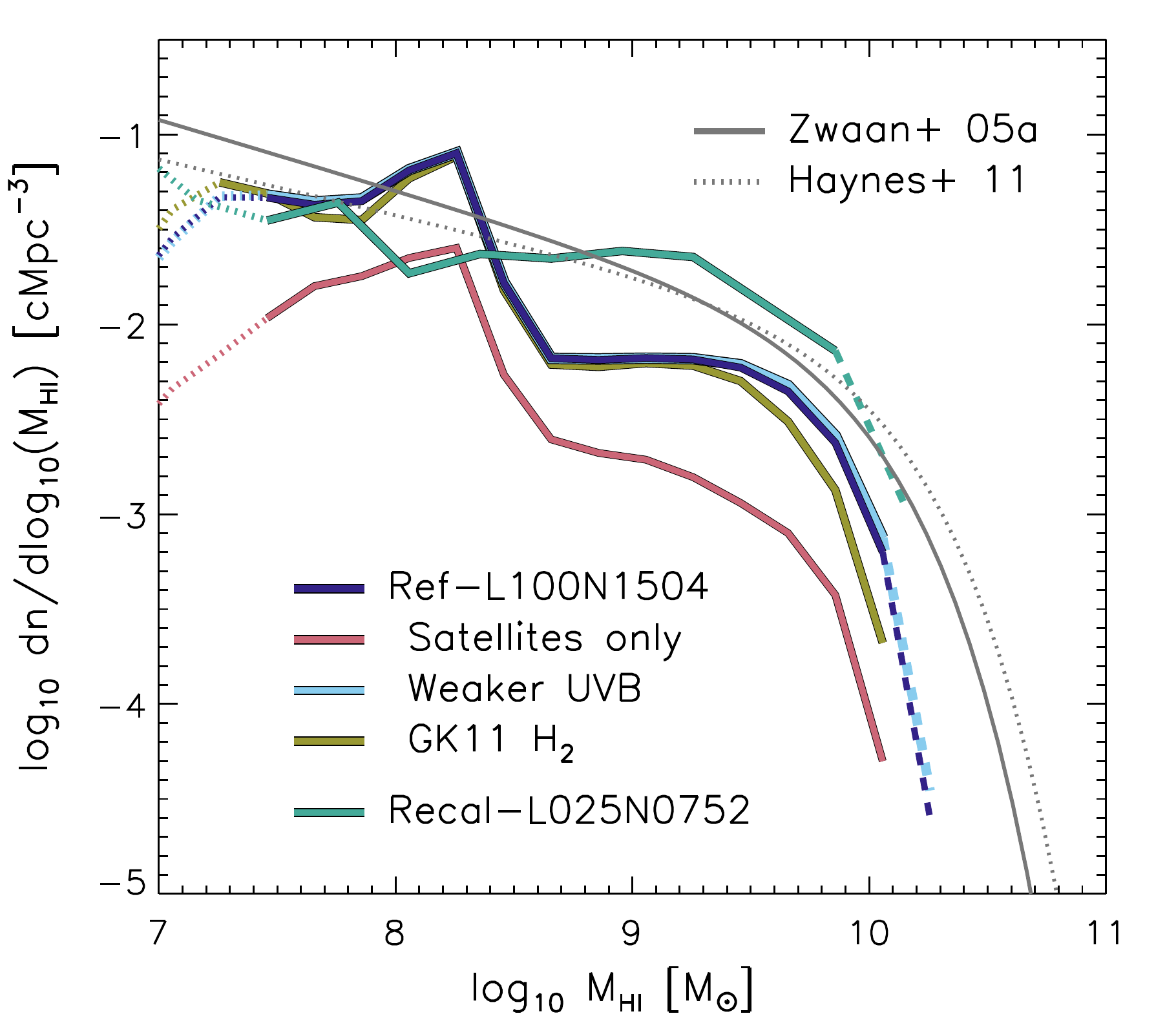}
\caption{The \HI\ mass functions (HIMFs) at $z=0$ of all (central and satellite) galaxies in Ref-L100N1504 (dark blue) and Recal-L025N0752 (green). The contribution of satellites to the Ref-L100N1504 HIMF is shown by the red curve. The Ref-L100N1504 HIMF assuming a weaker UVB (cyan), and that derived using the alternative GK11 prescription for partitioning neutral hydrogen into \HI\ and \Hmol\ (gold), are also shown. Curves are drawn with dashed lines where sampled by fewer than 10 galaxies per bin, and dotted lines below the median $M_{\rm HI}$ associated with central galaxies of $M_\star = 100m_{\rm g}$. The solid and dotted grey curves represent Schechter function fits to the HIMF derived from HIPASS and ALFALFA, respectively. The HIMF of Ref-L100N1504 (and all standard-resolution EAGLE simulations) exhibits an over-abundance of galaxies with $M_{\rm HI}\simeq 10^{8.2}\Msun$ and an under-abundance of galaxies with greater $M_{\rm HI}$. At high resolution, the correspondence with the observed HIMF is markedly improved.}
\label{fig:HIMF_convergence}
\end{figure}

The volumetric mass function of \HI\ sources is a fundamental diagnostic that is well-characterized by current 21 cm surveys of the local Universe \citep{Zwaan_et_al_03_short,Zwaan_et_al_05a,Martin_et_al_10,Haynes_et_al_11_short}, making it a valuable discriminator of galaxy formation models. Fig. \ref{fig:HIMF_convergence} shows the $z=0$ HIMF recovered from Ref-L100N1504 (dark blue) and Recal-L025N0752 (green). Since observational measurements of the HIMF are inferred from blind 21 cm surveys, we consider both central and satellite galaxies here; the contribution of satellites to the Ref-L100N1504 HIMF is shown by the red curve. As with the \HI\ CDDF and the $M_{\rm HI}-M_\star$ relation, we also show the HIMF of Ref-L100N1504 computed in the presence of a UVB photoionization rate a factor of 3 lower than the fiducial value (cyan), and for the case where neutral hydrogen is partitioned into \HI\ and \Hmol\ using the theoretically-motivated GK11 scheme (gold). The solid and dotted grey curves represent Schechter function fits to the HIMF derived from the HIPASS \citep{Zwaan_et_al_05a} and ALFALFA \citep{Haynes_et_al_11_short} surveys, respectively.

\begin{figure*}
\includegraphics[width=0.49\textwidth]{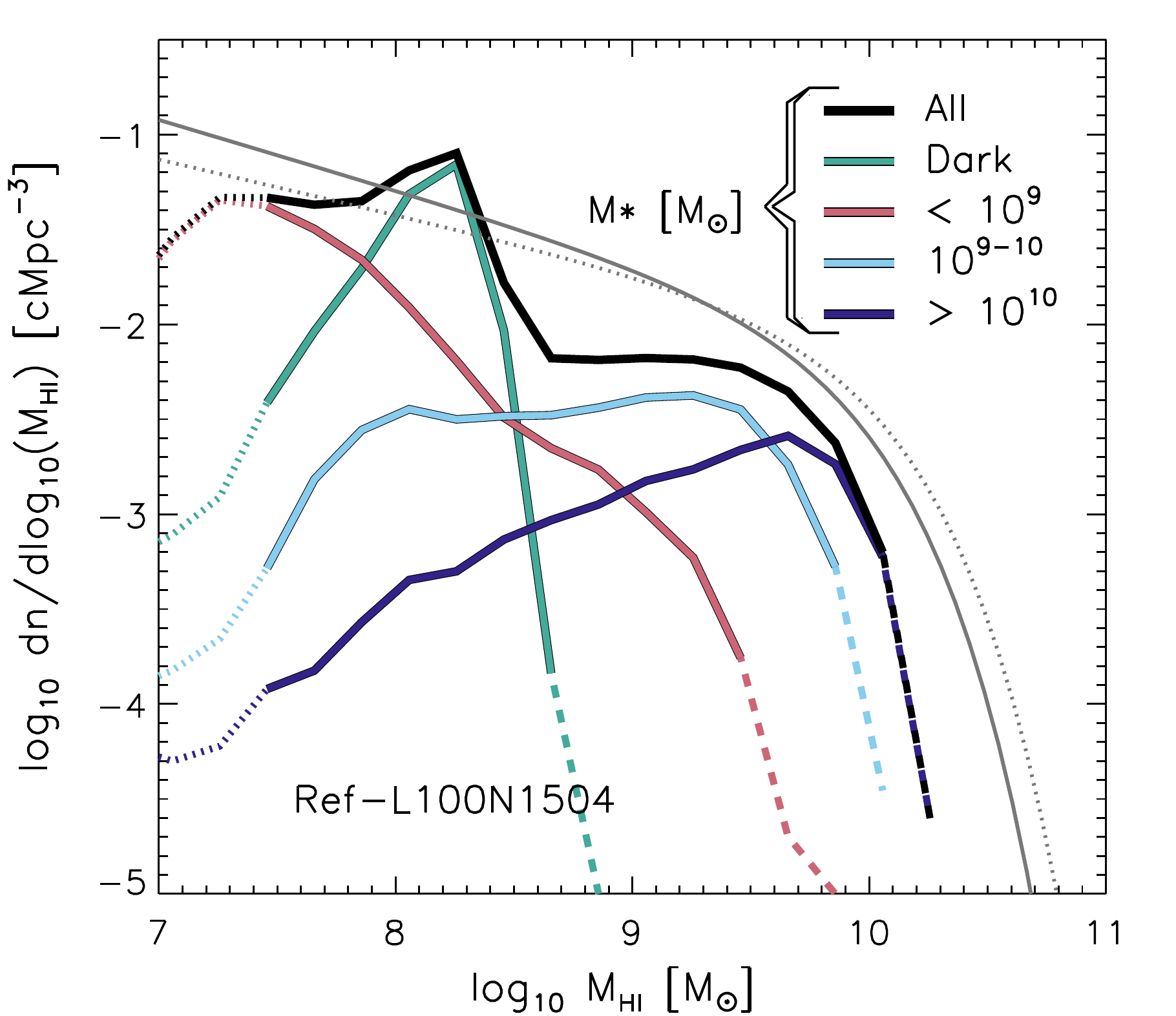}
\includegraphics[width=0.49\textwidth]{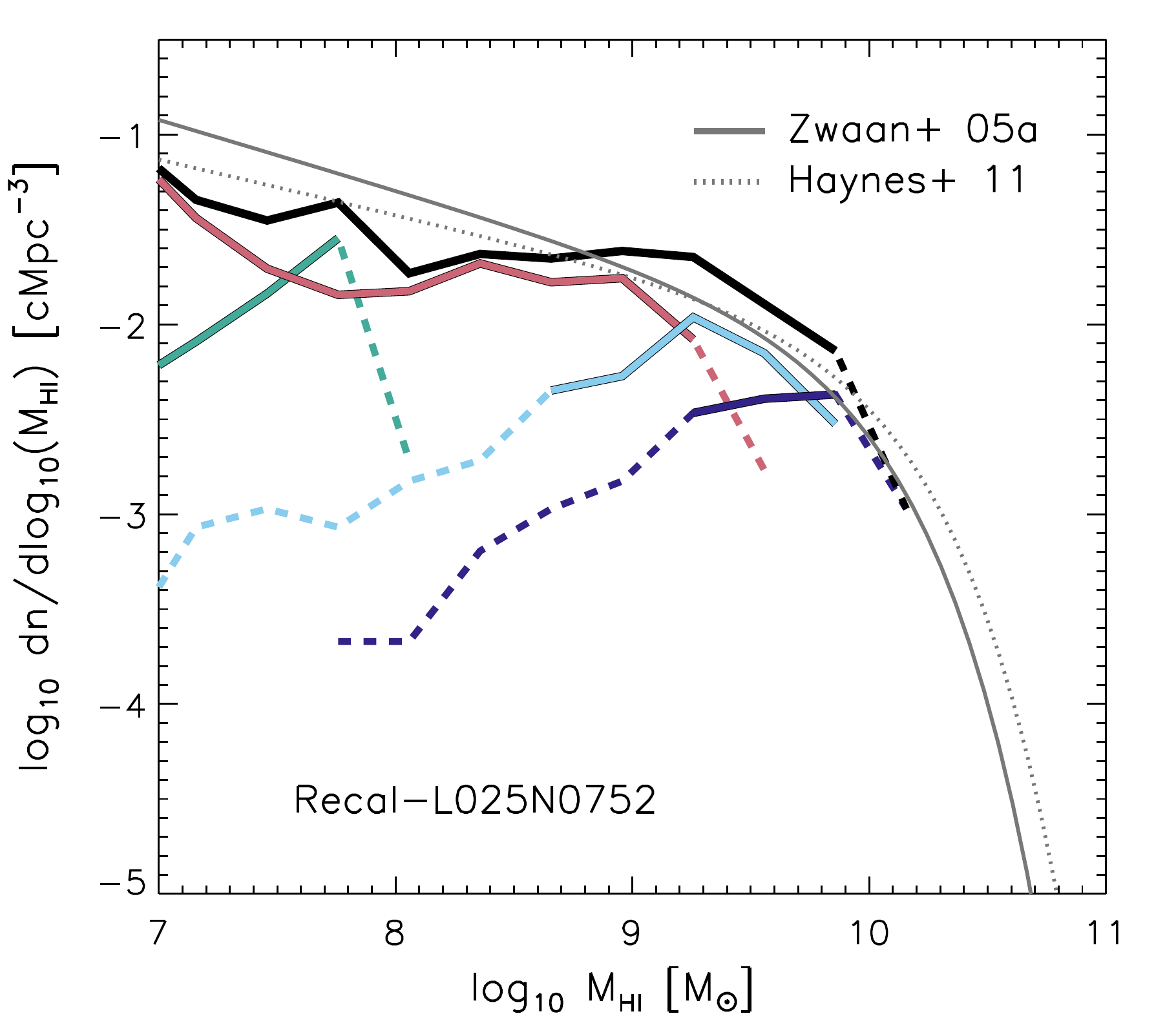}
\caption{The HIMF at $z=0$ of all (central and satellite) galaxies from Ref-L100N1504 (\textit{left}) and Recal-L025N0752 (\textit{right}), decomposed into contributions from galaxies of different stellar mass: all galaxies (black), $M_\star > 10^{10}\Msun$ (dark blue), $M_\star = 10^9 - 10^{10} \Msun$ (light blue), $M_\star < 10^9 \Msun$ (but non-zero, red) and `dark' galaxies (green), i.e. subhaloes yet to form stars. The peak in the HIMF of Ref-L100N1504 at $M_{\rm HI} \simeq 10^{8.3}\Msun$ is associated with a population of dark subhaloes that is largely absent from Recal-L025N0752. The population of \HI\ sources with $10^{8.5} \lesssim M_{\rm HI} \lesssim 10^{9.5} \Msun$ corresponds to galaxies of noticably different stellar mass in the two simulations: in Ref-L100N1504, it is dominated by galaxies of mass $10^9 \lesssim M_\star \lesssim 10^{10} \Msun$, whilst in Recal-L025N0752 it is comprised of galaxies with a wide range of stellar masses.}
\label{fig:HIMF_Mstar}
\end{figure*}

For $M_{\rm HI} \gtrsim 10^{8.5} \Msun$, and irrespective of the assumed UVB photoionization rate or the choice of \Hmol\ correction, the amplitude of the HIMF recovered from the Ref-L100N1504 simulation is lower than that of the observationally-inferred functions, signifying that the offset of the $M_{\rm HI}-M_\star$ relation with respect to observational measurements translates into a poor reproduction of the observed HIMF. At the characteristic break scale of the Schechter functions describing the latter, the Ref-L100N1504 HIMF is offset by $\simeq -0.2$ to $-0.3\,{\rm dex}$, and the discrepancy is considerably worse at both lower and higher $M_{\rm HI}$.

The Ref-L100N1504 HIMF also exhibits a second unrealistic feature. At $M_{\rm HI} \simeq 10^{8} - 10^{8.5}\Msun$, corresponding to $\simeq 50 - 175m_{\rm g}$, the number density of \HI\ sources sharply increases by approximately one decade. Environmental effects can be ruled out as the origin of this feature, since it is clear from Fig. \ref{fig:HIMF_convergence} that it is not dominated by satellite galaxies. The feature is absent from the HIMFs of the high-resolution Recal-L025N0752 and Ref-L025N0752 simulations (for clarity, the latter is not shown), both of which also exhibit a space density of \HI\ sources with $M_{\rm HI} \gtrsim 10^{8.5} \Msun$ that more closely reflects the HIPASS and ALFALFA measurements, though its relatively small volume precludes satisfactory sampling of high-$M_{\rm HI}$ regime. We return to the origin of the feature in the following section. 

The HIMF is therefore poorly converged in EAGLE, and it is clear that numerical resolution influences this diagnostic much more severely than systematic uncertainties such as the assumed UVB photoionisation rate and the \HI-to-\Hmol\ partitioning scheme. The latter however also has a significant influence, with the number density of galaxies of $M_{\rm HI} = 10^{10}\Msun$ recovered using the GK11 \Hmol\ correction being $0.3\,{\rm dex}$ lower than the fiducial case.

\subsubsection{The contribution of different stellar masses to the HIMF}
\label{sec:HIMF_Mstar}

The poor numerical convergence of the \HI\ CDDF, the $M_{\rm HI}-M_\star$ relation and the HIMF in EAGLE \citep[relative to the typical convergence of the stellar properties of EAGLE galaxies, e.g.][]{S15} signals a shortcoming of the ability of the standard-resolution simulations to model the neutral hydrogen component of low- and intermediate-mass galaxies. We therefore briefly explore the origin of the sources responsible for the `bump' in the HIMF at $M_{\rm HI} \simeq 10^{8}-10^{8.5} \Msun$, and the cause of the dearth of more massive \HI\ sources with respect to the high-resolution simulations, by decomposing in Fig. \ref{fig:HIMF_Mstar} the HIMFs of Ref-L100N1504 (left) and Recal-L025N0752 (right) into contributions from galaxies in 4 bins of stellar mass, $M_\star > 10^{10}\Msun$ (dark blue), $M_\star = 10^9 - 10^{10}\Msun$ (light blue), $M_\star < 10^9\Msun$ (but non-zero; red), and those subhaloes yet to form a stellar particle, which we term `dark subhaloes'. 

The unrealistic `bump' in the HIMF of the Ref-L100N1504 simulation is contributed almost entirely by dark subhaloes, within which \HI\ is able to accumulate under gravity without regulation by SF feedback. The feature is therefore purely a numerical artefact, and the \HI\ mass scale that it appears at is a function of the numerical resolution of the simulation. At fixed resolution, the associated \HI\ mass scale is dependent on the polytropic exponent of the Jeans-limiting temperature floor ($4/3$ in all runs featured here), because it determines the characteristic density of the gas whose pressure balances the gravitational potential of the galaxy. The associated space density of the sources is dependent on the SF law (i.e. the normalization of the adopted Kennicutt-Schmidt law), since a higher normalization increases the probability of star-forming particle at fixed pressure converting into a stellar particle and hence triggering self-regulation. In the standard-resolution Ref simulations, the dark subhaloes have characteristic masses $M_{\rm HI} \simeq 10^{8.0}-10^{8.5}\Msun$, and a space density nearly an order of magnitude greater than that of any other stellar mass bin. In the high-resolution simulations, the bump is barely visible in the overall mass function, since the space density of the dark subhaloes HIMF is reduced by $\simeq 0.5\,{\rm dex}$, and it is shifted to $M_{\rm HI} \simeq 10^{7.7}\Msun$ (Ref-L025N0752 and Recal-L025N0752 are near-identical in this regard). The simple explanation for this shift is that, at higher resolution, less gas needs to be assembled in order to realize a fixed probability of converting a gas particle into a stellar particle. However, the offset between the peak mass of the dark subhalo HIMF of Ref-L100N1504 and Recal-L025N0752 is a factor of $\simeq 4$, not a factor of $8$ as might be expected from scaling the particle mass, indicating that the offset is not exclusively a consequence of particle sampling associated with the SF implementation.

The contribution to the HIMF of galaxies in different stellar mass bins also differs significantly between the standard- and high-resolution simulations, as can also be inferred from comparison of their $M_{\rm HI}-M_\star$ relations. In the former, the HIMF above the \HI\ mass associated with dark subhaloes ($M_{\rm HI}\simeq 10^{8.5}\Msun$) is dominated by galaxies in the stellar mass range $M_\star = 10^9 - 10^{10}\Msun$. These galaxies exhibit a supra-linear $M_{\rm HI} - M_\star$ relation and \HI-reservoirs significantly less massive than inferred by the 21 cm stacking analysis of \citet[][see Fig. \ref{fig:M_HI_obs}]{Brown_et_al_15}, and so contribute a nearly-constant and relatively low number density across the interval $M_{\rm HI} \simeq 10^{8} - 10^{9.5}\Msun$. In contrast, galaxies of the same stellar mass exhibit a sub-linear $M_{\rm HI} - M_\star$ relation in the high-resolution simulations, resulting in a steady transition of the low-to-high ends of the HIMF being dominated by galaxies of low-to-high stellar mass. The higher normalization of the $M_{\rm HI}-M_\star$ relation in the high-resolution simulations results in a good match to the observed HIMFs where adequately sampled by $L=25\cMpc$ volumes, but we caution that the relatively small scatter in $M_{\rm HI}$ at fixed $M_\star$ would likely preclude the reproduction of the HIMF at the very highest \HI\ masses, even if we were to simulate a volume of $L=100\cMpc$ at high resolution. 

It is tempting to conclude that the difference in the characteristic number density of the HIMF between the standard- and high-resolution simulations is a consequence of the liberation of the cold gas `locked up' in dark subhaloes. Integration of the HIMF components reveals that such haloes do host nearly 20 percent of the $z=0$ cosmic \HI\ mass density in the intemediate-resolution simulations but, owing to the poor resolution convergence of the HIMF, the difference in the total cosmic \HI\ mass density between the standard- and high-resolution simulations can be much greater than this. For example, a strong convergence test applied to the Ref-L025 simulations (to eliminate box size effects) reveals $\Omega_{\rm HI}(z=0) = 1.24\times10^{-4}$ in Ref-L025N0376 and $\Omega_{\rm HI}(z=0) = 2.38\times 10^{-4}$ in Ref-L025N0752, a difference of nearly a factor of two. We stress that the differences are not a consequence of the \HI-to-\Hmol\ partitioning; similar trends are seen in the total neutral component of galaxies, and also in the \HI\ CDDF at column densities significantly lower than those associated with the formation of \Hmol. Therefore differences in the form of the HIMF are symptomatic of more complex numerical effects. 

\begin{figure*}
\includegraphics[width=0.49\textwidth]{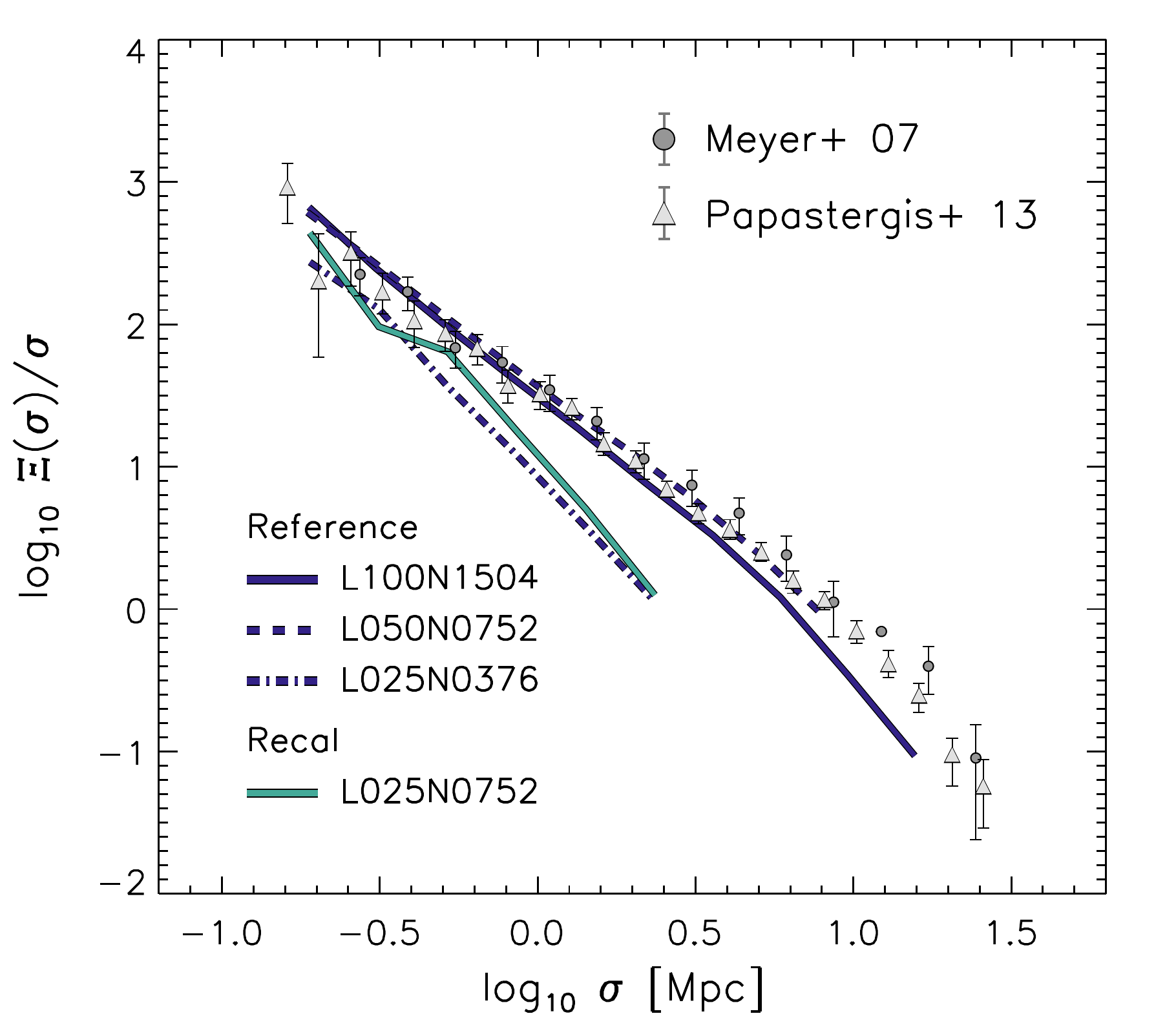}
\includegraphics[width=0.49\textwidth]{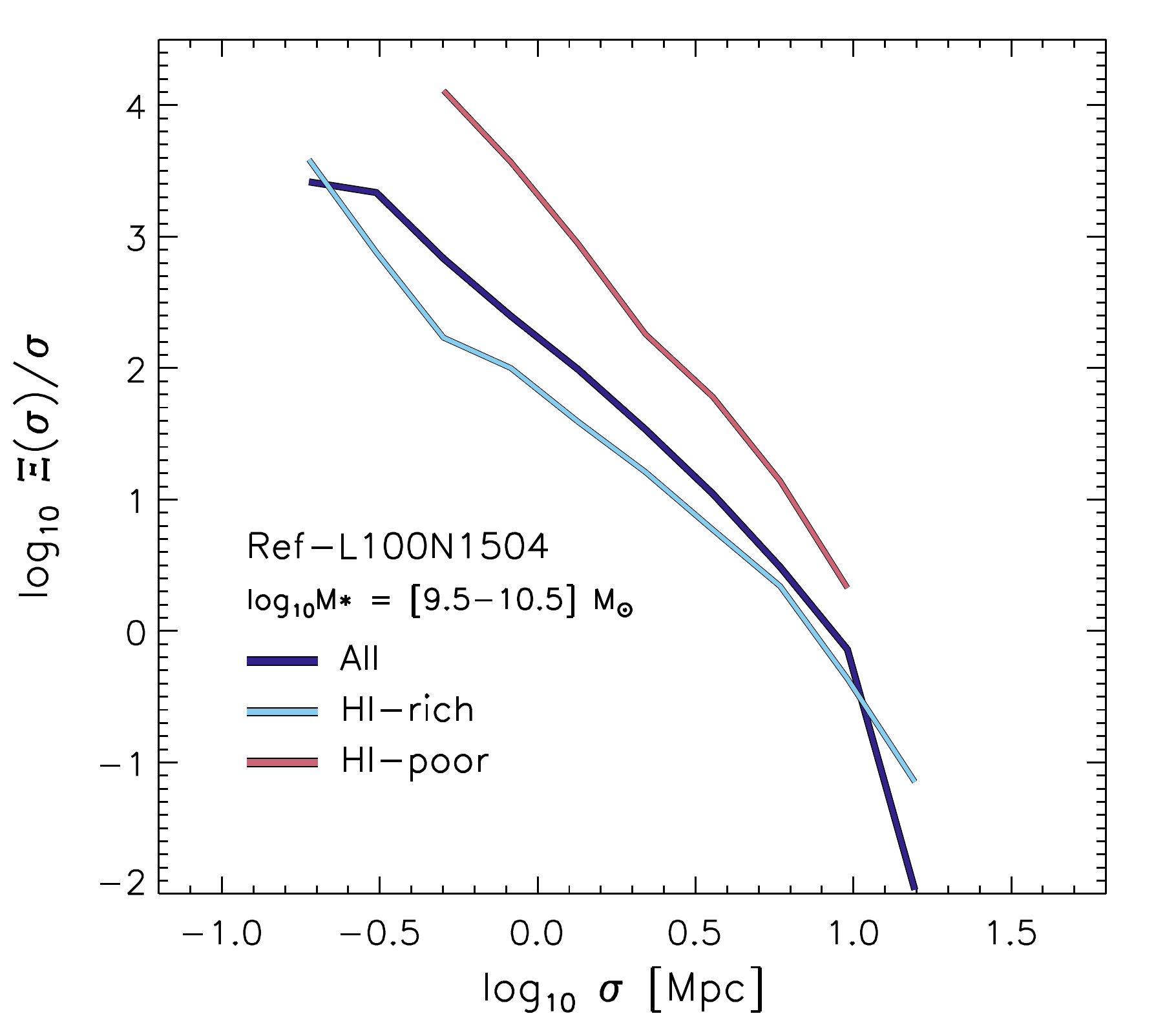}
\caption{\textit{Left:} The projected two-point correlation function of atomic hydrogen sources ($M_{\rm HI} > 10^{7.5}\Msun$) at $z=0$. Correlation functions are integrated in the radial coordinate over a distance corresponding to half of the simulation box size. Measurements are shown for Ref-L100N1504 (dark blue), Ref-L050N0752 (dark blue, dashed) and Ref-L025N0376 (dark blue, dot-dashed), to highlight the impact of absent large-scale power. Also shown is Recal-L025N0752 (green), which can be compared with Ref-L025N0376 to assess weak convergence on the spatial scales probed by the $L=25\cMpc$ simulations. Grey data points represent measurements from the HIPASS (circles) and ALFALFA (triangles) surveys. On scales well-sampled by the $L=100\cMpc$ volume, Ref-L100N1504 is consistent with the observationally-inferred correlation functions, indicating that the distribution of \HI\ in the simulation is realistic. \textit{Right:} The projected correlation function of galaxies from Ref-L100N1504 with stellar mass in the interval $10^{(9.5-10.5)} \Msun$ (dark blue), together with those of counterparts that are rich (pale blue) and poor (red) in \HI\, the two populations being delineated by the threshold $\log_{10}M_{\rm HI}/M_\star=-3$. The latter are markedly more clustered than the former, signalling a correlation between environment and the \HI\ richness of galaxies.}
\label{fig:clustering}
\end{figure*}

\section{The clustering of atomic hydrogen sources}
\label{sec:clustering}

The spatial distribution of galaxies encodes information concerning the processes underpinning galaxy formation, and also the underlying cosmology. Characterizing the distribution, for example via the clustering of galaxies, is therefore a key objective of large optical galaxy surveys \citep[e.g.][]{Zehavi_et_al_11_short,Christodoulou_et_al_12_short}. Next-generation wide-field radio telescopes promise to enable the collection of galaxy redshifts out to $z\simeq1$, and enabling clustering measurements of the gas-rich galaxy population with higher-fidelity than is possible with current optical surveys. Such measurements offer a means of examining the connection between 21 cm sources and their dark matter haloes. Although the largest EAGLE simulation follows a cosmic volume of `only' $(100 \cMpc)^3$, dwarfed by that expected to be mapped by next-generation radio telescopes, it is sufficient to enable a meaningful comparison with measurements from extant 21 cm surveys. Since HI sources are known to be weakly clustered \citep[e.g.][]{Meyer_et_al_07,Papastergis_et_al_13}, the dearth of rare cosmic structures found within relatively small simulation volumes is partially mitigated. 

Clustering is typically quantified using the two-point correlation function, $\xi(r)$, the excess probability of finding a pair of galaxies at a separation $r$, relative to an unclustered distribution. Most commonly, this diagnostic is computed using the \citet{Landy_and_Szalay_93} estimator, $\xi(r) = [DD(r) - 2DR(r) + RR(r)]/RR(r)$, where $DD$, $RR$ and $DR$ correspond to data-data, random-random and data-random pairs, respectively. For each simulation, the `$D$' catalogue comprises central and satellite galaxies with $M_{\rm HI} > 10^{7.5}\Msun$, a definition motivated by the minimum \HI\ mass adopted in the clustering analysis of ALFALFA 21 cm sources by \citet{Papastergis_et_al_13}. The `$R$' catalogue is created in each case by randomizing the spatial coordinates of galaxies in the corresponding $D$ catalogue.

Since the line-of-sight component of the separation must be estimated from a redshift, and is hence subject to a peculiar velocity contribution besides the expansion velocity (commonly known as `redshift-space distortions'), $\xi(r)$ is often presented as the 2D function $\xi(\sigma,\pi)$, where $\sigma$ and $\pi$ are the tangential (`on-sky') and radial (`line-of-sight') separations, respectively. A useful simplification, which minimizes the impact of redshift-space distortions, and eliminates the need to measure accurate redshifts\footnote{Redshift-space distortions can still induce a minor effect on large spatial scales, as measurement error on $\pi$ can scatter galaxies in/out of the integral.}, is to integrate over the radial component and thus define a projected correlation function, 
\begin{equation}
\frac{\Xi(\sigma)}{\sigma} = \frac{2}{\sigma}\int_0^{\pi_{\rm max}}\xi(\sigma,\pi){\rm d}\pi,
\end{equation}
where $\pi_{\rm max}$ is the radial separation at which the integral converges. Note that the inclusion of a factor $1/\sigma$ makes the quantity unitless. 

Fig. \ref{fig:clustering} shows the projected correlation functions of several EAGLE simulations at $z=0$. In each panel, grey triangles denote the observationally-inferred correlation function of \citet{Papastergis_et_al_13}. Measurements from a similar analysis applied to the HIPASS survey by \citet{Meyer_et_al_07} are shown as grey circles; their analysis adopted a higher mass threshold of $M_{\rm HI} \simeq 10^9 \Msun$ for inclusion, but \citet{Papastergis_et_al_13} show that the clustering is not significantly sensitive to this choice (see their Fig. 8). In the left-hand panel, the Ref-L100N1504 (dark blue), Ref-L050N0752 (dark blue, dashed) and Ref-L025N0376 (dark blue, dot-dashed) simulations are shown to isolate the impact of the simulation box size on clustering. Also shown is Recal-L025N0752 (green), which can be compared with Ref-L025N0376 to assess weak convergence of the clustering on the spatial scales probed by these volumes.

The projected correlation function is obtained by integrating $\xi(\sigma,\pi)$ over one half of the simulation box size, i.e. $\pi_{\rm max}= 12.5$, $25$ and $50\cMpc$ for $L=25$, $50$ and $100\cMpc$ volumes, respectively. In the latter case, this is a slightly smaller scale than the $\simeq 66\cMpc$ adopted by \citet{Papastergis_et_al_13}, however the integral is relatively insensitive to this upper bound. If $\pi_{\rm max}=25\cMpc$ (i.e. one-quarter of the simulation box size) is adopted, the projected correlation function of Ref-L100N1504 increases by $<0.3\,{\rm dex}$ on the most massive scales, where the systematic uncertainty is in any case dominated by the absence of large-scale modes.

On the scales well-sampled by the $L=100\cMpc$ volume ($\sigma \lesssim 3 \Mpc$), Ref-L100N1504 is consistent with the correlation functions recovered from HIPASS and ALFALFA. This indicates that, despite the relatively poor convergence of \HI\ masses for galaxies of $M_\star \lesssim 10^{10}\Msun$, the simulation broadly reproduces the true spatial distribution of \HI\ sources of $M_{\rm HI} > 10^{7.5}\Msun$. This is perhaps not entirely surprising, since the mean space density of such \HI\ sources is similar in Ref-L100N1504 and ALFALFA. However, despite being the largest-volume EAGLE simulation, Ref-L100N1504 is too small to accommodate the large-scale modes of the power spectrum that foster the formation of the most massive structures and deepest voids observed in the local Universe. Comparison of the Ref model realized in progressively larger volumes indicates that the absence of large scale modes within domains of box size $L=25\cMpc$ results in a strong under-prediction of the projected correlation function on all but the very smallest scales probed by 21 cm surveys. The inclusion of the larger scale modes within Ref-L050N0752 results in a correlation function much closer to that of Ref-L100N1504, indicating that the box size convergence is good for $L\gtrsim 50\cMpc$. In spite of the inability of $L=25\cMpc$ simulations to satisfactorily sample similar scales to the HIPASS and ALFALFA surveys, comparison of Ref-L025N0376 with Recal-L025N0752 demonstrates that the weak convergence of the correlation function is good; a comparison with Ref-L025N0752 (not shown) highlights that the strong convergence is also good. Note that we do not show the effect of varying the UVB photoionisation rate or the \Hmol\ correction scheme here, since they do not influence the spatial distribution of the \HI\ sources, and their influence on their \HI\ mass (which is relevant, since it can shift sources in/out of the selection criterion) is less significant than the shift seen in weak convergence tests (Fig. \ref{fig:M_HI_convergence}).

A detailed examination of the physical processes that shape the \HI\ content of EAGLE galaxies via the cosmic environment is presented by \citet{Marasco_et_al_16}. Here, a simple demonstration of the impact of environment on the \HI\ content of galaxies is afforded by comparison of the projected correlation functions of galaxies that are relatively rich and poor in \HI. Since, as discussed in \S\ref{sec:obs_measurements}, the \HI-richness of galaxies is a strong function of $M_\star$, it is necessary to conduct such a test within a sufficiently narrow range in $M_\star$ that the \HI-rich and \HI-poor samples have similar $M_\star$ distributions. The right-hand panel of Fig. \ref{fig:clustering} shows the projected correlation function of galaxies, drawn from Ref-L100N1504 at $z=0$, with stellar mass in the range $M_\star = 10^{9.5} = 10^{10.5} \Msun$. The dark blue curve corresponds to the clustering of all such galaxies, whilst the light blue and red curves correspond to galaxies with atomic hydrogen-to-stellar mass ratios greater than, and poorer than, ${\rm log}_{10} (M_{\rm HI}/M_\star) = -3$, respectively. \citet{Marasco_et_al_16} show that the population is approximately bimodal in this diagnostic, and effectively separated by this threshold. Ref-L100N1504 affords adequate sampling of the two populations, with 4799 HI-rich and 1135 HI-poor galaxies within the stellar mass interval. The \HI-poor population is markedly more clustered than the \HI-rich population. This offset is analogous to the well-established dependence of clustering amplitude on the optical colour or sSFR of galaxies \citep[e.g.][]{Madgwick_et_al_03,Zehavi_et_al_11_short}. This sensitivity of the projected correlation function to HI-richness highlights that 21 cm surveys trace the cosmic large scale structure in a markedly different fashion to optical surveys \citep[as previously demonstrated by][using semi-analytic galaxy formation simulations]{Kim_et_al_11}, particularly those focusing on luminous red galaxies \citep[e.g.][]{Eisenstein_et_al_05_short,Cannon_et_al_06_short}. 

\section{The accretion and evolution of the HI of present-day galaxies}
\label{sec:reservoir_accretion_and_evolution}

\begin{figure*}
\includegraphics[width=0.45\textwidth]{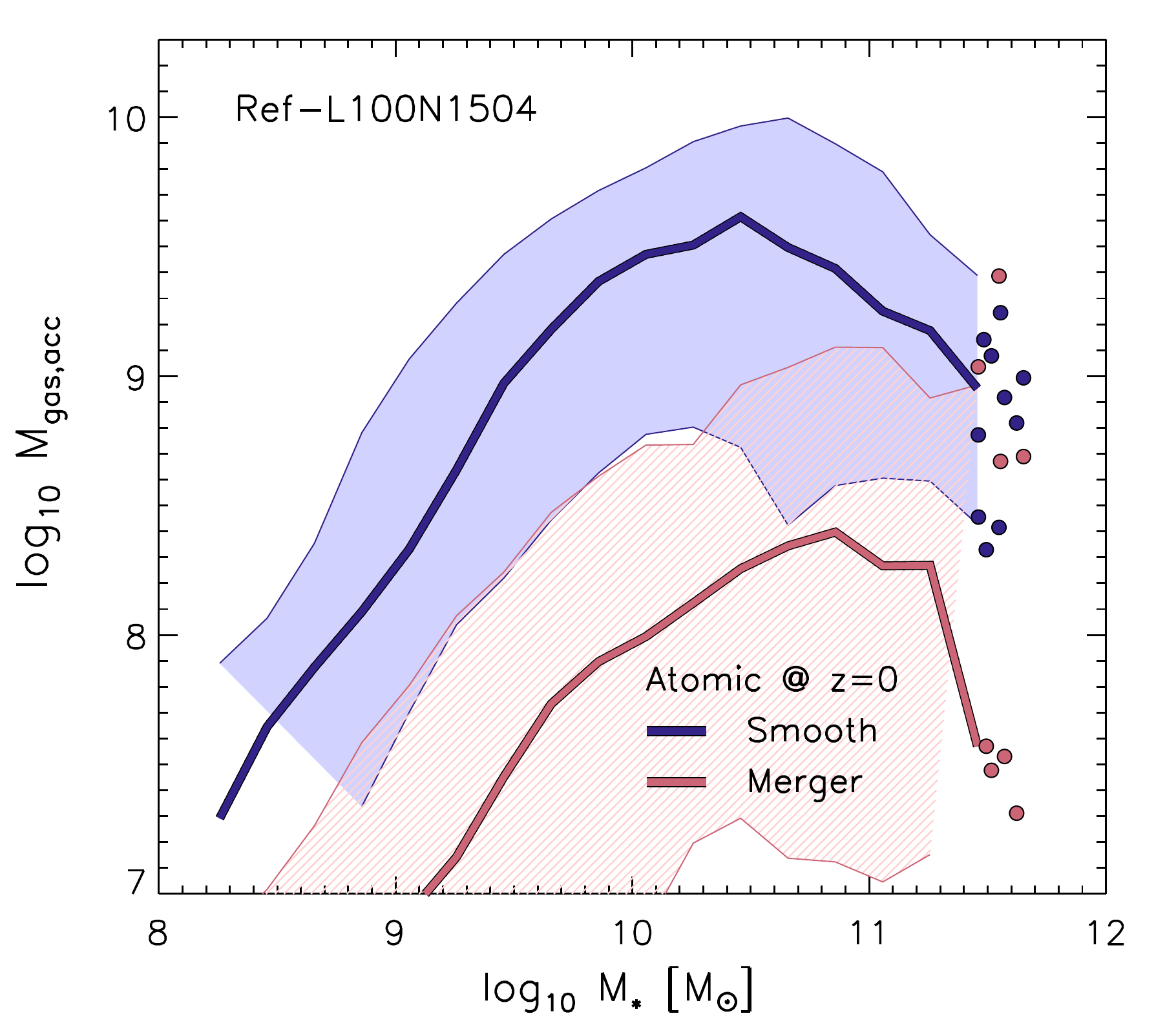}
\includegraphics[width=0.45\textwidth]{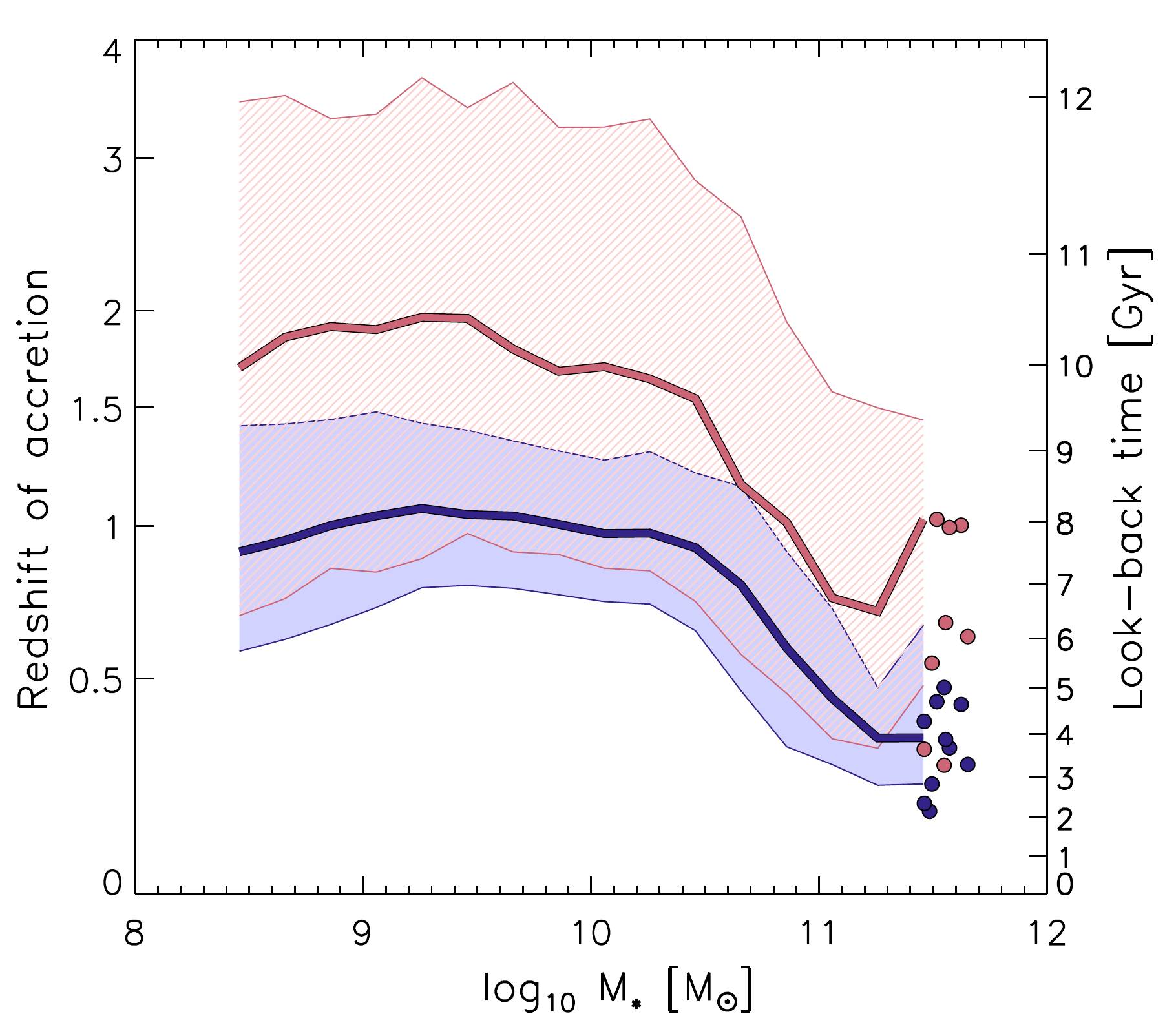}
\caption{The accreted mass (\textit{left}), and the redshift of accretion (\textit{right}), of gas that at $z=0$ is bound to central galaxies, resides within $r=70\pkpc$ of their centre, and is in the form of \HI. Gas is categorized into smooth accretion (dark blue) and mergers (red), and plotted as a function of $z=0$ stellar mass. Particles are considered to have accreted in a merger if they comprised a self-bound substructure with mass greater than 10 percent of the descendant structure at the time of accretion. Curves represent the median accreted masses and are drawn only where the $z=0$ galaxies are well resolved ($M_\star > 100 m_{\rm g}$). Individual galaxies are plotted where bins are sampled by fewer than 10 galaxies. Shaded regions denote the $1\sigma$ scatter about the median. Present-day \HI\ was mostly accreted in diffuse form by the main progenitor subhalo of the host. For galaxies with $M_\star \lesssim 10^{10.5}\Msun$ the accretion occurred at $z\simeq1$, and was more recent in more massive systems.}\label{fig:reservoir_accretion}
\end{figure*}

Having established the degree to which the simulations reproduce the observed masses and spatial distribution of \HI\ sources, we turn to an examination of how and when the gas comprising these reservoirs was accreted. This fundamental question cannot at present be authoritatively addressed with the available observational evidence, and is well-suited to study with Lagrangian simulations such as EAGLE, because the individual fluid elements representing the \HI\ of $z=0$ galaxies can be tracked to earlier times. 

For example, the accretion rate of galaxies in the Local Group and the very local Universe, for which 21 cm surveys are able to map \HI\ clumps \citep[often in the guise of high velocity clouds, HVCs, e.g.][]{Blitz_et_al_99,Thilker_et_al_04,Wakker_et_al_07_short,Miller_Bregman_and_Wakker_09}, is estimated to represent only a small fraction of their SFR  \citep{Sancisi_et_al_08,Prochaska_and_Wolfe_09,Bauermeister_Blitz_and_Ma_10,Di_Teodoro_and_Fraternali_14}. The accretion of \HI-rich satellites is mostly deemed to be an unlikely means of building \HI\ reservoirs, as the majority of satellites accreted by star-forming galaxies are stripped of their gas at large radii \citep[e.g.][]{Bland-Hawthorn_et_al_07,Grcevich_and_Putman_09,Peek_09}.

A simple comparison of the evolving cosmic densities of SFR and \HI, in combination with estimates of the \Hmol\ depletion rate of local galaxies, indicates that galaxies' cold gas must be replenished by the accretion of ionized gas \citep[e.g.][]{Bauermeister_Blitz_and_Ma_10}. Although their satellites can host ionized gas, the CGM of central galaxies represents the major reservoir of \HII, with the potential to cool, become shielded and contribute to the build-up and maintenance of a galaxy's \HI\ mass. This raises the possibility that much of the \HI\ associated with galaxies was not delivered by mergers with other galaxies, but was accreted smoothly in diffuse form from the ionized IGM. Recent observations using the Cosmic Origins Spectrograph (COS) spectrograph indicate that most of the gas accreting on to galaxy discs indeed arrives in an ionized state \citep{Lehner_and_Howk_11}.

In \S\ref{sec:reservoir_accretion} we compute the fraction of the \HI\ mass of $z=0$ galaxies that was accreted by the main progenitor subhalo of central galaxies i) smoothly from the IGM or ii) in mergers with other progenitor subhaloes. This procedure requires the construction and analysis of high-temporal resolution merger trees, details of which are presented in the Appendix. In \S\ref{sec:reservoir_evolution} we examine the evolving thermodynamic state of the gas comprising the \HI\ reservoirs of all (central and satellite) galaxies at a range of redshifts, in order to establish its `life cycle'. For these tests, we focus on Ref-L100N1504 in order to sample the full range of stellar masses associated with \HI\ reservoirs; the outcomes of these tests are reasonably robust to numerical resolution.

\subsection{How and when was the HI of present-day galaxies accreted?}
\label{sec:reservoir_accretion}

The left-hand panel of Fig. \ref{fig:reservoir_accretion} shows, for Ref-L100N1504, the mass of gas comprising the \HI\ reservoirs of present-day central galaxies that accreted on to the galaxy's main progenitor subhalo smoothly ($M_{\rm HI,smooth}$, \textit{blue curve}) or in a merger with another progenitor subhalo ($M_{\rm HI,merger}$, \textit{red}). The right-hand panel shows the \HI\ mass-weighted redshift of the accretion events (see the Appendix for a description of how these quantities are calculated). The \HI\ of galaxies of all $M_\star$ was acquired almost entirely via the smooth accretion of intergalactic gas, mirroring the nature of the accretion of the \textit{total} gas content of haloes (i.e. irrespective of ionization state) reported by \citet{van_de_Voort_et_al_11a} from analysis of the OWLS simulations. The $M_{\rm HI, smooth}-M_\star$ relation therefore closely resembles the overall $M_{\rm HI}-M_\star$ relation, and mergers typically contribute $\ll 10$ percent of the \HI\ mass.

Relaxation of the criterion that the gas must be atomic at $z=0$, such that the accretion of \HII\ (and \Hmol) is also considered (for brevity, these curves are not drawn on Fig. \ref{fig:reservoir_accretion}) demonstrates that the total mass accreted, both smoothly and in mergers, continues to rise with $M_\star$. This result can also be inferred by inspection of the baryon fractions of the haloes of massive galaxies, which increase monotonically with virial mass \citep[see Fig. 3 of][]{Schaller_et_al_15a}. The downturn of the $M_{\rm HI, smooth}-M_\star$ relation at high masses is therefore a consequence of the gas accreted on to massive galaxies becoming dominated by \HII, as a consequence of the majority of it heating to $\sim T_{\rm vir}$ via accretion shocks and AGN heating \citep[e.g.][]{van_de_Voort_et_al_11a,van_de_Voort_et_al_11b}.

The accretion epoch of the gas that will subsequently comprise present-day \HI\ reservoirs differs significantly for the two accretion channels, with the gas accreted in mergers typically being acquired earlier. For galaxies with $M_\star \lesssim 10^{10.5}\Msun$, the smoothly-accreted component is acquired at $z\sim 1$ ($t_{\rm lookback}\sim 7-8\Gyr$), whilst that delivered in mergers was acquired at $z=1.5-2$ ($t_{\rm lookback}\sim 10\Gyr$). For more massive galaxies, the characteristic epoch of accretion is more recent, for both channels, the smoothly-accreted component arriving between $z=0.3$ and $z=1$, and the merger component between $z=1$ and $z=1.5$.

\begin{figure*}
\includegraphics[width=0.95\textwidth]{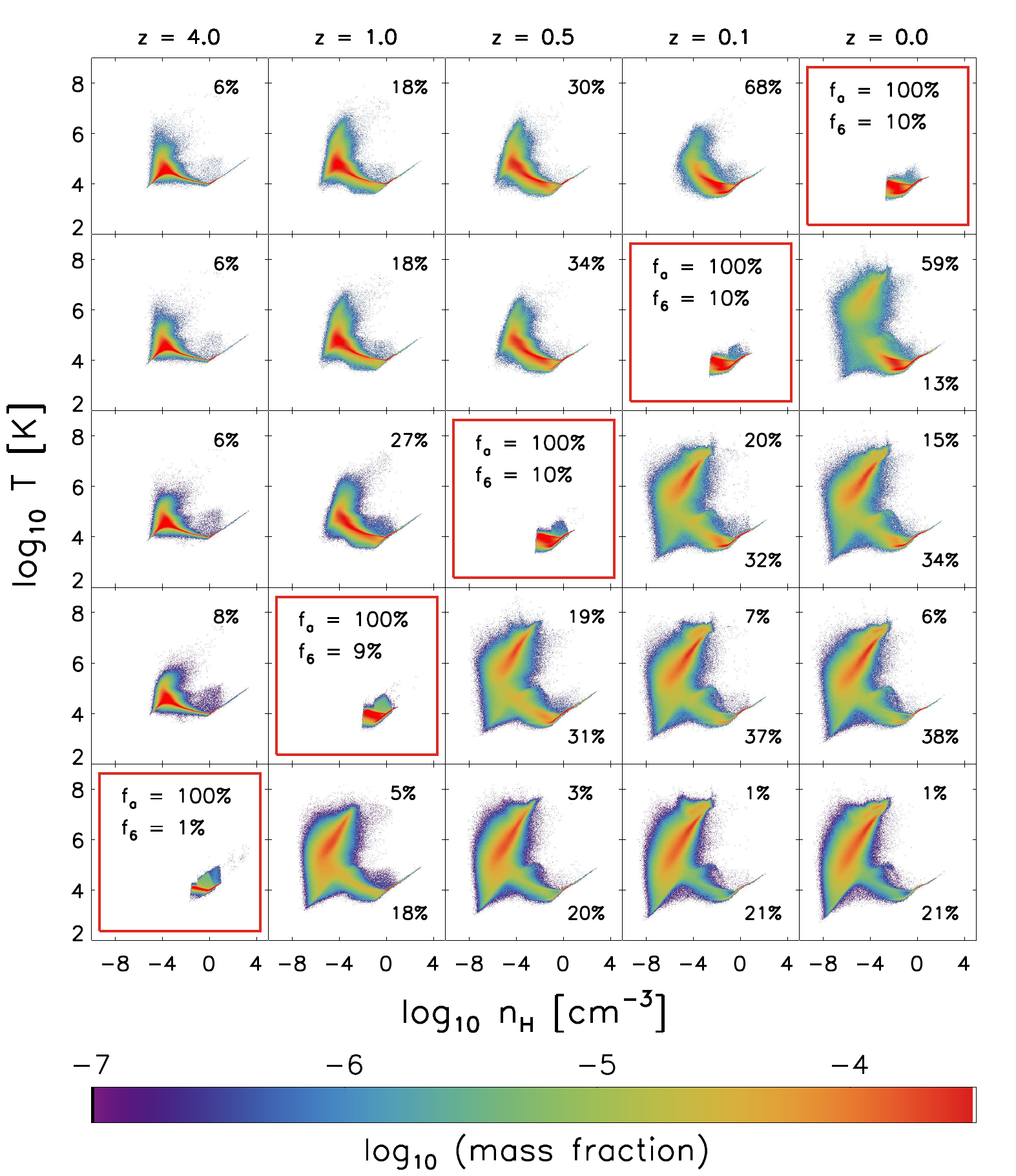}
\caption{The temperature-density distribution of \HI\ associated with galaxies at the five epochs $z=4, 1, 0.5, 0.1, 0$. Panels along the diagonal (with a red outline) denote this `selection epoch', and the panels of each row show the distribution of the same gas at the other epochs. For example, the bottom row shows the future state of \HI\ associated with $z=4$ galaxies, whilst the top row shows the past history of \HI\ associated with $z=0$ galaxies. The value in the upper-right corner of each panel denotes the fraction of the selected gas that is atomic at each epoch; by construction this is $100\%$ along the diagonal. The value in the lower-right of panels right of the diagonal denotes the fraction of the gas that has been converted to stars by that time. The second value in panels along the diagnonal specifies the fraction of the gas that has a maximum past temperature greater than $10^6\K$, as a crude estimate of the fraction of the \HI\ at that epoch that has been reincorporated by galaxies following strong heating by feedback at an earlier time. The \HI\ of galaxies at any given epoch is sourced from similar regions of temperature-density space, avoiding the high-temperature plume, but has a `diverse' future and can be quickly redistributed by feedback. For example, $<60$ percent of the \HI\ of $z=0.1$ galaxies remains atomic at $z=0$ ($\Delta t \simeq 1.3\Gyr$); $13$ percent has been converted to stars and $28$ percent has been heated and ionized by feedback.}
\label{fig:matrix}
\end{figure*}

Since the bulk of cosmic \HI\ at the present-day is associated with galaxies of $M_\star \sim 10^{9.5}-10^{10.5}\Msun$ \citep[e.g.][]{Zwaan_et_al_05a,Meyer_et_al_07}, the simulations indicate that the smooth accretion of diffuse gas from the IGM at $z \sim 1$ is the dominant mechanism by which present-day \HI\ reservoirs are built-up. It is important to stress that this conclusion specifically concerns the accretion of gas on to the `environment' of galaxies defined by their main progenitor subhalo. This description may not be an ideal proxy for, for example, the disc-halo interface, so the result does not preclude a significant or even dominant role for HVCs as a means of replenishing the gas consumed by SF. Such structures may form after accretion on to the main progenitor subhalo, either due to thermal instability and fragmentation of the warm/hot ionized phase of the CGM \citep[e.g.][]{Maller_and_Bullock_04}, or within the ISM itself prior to ejection into the CGM as a galactic fountain \citep[e.g.][]{Booth_and_Theuns_07}.

\subsection{The history and fate of gas of the HI associated with galaxies at different epochs}
\label{sec:reservoir_evolution}

A complementary means of exploring the history of $z=0$ \HI\ reservoirs is to examine its thermodynamic state at earlier epochs. The Lagrangian nature of the simulations enables the particles comprising the \HI\ reservoirs of galaxies at \textit{any} epoch to be identified, and the state of these particle sets tracked back or forth in time. Hence the evolving properties of the gas comprising high-redshift \HI\ reservoirs can be compared with that observable around local galaxies via its 21 cm emission. 

Fig. \ref{fig:matrix} presents such an analysis. Each row shows the evolving mass-weighted 2D PDF in temperature-density space of the \HI\ associated with galaxies at the `selection epoch' of the row, denoted by the red-outlined boxes along the diagonal. From top-to-bottom, the selection epochs of the row correspond to $z = [0, 0.1, 0.5, 1, 4]$. Here, for gas to be considered as part of a galaxy's \HI\ reservoir, it need only be atomic and be bound to a subhalo (central or satellite), in order to exclude \HI\ in the IGM. The requirement to reside within an aperture of radius $70\pkpc$ is relaxed since this scale was motivated by the typical beam size of 21 cm surveys of the local Universe (i.e. $z\simeq0$).  

At each selection epoch, the 2D PDF resembles the upper-right panel of Fig. \ref{fig:phase_diagrams}, without the low-density tail of gas contributed by the IGM. The columns correspond to the same set of redshifts as the selection epochs; for example the middle row shows the past history ($z=4, z=1$) and future evolution ($z=0.1$, $z=0$) of the \HI\ associated with galaxies at $z=0.5$. The value in the upper right of each panel specifies the fraction, $f_{\rm a}$, of the selected gas remaining in atomic form at each epoch; specifically, this is the fraction (by number) of the selected particles that exhibit $f_{\rm HI} > 0.5$ at each epoch (see the Appendix; by construction, this is 100 percent at the selection epochs). The value in the lower-right corner specifies the fraction of the gas converted to stars (again, by particle number), which can only be non-zero to the right of the diagonal. Along the diagonal, the second value specifies the fraction, $f_6$, of the gas with a maximum past temperature $T_{\rm max}>10^6\K$, which provides a crude estimate of the fraction of the gas previously heated by feedback and subsequently reincorporated into a galaxy's \HI\ reservoir. This should be considered an upper limit, since accretion shocks can also heat gas to such temperatures. We do not adopt a fiducial threshold temperature closer to the heating temperature increment of SF feedback events, $\Delta T_{\rm SF}=10^{7.5}\K$, since the pressure gradient established by stochastically-heated particles entrains neighbouring particles into outflows, which might also be heated to temperatures $T\gg 10^5\K$ as they shock against ambient gas. Note that it is also possible for gas to become entrained in outflows without heating significantly.

The top row of Fig. \ref{fig:matrix} shows the evolution in temperature-density space of the \HI\ associated with present-day galaxies. At $z=4$ much of the gas exhibits the positive $T(n_{\rm H})$ relation characteristic of the photoionized IGM but, by $z=1$, almost all of the gas has evolved to the higher densities at which radiative cooling becomes efficient. Thus, the gas resides along either the negative $T(n_{\rm H})$ track at intermediate densities, or even the $T\propto n_{\rm H}^{1/3}$ slope of the Jeans-limiting temperature floor (\S\ref{sec:simulations}) at the high densities associated with star-forming gas. As a result, a significant fraction of the gas was already in atomic form at earlier times; for example, $f_{\rm a} = 18~{\rm percent}$ at $z=1$ and $30~{\rm percent}$ at $z=0.5$. Recall from Fig. \ref{fig:reservoir_accretion} that the majority of this gas accreted on to the main progenitor subhalo of its host galaxy at $z\simeq1$. Hence, although a significant fraction of the gas was already in atomic form, we can conclude that the majority of the gas comprising \textit{present-day} \HI\ reservoirs was ionized when it was accreted. Note that the picture is likely to differ for \HI\ associated with high-redshift galaxies, since the accretion of gas by haloes at high redshift is more strongly dominated by cold accretion \citep[e.g.][]{van_de_Voort_et_al_11a}.

A second notable characteristic of the panels to the left of the selection diagonal is the absence of the high-temperature plume of gas that is clearly visible in the panels to the right of the selection diagonal. This is perhaps unsurprising since, as discussed in \S\ref{sec:temp_dens}, the approximate $T\propto n_{\rm H}^{2/3}$ slope of this feature is characteristic of adiabatic, and hence inefficient, cooling. Therefore it seems little gas heated to $T\gg 10^5\K$ subsequently cools to become reincorporated into the \HI\ reservoir of a galaxy. This conclusion is corroborated by our crude quantitative estimate, $f_6$, which indicates that the fraction of \HI\ associated with galaxies that had a maximum past temperature of $T > 10^6\K$ is approximately $1$ percent at $z=4$, and $\simeq 10$ percent for $z\lesssim 1$. 

We stress that this does not mean that feedback has not influenced the \HI\ associated with EAGLE galaxies. Although it is clear that this subset of particles has not been heated directly by the stochastic feedback implementation, their dynamics may still have been influenced by pressure gradients established in response to the heating of neighbouring particles. In this fashion, some fraction of the particles comprising the \HI\ reservoir of a galaxy may be entrained in pressure-driven outflow without heating to high temperatures. In the absence of strong energy driving, such outflows may stall, sink and reincorporate into the \HI\ reservoir. A comprehensive analysis of the dynamics of outflowing gas is beyond our current scope and will be the focus of a separate study (Crain et al. in prep). However, we note that the dynamics of gas influenced by feedback are clearly sensitive to the implementation of the feedback model. For example, the simulations of \citet{Dave_et_al_13} implement outflows by temporarily decoupling ISM particles from the hydrodynamics scheme and then imparting to them a velocity kick. By construction, such outflows are cold until the particles are reconnected to the hydrodynamics scheme, at which point they are able to shock heat against the ambient medium. \citet{Oppenheimer_et_al_10} demonstrated that the formation of stars from particles that had previously been kicked was the key aspect enabling those simulation to successfully reproduce the faint end of the GSMF. 

Inspection of the panels corresponding to selection epochs $z>0$ highlights that the \HI\ of high-redshift galaxies is largely unconnected to that associated with present-day counterparts. For example, examination of the two right-most panels of the second row demonstrates that $41$ percent of the \HI\ associated with $z=0.1$ galaxies is no longer atomic by $z=0$ (only $\simeq 1.3\Gyr$ later); $13$ percent of the $z=0.1$ \HI\ has been converted into stars and $28$ percent has been ionized. The emergence of the $T\propto n_{\rm H}^{2/3}$ plume in the right-most panel indicates that the ionization is primarily as result of heating by feedback, though since we include satellite galaxies in this analysis, some fraction may also have been stripped during infall and heated by shocking against the hot circumgalactic/intracluster medium. Examining more extreme time-scales, we find that only $6$ percent ($1$ percent) of the \HI\ associated with galaxies at $z=1$ ($z=4$) remains atomic at $z=0$.

\section{Summary and discussion}
\label{sec:summary}

We have examined the properties of atomic hydrogen associated with galaxies in the EAGLE suite of cosmological hydrodynamical simulations. The simulations do not save the ionization state of hydrogen nor do they account for self-shielding, so we estimate in post-processing the neutral fraction of gas particles using a fitting function calibrated against cosmological radiative transport simulations \citep{R13a}. We partition this neutral gas ito atomic and molecular components using the empirical pressure law of \citet{BR06} or the theoretical prescription of \citet{GK11}.

Results are drawn primarily from a set of simulations for which the parameters of the subgrid models governing feedback were calibrated to ensure reproduction of the $z=0.1$ galaxy stellar mass function (GSMF), subject to the requirement that the sizes of disc galaxies must be reasonable. The adopted calibration has been shown to reproduce the colours and luminosities of galaxies at $z=0.1$ \citep{Trayford_et_al_15_short}, and the observed evolution of the stellar mass function and galaxy sizes \citep{Furlong_et_al_15_short,Furlong_et_al_16}. 

The gaseous properties of galaxies were not considered during the calibration. The primary aim of this study was therefore to assess the degree to which reproduction of key aspects of the present-day stellar component of galaxies also results in the reproduction of realistic \HI\ properties, such as the $z=0$ \HI\ column density distribution function (CDDF), the $M_{\rm HI} - M_\star$ relation and the \HI\ mass function (HIMF). We have confronted these outcomes from EAGLE with observational measurements, and explored their sensitivity to ill-constrained aspects of galaxy formation physics, using simulations that adopt subgrid parameter values different to those of Ref. We also assessed the sensitivity of these diagnostics to uncertainties in our post-processing procedures, by varying the UVB photoionization rate and by adopting the alternative \citet{GK11} prescription for partitioning the neutral hydrogen into \HI\ and \Hmol. 

An additional aim of this study was to quantify the contribution of smooth accretion and mergers to the build-up of present-day \HI\ reservoirs, and to chart the history and fate of fluid elements comprising \HI\ reservoirs associated with galaxies at different epochs. These questions cannot be authoritatively addressed with extant observations, and are well-suited to address via calibrated numerical simulations such as EAGLE. \\

\noindent Our results can be summarized as follows:

\begin{enumerate}
\item The $z=0$ \HI\ CDDF of Ref-L100N1504 is systematically offset to lower column density with respect to the CDDF derived from WHISP survey measurements, by $\simeq -0.2$ to $-0.3$ dex. The CDDFs of the high-resolution Ref-L0025N0752 and Recal-L025N0752 simulations are shifted to higher $N_{\rm HI}$ at fixed number density (by $0.28$ and $0.26\,{\rm dex}$, respectively, at $\log_{10} f(N_{\rm HI})[\cmsquared]= -22$), yielding better agreement with observations (Fig. \ref{fig:CDDF}). 

\item The similarity of the Ref-L025N0752 and Recal-L025N0752 \HI\ CDDFs indicates that numerical resolution influences column densities more than the recalibration of the feedback parameters required to reproduce the GSMF at high resolution. The factor $\simeq 3$ uncertainty on the UVB photoionization rate mildly impacts the \HI\ column density of systems with $\log_{10} N_{\rm HI}\,[\cmsquared] \lesssim 20$, whilst the uncertainty on the atomic-to-molecular transition can significantly influence $\log_{10} N_{\rm HI}\,[\cmsquared] \gtrsim 21$ systems (Fig. \ref{fig:CDDF}).

\item In Ref-L100N1504, the \HI\ mass of central galaxies is a roughly linear function of $M_\star$ for $10^8 \lesssim M_\star \lesssim 10^{10.5}\Msun$. The relation turns over at higher masses, as the galaxy population becomes dominated by passive early-type galaxies hosting BHs that drive efficient AGN feedback. The high-resolution simulations are offset to higher $M_{\rm HI}$ at fixed $M_\star$ with respect to Ref-L100N1504, by up to a decade at $M_\star \simeq 10^{8.5}\Msun$ (Fig. \ref{fig:M_HI_convergence}, \textit{left}). This poor convergence behaviour is consistent with that of the \HI\ CDDF and is similar to the convergence behaviour of the gas-phase metallicity - stellar mass relation presented by \citet{S15}, with high $Z_{\rm gas}$ corresponding to low $M_{\rm HI}$.

\item The factor $\simeq 3$ uncertainty on the UVB photoionization rate has a negligible effect on the predicted \HI\ mass of galaxies, consistent with its influence being confined to low column density gas. The prescription for partitioning \HI\ and \Hmol, however, affects galaxies of all $M_\star$, particularly massive galaxies for which cold gas reservoirs exhibit high pressure and high metallicity (Fig. \ref{fig:M_HI_convergence}, \textit{right}).

\item The $M_{\rm HI}-M_\star$ relation of Ref-L100N1504 at $M_\star > 10^{10}\Msun$ is in good agreement with direct measurements from the GASS survey \citep[as shown by][]{Bahe_et_al_16_short}, but it declines more steeply than the relation recovered by the stacking analysis of \citet{Brown_et_al_15} over the interval $M_\star = 10^9 - 10^{10}\Msun$. The linear mean $M_{\rm HI}-M_\star$ relation of Ref-L100N1504 is offset by $-0.55\,{\rm dex}$ from Brown et al. measurement at $M_\star \simeq 10^9\Msun$. The high-resolution simulations are however in good agreement with the Brown et al. measurements in the region of overlap (Fig. \ref{fig:M_HI_obs}). 

\item The $M_{\rm HI}-M_\star$ relation evolves similarly for Ref-L100N1504 and Recal-L025N0752 at $z\gtrsim 1$. At later times, $M_{\rm HI}$ continues to decline at fixed $M_\star$ in Ref-L100N1504, but it barely evolves in Recal-L025N0752, indicating that the poor convergence of the $M_{\rm HI}-M_\star$ relation mostly develops at late cosmic epochs. Recal-L025N0752 therefore indicates that the CHILES survey `deep field', which probes the redshift interval $z\simeq0-0.5$, should not detect evolution of the characteristic \HI\ mass of galaxies. Owing primarily to the growth of the overall neutral gas mass, the \HI\ mass of galaxies increases significantly at $z>1$. In Recal-L025N0752, galaxies of $M_\star = 10^9\Msun$ $(10^{10}\Msun)$ exhibit \HI\ masses a factor of $4.2$ $(2.0)$ greater at $z=4$ than at $z=0$ (Fig. \ref{fig:M_HI_evol}).

\item At fixed $M_\star$, the \HI\ mass of star-forming galaxies is proportional to the feedback efficiency and inversely proportional to the SF efficiency, findings readily interpreted in the self-regulation framework. Greater feedback efficiency shifts galaxies of fixed $M_\star$ into more massive haloes with higher infall rates, requiring the accrual of more cold gas to generate the rates of SF and ejective feedback required to balance the infall. This shifts the $M_{\rm HI}-M_\star$ relation to the lower $M_\star$, as a fixed $M_{\rm HI}$ becomes associated with less massive galaxies. Increasing the SF efficiency enables a given infall rate to be balanced with less gas, shifting the relation to lower $M_{\rm HI}$. Whilst the SF (i.e. Kennicutt-Schmidt) law is well constrained by observations, the feedback efficiency is not, and must be calibrated (Fig. \ref{fig:M_HI_subgrid}).

\item The offset of the Ref-L100N1504 $M_{\rm HI}-M_\star$ relation with respect to observations translates into a poor reproduction of the observed HIMF in standard-resolution EAGLE simulations. The function is offset with respect to the HIPASS and ALFALFA Schechter functions by $\simeq -0.2$ to $-0.3\,{\rm dex}$ at the break scale, and the offset is greater at lower and higher masses. Due to the steep $M_{\rm HI}-M_\star$ relation, \HI\ systems of $M_{\rm HI} \gtrsim 10^{8.5}\Msun$ are associated with galaxies in a narrow stellar mass range, $M_\star \simeq 10^{9}-10^{10}\Msun$. The Ref-L100N1504 HIMF also features an unrealistic `bump' at $M_{\rm HI}\gtrsim 10^{8}-10^{8.5}\Msun$, contributed by `dark' subhaloes yet to begin self-regulating. The HIMF of the high-resolution simulations more closely resembles the observed functions and does not exhibit the unrealistic bump (Figs. \ref{fig:HIMF_convergence} \& \ref{fig:HIMF_Mstar}).

\item In spite of the relatively poor convergence of galaxy \HI\ masses at low redshift, on scales well-sampled by Ref-L100N1504 ($\sigma \lesssim 3 \Mpc$, where $\sigma$ is the `on-sky' separation of galaxies), the $z=0$ projected \HI\ two-point correlation function ($M_{\rm HI} > 10^{7.5}\Msun$) is broadly consistent with results from HIPASS and ALFALFA. The spatial distribution of (high column density) \HI\ in the simulations is therefore realistic. \HI-poor galaxies are found to cluster more strongly than \HI-rich counterparts, particularly on small scales, similar to the well-established dependence of clustering amplitude on the optical colour or sSFR of galaxies (Fig. \ref{fig:clustering}).

\item The \HI\ of present-day galaxies was mostly acquired via the smooth accretion of gas from the IGM; the contribution by mass of mergers (of subhaloes with mass at least 10 percent of the descendant) is typically $\ll 10$ percent. In galaxies of $M_\star \lesssim 10^{10.5}\Msun$, the gas that was accreted smoothly on to the main progenitor subhalo of present-day galaxies did so at $z\simeq 1$, whilst the small fraction delivered by mergers was typically acquired between $z=1.5$ and $z=2$. In more massive galaxies, accretion via both channels tends to occur at later epochs. Since the bulk of cosmic \HI\ at the present-day is associated with galaxies of $M_\star \sim 10^{9.5}-10^{10.5}\Msun$ \citep[e.g.][]{Zwaan_et_al_05a,Meyer_et_al_07}, the simulations indicate that the smooth accretion of diffuse gas from the IGM at $z \sim 1$ dominates the build-up of present-day \HI\ reservoirs (Fig. \ref{fig:reservoir_accretion}).

\item The \HI\ associated with galaxies at all epochs occupies a similar distribution in temperature-density space at earlier times, that avoids the high-entropy part of the diagram that is characteristic of adiabatically-cooling gas heated by accretion shocks and feedback. Very little of the \HI\ of present-day galaxies was therefore contributed by the reincorporation of strongly-heated gas, but this finding does not preclude the reincorporation of cooler gas entrained in outflows \citep[SF from ISM gas previously ejected in a cold state is a key feature of some galaxy formation simulations, e.g.][]{Oppenheimer_et_al_10}. The \HI\ of present-day galaxies was mostly in an ionized state at $z=1$, the epoch when the majority of it accreted on to the main progenitor subhalo of its present-day host galaxy. The \HI\ associated with galaxies at all epochs constitutes a dynamic reservoir that can be rapidly redistributed in temperature-density space. (Fig. \ref{fig:matrix}).

\end{enumerate}

The results presented here demonstrate that cosmological hydrodynamical simulations, when calibrated to reproduce the GSMF and disc sizes at $z\sim0$ (both properties of the stellar component of galaxies), do not automatically reproduce the observed \HI\ properties of galaxies. In comparison to scaling relations describing the stellar components of galaxies, the \HI\ properties of EAGLE galaxies are poorly converged with numerical resolution, in both the strong and weak regimes. Rather than signalling a major problem with the galaxy formation astrophysics implemented in the simulations, the shortcomings appear to be primarily a consequence of the limited resolution, because the key discrepancies with respect to observational measurements are much less severe in the high-resolution EAGLE simulations. Both high-resolution simulations studied here, Ref-L025N0752 and Recal-L025N0752, yield galaxies with reasonably realistic \HI\ properties, indicating that the difference in numerical resolution between the standard- and high-resolution simulations impacts upon the properties of cold gas in galaxies more strongly than does the recalibration of the subgrid feedback parameters necessary to reproduce the $z=0.1$ GSMF at high resolution. 

The particle mass of the standard-resolution simulations ($m_{\rm g} \sim 10^6\Msun$) was chosen to ensure that the Jeans scales are (marginally) resolved in the diffuse, photoionized phase of the ISM ($n_{\rm H} \lesssim 0.1\,\cmsquared, T \sim 10^{4}\K$). It is perhaps unsurprising then that the high-resolution simulations fare better in terms of modelling \HI\ reservoirs, particularly so for the low-mass galaxies that are the least well sampled by Lagrangian methods, since they can resolve mass (length) scales a factor of 8 (2) smaller than standard-resolution counterparts. Moreover, the implementation of feedback associated with the formation of stars and the growth of black holes is better sampled at high resolution, since individual heating events are a factor of 8 less energetic. Both factors are likely to influence the ability of the simulations to model the formation and evolution of \HI\ reservoirs faithfully. Simulating large cosmological volumes at equivalent resolution to the L025N0752 simulations is clearly extremely demanding (a volume with $L=100\cMpc$ would need to be sampled by $\simeq 3000^3 = 2.7\times 10^{10}$ fluid elements), but it is encouraging that the fundamental properties of observed \HI\ reservoirs are reproduced by simulations of that resolution, incorporating relatively simple treatments of the astrophysical processes thought to be relevant. For example, the EAGLE simulations do not explicitly model the cold interstellar gas phase, RT or non-equilibrium chemistry, and they only incorporate a single mode of feedback associated with SF, and a single mode of feedback associated with black hole growth.

The EAGLE simulation suite therefore represents an attractive means of examining the cold gas properties of galaxies, since the simulations rely on relatively few simplifying assumptions, and track the full 3D Lagrangian history of the gas. Besides enabling the examinations of the build-up and maintenance of \HI\ reservoirs presented here, we show in a companion paper \citep{Marasco_et_al_16} that it is necessary to account for the detailed interaction of galaxies with the neighbours in order to accurately reproduce key observed trends between the gas content of galaxies and their cosmic environment. The ability to trace self-consistently the life cycle of the gas participating in the galaxy formation process enables a number of questions to be addressed that are inaccessible to simplified modelling techniques. In particular, the results presented in Fig. \ref{fig:matrix} indicate that any gas entrained in feedback-driven outflows but subsequently reincorporated into the \HI\ reservoir of a galaxy cannot be strongly heated. This is conceptually at odds with the implementation of feedback in several popular semi-analytic models. In a follow-up study, we intend to investigate the baryon life cycle of EAGLE galaxies in greater detail to examine whether the continuity equations used by such models warrant revision.


\section*{Acknowledgements}  
\label{sec:acknowledgements}

RAC is a Royal Society University Research Fellow. CL is funded by a Discovery Early Career Researcher Award (DE150100618). The research was supported in part by the European Research Council under the European Union's Seventh Framework Programme (FP7/2007-2013) / ERC Grant agreements 278594-GasAroundGalaxies (JS) and 291531-HIStoryNU (AM and JMvdH) and the Interuniversity Attraction Poles Programme initiated by the Belgian Science Policy Office ([AP P7/08 CHARM]). We gratefully acknowledge PRACE for awarding us access to the Tier-0 Curie facility in Cycle-6, and the Dutch National Computing Facilities Foundation (NCF), with financial support from the Netherlands Organization for Scientific Research (NWO), for the use of supercomputer facilities. This study used the DiRAC Data Centric system at Durham University, operated by the Institute for Computational Cosmology on behalf of the STFC DiRAC HPC Facility (www.dirac.ac.uk). This equipment was funded by BIS National E-infrastructure capital grant ST/K00042X/1, STFC capital grant ST/H008519/1, and STFC DiRAC Operations grant ST/K003267/1 and Durham University. DiRAC is part of the National E-Infrastructure. The simulation data used in this study are available through collaboration with the authors.



\bibliographystyle{mnras}
\bibliography{bibliography} 



\begin{appendix}
\section{Methods for examining how and when gas accreted on to galaxies.}
\label{sec:tree_methods}
  
The examination how and when the \HI\ associated with present-day galaxies was accreted (see \S\ref{sec:reservoir_accretion}) requires the construction of merger trees. This enables the identification of the subhaloes (and galaxies) that, at each epoch, are the progenitors of those identified at later times. The procedure by which the standard EAGLE galaxy and subhalo merger trees\footnote{These data are described by \citet{McAlpine_et_al_16_short} and are available online at http://www.eaglesim.org/database.html} are generated is described in detail by \citet{Qu_et_al_16_short}. The procedure is based on the D-Trees algorithm \citep{Jiang_et_al_14}, the principle step of which is the tracking of subhaloes via the identification of the descendant subhalo comprising the greatest fraction of some number of a progenitor subhalo's most-bound particles. 

The standard EAGLE trees connect the 29 epochs for which snapshots, i.e. records of all particle properties, are saved. The examination of gas accretion on to galaxies, however, requires markedly superior temporal resolution than is afforded by this sampling. Fortunately, as described by \citet{S15}, the EAGLE simulations each also record 400 `snipshots', between redshifts 20 and 0, comprising fewer variables than found in snapshots. These data are sufficient to enable the identification of subhaloes with SUBFIND, and the partitioning of gas particles into \Hmol, \HI\ and \HII. Owing to the computational expense of applying SUBFIND to the outputs of very large simulations, only the 200 even-numbered snipshots of the Ref-L100N1504 simulation have been catalogued. This nevertheless enables the construction of merger trees with a mean (median) temporal resolution of $\simeq 67\,(55) \Myr$, significantly better than the $\simeq 470\,(350) \Myr$ afforded by the standard merger trees. 

The accretion of gas on to a galaxy can therefore be tracked by identifying the snipshot in which a gas particle first became associated with the galaxy's main progenitor subhalo. Examination of the accretion of the gas comprising the \HI\ reservoirs of present-day galaxies therefore proceeds by identifying the set of particles bound to central galaxies at $z=0$, residing within a spherical aperture of $70\pkpc$ radius, and with atomic fraction $f_{\rm HI} > 0.5$. Such particles account for $94$ percent of the total \HI\ mass density associated with galaxies at $z=0$ in both Ref-L100N1504 and Recal-L025N0752, and the mean \HI\ fraction of these particles is $86$ and $87$ percent, respectively, in those simulations. The aperture mimics the beam size of the ALFA instrument at the median redshift of the GASS survey (see \S\ref{sec:id_galaxies}). Those particles that, at the epoch immediately prior to that at which they first became bound to the main progenitor subhalo, were bound to a subhalo with mass at least $10$ percent of the total mass of the descendent, are classified as merger events. Particles that accreted whilst bound to a subhalo of less than $10$ percent of the descendent mass, or no subhalo at all, constitute smooth accretion. We exclude mergers of mass ratio $< 0.1$ in order to minimize the sensitivity of the calculation to numerical resolution. The main progenitor subhalo of a galaxy is defined as the subhalo with the most massive `main branch' since, in the case of near-equal mass mergers and as discussed by \citet{DeLucia_and_Blaizot_07}, this definition is not subject to the `noise' associated with definitions such as the instantaneous subhalo mass. 

\end{appendix}


\bsp	
\label{lastpage}
\end{document}